\documentclass[%
reprint,onecolumn,
superscriptaddress,
amsmath,amssymb,
aps,
,floatfix
]{revtex4-1}

\usepackage{dcolumn}
\usepackage{bm}

\usepackage{amsmath}
\usepackage{amssymb}
\usepackage{amsthm}
\usepackage{mathtools}
\usepackage{mathrsfs}
\usepackage{bm}
\usepackage{slashed} 
\usepackage{graphicx}
\usepackage{multirow}
\usepackage{tikz}
\usepackage[caption=false]{subfig}
\usepackage{relsize}	
\usepackage{array}
\usepackage{float}
\usepackage{color}
\usepackage{xcolor}
\usepackage{soul}
\usepackage{verbatim} 

\allowdisplaybreaks

\newcommand{\seq}{\begin{subequations}}
	\newcommand{\sen}{\end{subequations}}
\newcommand{\be}{\begin{eqnarray}}
	\newcommand{\ee}{\end{eqnarray}}
\newcommand{\eq}{\begin{eqnarray}}
	\newcommand{\en}{\end{eqnarray}}

\newcommand{\nn}{\nonumber}

\usepackage{booktabs}
\AtBeginDocument{
	\heavyrulewidth=.08em
	\lightrulewidth=.05em
	\cmidrulewidth=.03em
	\belowrulesep=.65ex
	\belowbottomsep=0pt
	\aboverulesep=.4ex
	\abovetopsep=0pt
	\cmidrulesep=\doublerulesep
	\cmidrulekern=.5em
	\defaultaddspace=.5em
}

\usepackage{relsize}
\begin{document}
	
	\preprint{APS/123-QED}
	\title{Real and virtual photons within basis light-front quantization}
	
	\author{Sreeraj Nair}
	\email{sreeraj@impcas.ac.cn}
	\author{Chandan Mondal}
	\email{mondal@impcas.ac.cn}
	\author{Xingbo Zhao}
	\email{xbzhao@impcas.ac.cn}
	\affiliation{Institute for Modern Physics, Chinese Academy of Sciences, Lanzhou-730000, China\\ 
		School of Nuclear Science and Technology, University of Chinese Academy of Sciences, Beijing 100049, China\\
		CAS Key Laboratory of High Precision Nuclear Spectroscopy, Institute of Modern Physics, Chinese Academy of Sciences, Lanzhou 730000, China}

	\author{Asmita Mukherjee }
	\email{asmita@phy.iitb.ac.in}
	\affiliation{Indian Institute of Technology Bombay, Powai, Mumbai 400076, India}
	
	\author{James P. Vary}
	\email{jvary@iastate.edu}
	\affiliation{Department of Physics and Astronomy, Iowa State University, Ames, Iowa 50011, USA}
	
	\collaboration{BLFQ Collaboration}
	
	\begin{abstract}

		We compute the structure function, transverse momentum dependent parton distributions (TMDs), and generalized parton distributions (GPDs) for the physical photon from the light-front quantum electrodynamics (QED) Hamiltonian, determined for its constituent bare photon and electron-positron Fock components within the  Basis Light-Front Quantization framework. After performing nonperturbative renormalization, we obtain a good quality description of the perturbative QED properties of the photon, as well as good agreement with the experimental data for the photon's structure function. We also investigate the TMDs and GPDs of the space-like and the time-like virtual photon by incorporating a nonzero photon mass.

	\end{abstract}

	\maketitle
	
	\section{Introduction }

	Unlike the \emph{bare} photon, which is the structureless gauge boson of QED, the \emph{resolved} photon can be endowed with a 
	structure owing to the energy-time uncertainty principle, which allows it to fluctuate into a charged fermion antifermion pair. 
	These quantum fluctuations allow the possibility of photon-photon  interactions. These interactions between two photons are experimentally accessed 
	by colliding energetic $e^{\pm}$ with $e^-$ beams~\cite{walsh}. The photon structure functions are probed in experiments 
	when one of the  photons has high virtuality ($Q^2 >> 0$) and the other is almost real ($P^2 \sim 0$). These scattering processes are analogous to deep inelastic scattering (DIS) experiments on a hadron but with a photon as the probed system. The point-like contribution to the photon structure 
	function can be calculated perturbatively~\cite{Nisius:1999cv}. On the other hand, the hadron-like part of the photon structure function involves nonperturbative contributions, which are usually parameterized using the vector meson dominance (VMD) model~\cite{Peterson:1982tt}. One of the seminal papers was written by Witten~\cite{Witten:1977ju},  which investigated the photon structure function in an asymptotically free gauge theory and studied the scale dependence. The photon structure function
	has been experimentally studied over a wide kinematic range starting from the first measurement by the PLUTO Collaboration~\cite{Berger:1981bh}.  The calculability of the photon observables provides a very good opportunity to use a photon as a testing ground for a light-front Fock-space based approach aimed at calculating detailed properties of the resolved hadron. In this work, we investigate a few key observables and  compare our results with estimates made in perturbative QED  as well as with available experimental data.
	
	For context, consider a hadron where GPDs were introduced in relation to  deeply virtual Compton scattering (DVCS)~\cite{Ji:1996nm,Radyushkin:1996nd,Mueller:1998fv}, and  are rich in information since they contain combined information about the form factors and the parton distribution functions (PDFs). GPDs reduce to ordinary PDFs in the forward limit of zero momentum transfer and, when integrated over the longitudinal momentum fraction $x$, they give the corresponding hadron form factor. These GPDs have been studied extensively both theoretically as well as experimentally from DVCS data on a hadron target~\cite{gpd_all}. 
	In a similar vein, the GPDs for the photon were introduced  in Ref.~\cite{Friot:2006mm}, where  the authors considered  DVCS ($\gamma^* \gamma \rightarrow \gamma \gamma$) taking the nearly real photon as a photon target. Their calculation was done at leading order in the electro-magnetic coupling $\alpha_{\mathrm{em}}$ and zeroth order in the strong coupling $\alpha_{\mathrm{s}}$ by considering all the amplitudes corresponding to the Born order diagrams for $\gamma^* \gamma \rightarrow \gamma \gamma$. 
	
	The first calculation for the photon GPDs with non-zero transverse momentum transfer was presented in Ref.~\cite{Mukherjee:2011bn}. This perturbative calculation was based on a light-front Hamiltonian framework, where the Fock space expansion of the photon state was truncated at the two-particle ($q\bar{q}$) sector. Both the polarized and the unpolarized photon GPDs were obtained using overlaps of photon light-front wave functions (LFWFs). The photon GPDs for non-zero skewness using the same approach were reported in Refs.~\cite{Mukherjee:2011an,Mukherjee:2013yf}.  
	
	Somewhat less explored are the transverse momentum dependent distribution functions (TMDs) for the photon. For the hadron, these are objects of substantial interest; these give the distribution of quarks and gluons in the hadron in three-dimensional momentum space~\cite{Angeles-Martinez:2015sea,Barone:2001sp,Accardi:2012qut}. Experimentally, the TMDs are accessed via  semi-inclusive reactions like the semi-inclusive deep inelastic scattering (SIDIS)~\cite{Brodsky:2002cx,Bacchetta:2017gcc} and Drell-Yan processes~\cite{Ralston:1979ys,Donohue:1980tn,Tangerman:1994eh}. At leading twist for a spin-half hadron, like the nucleon, there are eight quark TMDs, out of which six are time-reversal even (T-even) and two are time-reversal odd (T-odd). For spin-one hadrons, like the deuteron or vector mesons, there are additional TMDs following the increase in the spin degrees of freedom~\cite{Hoodbhoy:1988am,Hino:1999qi,Bacchetta:2000jk}. A model calculation with a covariant formalism for the leading-twist time-reversal even TMD for a spin-one target was shown in Ref.~\cite{Ninomiya:2017ggn}.
	
	In this work, we utilize the Basis Light-Front Quantization (BLFQ) approach for calculating the observables for the photon and comparing them with the results from the perturbative approach. BLFQ is a nonperturbative approach for solving bound state problems \cite{blfq1}, which uses the \emph{front-form} of relativistic dynamics.
	BLFQ employs the light-front Hamiltonian~\cite{brodsky1}, which is diagonalized in a truncated Fock space to obtain the stationary states. So far, BLFQ has been successfully applied to problems within QED such as the electron anomalous magnetic moment~\cite{maris,zhao}, the strong coupling bound-state positronium problem~\cite{li}, and the GPDs~\cite{zhao2} and the TMDs~\cite{Hu:2020arv} of the physical electron and the real photon~\cite{Nair:2022evk}. More recently, BLFQ has been employed to solve light mesons~\cite{Jia:2018ary,Lan:2019vui,Lan:2019rba,Adhikari:2021jrh,Lan:2021wok,Mondal:2021czk}, heavy quarkonia~\cite{Li:2015zda,Li:2017mlw,Li:2018uif,Lan:2019img}, heavy-light mesons~\cite{Tang:2018myz,Tang:2019gvn}, nucleon 
	\cite{Xu:2019xhk,Xu:2021wwj,Liu:2022fvl,Hu:2022ctr}, heavy baryons~\cite{Peng:2022lte}, and the
	all-charm tetraquark~\cite{Kuang:2022vdy}  as QCD bound states.
	One of the primary goals of this work is to compare the results of the BLFQ approach with the perturbative results for the photon, in order to determine how well this approach can describe a relativistic composite system and to verify the renormalization procedure. We utilize the sector-dependent renormalization technique \cite{Karmanov:2008br,Karmanov:2012aj} and the wavefunction scaling approach to mitigate the effects of the artifacts originating with Fock space truncation. Thus our work serves as a test for the methodology employed in solving bound state problems within the BLFQ approach. In addition, we anticipate that our approach will be useful for comparing theory with future experiments such as those at the Electron-Ion Collider (EIC).
	
	We organize the paper by first introducing in Sec.~\ref{secblfq} the basic methodology involved in solving any stationary state problem in BLFQ. Then in Sec.~\ref{secrenorm}, we discuss briefly the strategy used for the renormalization and wavefunction rescaling. In Sec.~\ref{secobs}, we show the formulation of various observables that we calculate within BLFQ and compare them with the corresponding perturbative formulation. In Sec.~\ref{secnum}, we present our numerical results for various observables of the real and the virtual photons and show the comparisons with the perturbative results and with experimental data. Finally, we summarize our work in Sec.~\ref{con}.
	
	\section{Basis Light-Front Quantization}
	\label{secblfq}
	
	BLFQ aims at solving the following eigenvalue equation to obtain the mass spectrum and the LFWFs:
	\be
	\mathrm{H_{LC}} \mid \psi \rangle &=& M^2 \mid \psi \rangle ,
	\label{eveq}
	\ee
	\noindent
	where $\mathrm{H_{LC}}  =  P^+ P^- - \left(P^{\perp}\right)^2 $ is the light-front Hamiltonian and the operators $P^+ $, $P^-$, and $P^{\perp}$ are the longitudinal momentum, the light-front quantized Hamiltonian, and the transverse momentum, respectively.
	The diagonalization of Eq.~(\ref{eveq}) using a suitable matrix representation for the Hamiltonian generates the invariant-mass spectrum ($M$) and the 
	light-front state vectors $\left(\mid \psi \rangle \right)$. The basis states for BLFQ are expanded in terms of Fock state sectors, where each basis state has longitudinal, transverse, and spin degrees of freedom. The longitudinal degrees of freedom are discretized by imposing an antiperiodic (periodic)
	boundary conditions for the fermions (bosons) within a box of length $2L$ such that $-L \le x^- \le +L$. $P^+$ is further parameterized as $K = \sum_i \kappa_{i}$ such that $P^+ = 2 \pi K/L$. The longitudinal momentum fraction of the $i^{\mathrm{th}}$ parton is then defined as $x_i = k_i^+/P^+ = \kappa_{i}/K$. The sums run over all partons in each many-parton Fock space basis state.
	The longitudinal momenta are discretized by fixing the value of $K$. The parameter $K$ can be seen as the ``resolution" in the longitudinal direction, and hence a resolution on the PDFs. When $K \rightarrow \infty$, we approach the continuum limit in the longitudinal direction. Two-dimensional (2D) harmonic oscillator (HO) modes are chosen as the basis states for the transverse degrees of freedom.
	The HO states are specified by $n$, $m$, and $\Omega$, where $n$ and $m$ are the principal and the orbital angular quantum numbers, respectively and $\Omega$ represents the HO energy. The HO wavefunctions are conveniently represented in terms of a dimensionless parameter $\rho = |p^{\perp}| /b_0$, where the scale parameter $b_0\equiv\sqrt{M_0 \Omega}$ and $M_0$ has mass dimension. We use an $x$-dependent scale parameter $b$ such that $b=b_0\sqrt{x(1-x)}$ consistent with the coordinate adopted in Refs.~\cite{li,maris} where $x$ is taken to be the electron's light-front momentum fraction.
	When expressed using polar coordinates $(\rho,\phi)$, the HO wavefunctions can be written in terms of generalized Laguerre polynomials $L_n^{|m|}\left(\rho^2\right)$,
	\be
	\phi_{nm}\left( p^{\perp} \right) = \sqrt{\frac{2\pi}{M_0\Omega}}\sqrt{\frac{2n!}{(|m|+n)!}}e^{im\varphi} \rho^{|m|}e^{-\rho^2/2}~L_n^{|m|}\left(\rho^2\right) .
	\label{wavefn}\ee

	The continuum limit in the transverse direction is dictated by the parameter $N_{\rm max} \rightarrow \infty$, where $N_{\mathrm{max}} \ge \sum_i \left(2n_i + |m_i|+1 \right)$. The ultra-violet (UV) and the infra-red (IR) cutoffs in the transverse direction are also determined by $N_{\rm max}$. In this work, we study the photon by truncating the Fock space expansion at the two-particle sector. At fixed light-front time, the photon
	state can be expressed schematically as follows:
	\be
	\mid \gamma_{\mathrm{phy}} ~\rangle =   \mid \gamma ~\rangle +  \mid e^+ e^- ~\rangle .
	\label{fock}
	\ee
	The full light-front QED Hamiltonian that we diagonalize can be written as
	\be
	\hat{H} = \hat{H}_{\mathrm{QED}}  + \hat{H'},
	\label{fullhami}
	\ee
	where $\hat{H}_{\mathrm{QED}} = P^+ \hat{P}^{-}_{\mathrm{QED}} - \left(\hat{P}^{\perp}\right)^2$. The light-front QED Hamiltonian $\hat{P}^{-}_{\mathrm{QED}}$ relevant to the photon's leading two Fock sectors in the light-cone gauge $A^+ = 0$ is~\cite{Nair:2022evk,Zhao:2014xaa}
	
	\be
	\hat{P}^{-}_{\mathrm{QED}} =
	\int dx^- d^2x^{\perp}\left[
	\frac{1}{2}\bar{\Psi}(x)
	\gamma^+ \frac{m_e^2 + \left(i\partial^{\perp}\right)^2}{i\partial^+}
	\Psi(x) + \frac{1}{2} A^k \left( i\partial^{\perp}\right)^2 A^k
	\right] + 
	V_{\mathrm{int}},
	\label{eff}\ee
	where $m_e$ is the electron mass, $\Psi$ and $A_{\mu}$ are the fermion and the gauge boson fields, respectively. Note that with the leading two Fock sectors, the instantaneous-fermion interaction does not contribute. On the other hand,
	the instantaneous-photon interaction (IPI)  contributes to the overall renormalization factor, thus the intrinsic structure of the physical photon remains unaffected.
	The IPI must be accompanied by the explicit photon exchange contribution from higher Fock sectors (which are not present in this work) in order to cancel the small-$x$ divergences \cite{Zhao:2014xaa}. Therefore, we choose to exclude the IPI term 
	from the Hamiltonian. The first two terms correspond to the kinetic energy of the electron and the photon, respectively, while the last term represents the interaction vertex ($V_{\mathrm{int}}$) for a photon creating a fermion anti-fermion pair.
	The interaction term $V_{\mathrm{int}}$ in Eq.~(\ref{eff}) is given by
	\be
	V_{\mathrm{int}} = e \int dx^- d^2x^{\perp} \bar{\Psi}(x) \gamma^{\mu} \Psi(x)A_{\mu}(x)\Big{|}_{x^+=0},
	\ee
	where $e$ is the physical electromagnetic coupling constant. Lastly, we need to introduce a Lagrange multiplier ($\lambda$) to the Hamiltonian to separate the center-of-mass (CM) motion from the intrinsic motion. The CM motion is involved because of the use of single-particle coordinates and the HO basis states coupled with $N_{\mathrm{max}}$ truncation allow for the factorization of CM and intrinsic motions. The factorization is achieved numerically by adding a Lagrange multiplier term proportional to the HO Hamiltonian that acts only on the CM, $H_{\rm CM}$, with positive coefficient $\lambda$~\cite{zhao},
	\be
	\hat{H'} = \lambda \left(\hat{H}_{\mathrm{CM}} - 2b^2 I\right).
	\label{hamifinal}
	\ee
	The CM excitations are shifted up by the Lagrange multiplier term. The value of $\lambda$ is chosen large enough such that the low-lying spectra of interest are not affected by the CM excitation~\cite{li}. More details on the CM factorization in BLFQ can be found in Ref.~\cite{Wiecki:2014ola}.
	\section{Photon Mass Renormalization and Rescaling}
	\label{secrenorm} 
\begin{figure}[htp!]
		\centering
		\includegraphics[width=8.4cm,height=6.5cm,clip]{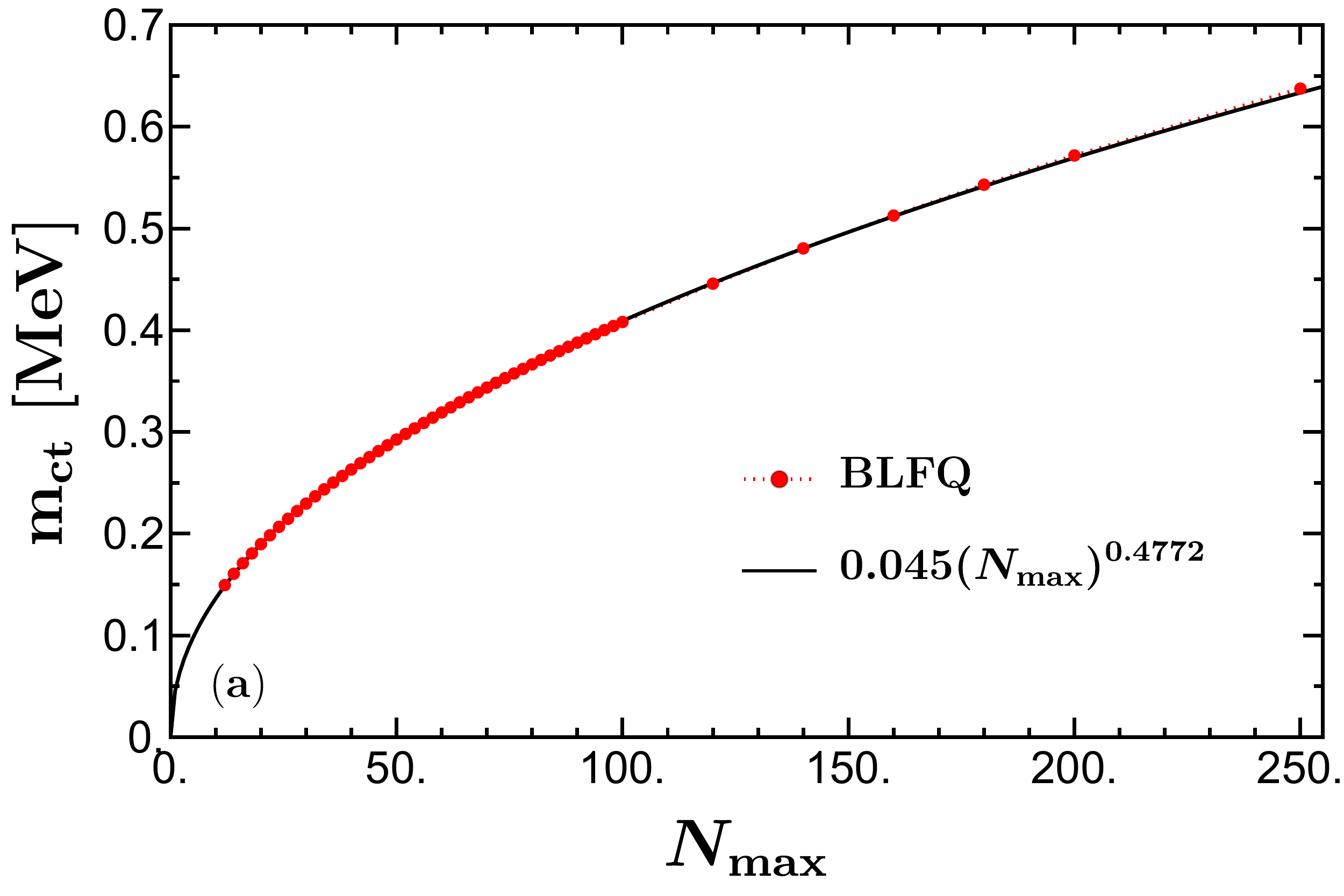}
		\includegraphics[width=8.4cm,height=6.5cm,clip]{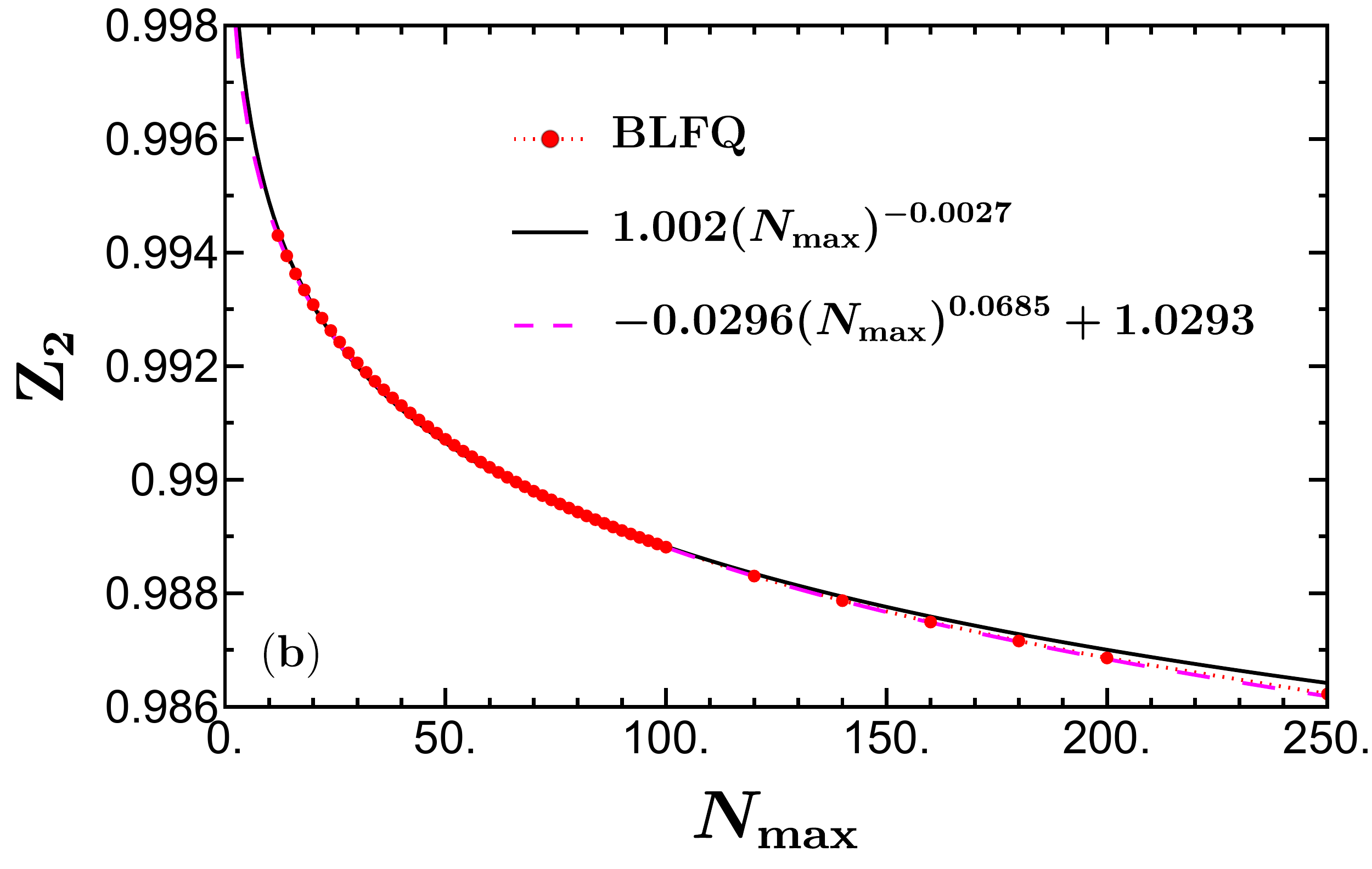}
		\caption{\label{fig1} Plot (a) is for the mass counter term $ {\mathrm{m_{ct}}}$ vs. the basis truncation parameter $N_{\mathrm{max}}$ and plot (b) is for the  wave function renormalization factor $Z_2$ vs. $N_{\mathrm{max}}$. Both results are calculated with $N_{\mathrm{max}} = K$ and $b_0=m_e=0.511 ~\mathrm{MeV}$.} 
	\end{figure}

The photon mass needs to be renormalized and this is achieved by employing a sector-dependent renormalization procedure~\cite{Karmanov:2008br,Karmanov:2012aj}. This renormalization procedure is performed numerically by implementing a root-finding algorithm to obtain the value of the mass counter term ($ {\mathrm{m_{ct}}}$), which is added to the bare photon mass ($m_0$) such that the ground state eigenvalue of the full Hamiltonian becomes equal to the physical photon mass ($m_\gamma$). This mass counter term ${\mathrm{m_{ct}}} = m_\gamma - m_0$ represents the numerical mass correction required to obtain the mass squared, $m_\gamma^2$, of the physical photon system.  For the real photon $m^2_\gamma=0$ and for the time-like (space-like) virtual photon $m^2_\gamma > 0 ~(m^2_\gamma<0)$.
	The mass counter-term is a function of the truncation parameters $N_{\mathrm{max}}$ and $K$. It is expected to increase \cite{brodsky1} with increasing value of the truncation parameters as can be seen in Fig.~\ref{fig1} (a).
	The act of truncating the Fock sector expansion results in the violation of the Ward-identity and hence, we introduce a rescaling factor to repair this violation following the similar procedure as employed for the physical electron~\cite{Zhao:2014hpa,Zhao:2014xaa,Chakrabarti:2014cwa,Brodsky:2004cx}.
	We rescale the naive photon observables $\mathcal{O}$ by the rescaling factor ($Z_2$) to obtain the rescaled observables $\mathcal{O_\mathrm{rs}}$ such that
	\be
	\mathcal{O_\mathrm{rs}} = \frac{\mathcal{O}}{Z_2},
	\label{rs}
	\ee
	where
	\be
	Z_2 = \sum_{|\gamma\rangle}  \Big{|} \Big{\langle} \gamma \Big{|} \gamma_{\mathrm{phys}} \Big{\rangle}  \Big{|}^2 .
	\label{z2}
	\ee
	The sum shown in Eq.~(\ref{z2}) runs over all basis states present in the photon Fock sector $\left(|\gamma\rangle \right)$. The photon rescaling factor $Z_2$ can be interpreted as the probability of finding a bare photon inside a physical photon. The behavior of the rescaling factor $Z_2$ with increasing basis truncation parameters is shown in Fig.~\ref{fig1} (b).
	
	The $N_{\mathrm{max}}$ values used in the fitting shown in Fig.~\ref{fig1} are over an unequally spaced grid where 
	the first grid starting from $N_{\mathrm{max}} = 10$ up to $N_{\mathrm{max}} = 100$ is in steps of one and the second grid from $N_{\mathrm{max}} = 100$ 
	up to $N_{\mathrm{max}} = 200$ is in steps of twenty and the final value being at $N_{\mathrm{max}} = 250$. 
	The fitting was done with equal weight for all the points on the grid. 
	A power law function of the form $f_0\left(N_{\mathrm{max}}\right) = a_0 \left(N_{\mathrm{max}}\right)^{\beta_0} $ was chosen as the fitting function for the mass-counter term values in  Fig.~\ref{fig1} (a). The wave-function renormalization factor $Z_2$ was fitted with two functions of the form $f_1\left(N_{\mathrm{max}}\right) = a_1 \left(N_{\mathrm{max}}\right)^{\beta_1}$ and $f_2\left(N_{\mathrm{max}}\right) = a_2 \left(N_{\mathrm{max}}\right)^{\beta_2} + c_2$. The best fit values for all these parameters were found to be $a_0 = 0.045$ , $\beta_0 = 0.4772$, $a_1 = 1.002$ , $\beta_1 = -0.0027$, $a_2 = -0.0296$ , $\beta_2 = 0.0685$ and $c_2 = 1.0293$.

	\section{Photon Observables}
	\label{secobs}
	
	We consider the physical (dressed) photon as a spin-one composite particle, with a bare photon and an electron-positron pair that appears from quantum fluctuation as its partons. We compute the TMDs and the GPDs of the electron (positron) inside the physical photon, as well as the structure function of the photon.  We further investigate the effect of nonzero photon mass on the observables. The unpolarized $f^1_{\gamma}$ and the polarized $g^{1L}_{\gamma}$ TMDs for a spin-one target, under the approximation that the gauge link is the identity operator, are given by the following correlators at uniform light-front time $y^+=0$, 
	\be
	f^1_{\gamma}\left(x,\left(k^{\perp}\right)^2\right) &=& \int \frac{dy^- d^2y^{\perp}}{(2\pi)^3}e^{ik.y} \Big{\langle} \gamma\left(P,\Lambda\right) \Big{|} \bar{\psi}(0)\gamma^+ \psi(y^-) \Big{|} \gamma\left(P,\Lambda\right) \Big{\rangle}\Big|_{y^+=0},
	\label{tmdeqf1} \nn \\ 
	g^{1L}_{\gamma}\left(x,\left(k^{\perp}\right)^2\right) &= &\int \frac{dy^- d^2y^{\perp}}{(2\pi)^3}e^{ik.y} \Big{\langle} \gamma\left(P,\Lambda\right) \Big{|} \bar{\psi}(0)\gamma^+ \gamma_5 \psi(y^-) \Big{|} \gamma\left(P,\Lambda\right) \Big{\rangle}\Big|_{y^+=0}.
	\label{tmdeqg1l}
	\ee
	The physical photon state $\mid \gamma(P,\Lambda) \rangle$ with momentum $P$ and light-front helicity $\Lambda$ can be expressed by considering all
	Fock components as~\cite{Brodsky:2000xy}
	\be
	\Big{|} \gamma(P,\Lambda) \Big{\rangle} &=& \sum_n \sum_{\lambda_1...\lambda_n} \int 
	\prod_{i=1}^n \left[ \frac{dx_i d^2k_i^{\perp}}{\sqrt{x_i}16\pi^3}\right] 16\pi^3 \delta \left( 
	1- \sum_{i=1}^n x_i
	\right) \nn \\ && \delta^2 \left( \sum_{i=1}^n k_i^{\perp}\right) \psi^{\Lambda}_{\lambda_1...\lambda_n}\left(
	x_i,k_i^{\perp}
	\right) \Big{|} n,x_i P^+, x_i P^{\perp} + k_i^{\perp}, \lambda_i\Big{\rangle} ,
	\label{multi}\ee
	where $x_i = \frac{k_i^+}{P^+}$ and $k_i^{\perp}$ are the longitudinal momentum fraction and the relative transverse momentum of the $i^{\mathrm{th}}$ parton, respectively. $\lambda_i$ is the light-front helicity of the parton and $n$ denotes the number of particles in a Fock state. The physical transverse momentum of the parton is given by $p_i^{\perp} = x_i P^{\perp} + k_i^{\perp}$ and the physical longitudinal momentum is $p_i^+ = k_i^+ = x_iP^+$. The LFWFs $\psi^{\Lambda}_{\lambda_1...\lambda_n}$ are boost-invariant and depend only on $x_i$ and $k_i^{\perp}$.
	The BLFQ LFWFs are generated by solving the eigenvalue problem of the Hamiltonian defined in Eq.~(\ref{fullhami}) using the plane wave and the harmonic oscillator basis functions with the truncated Fock sectors mentioned in Eq.~(\ref{fock}). For the second Fock sector,  the corresponding component of the  LFWF in the momentum space can then be expressed as follows:
	\be
	\Psi_{\lambda_e,\lambda_{\bar{e}}}^{\Lambda}\left(x_e,p_e^{\perp},x_{\bar{e}},p_{\bar{e}}^{\perp}\right) =
	\sum_{\substack{n_e,m_e \\ n_{\bar{e}},m_{\bar{e}}}} \Big[\Psi\left( \alpha_e,\alpha_{\bar{e}}\right) \phi_{n_e,m_e }
	\left( p_e^{\perp} \right)\phi_{n_{\bar{e}},m_{\bar{e}}}\left( p_{\bar{e}}^{\perp} \right)\Big],
	\label{single}\ee 
	where $\alpha_i = \left( \kappa_i,\,n_i,\,m_i,\,\lambda_i\right)$ is the compact notation which encompasses all the four quantum numbers associated with a single particle basis state namely the longitudinal degrees of freedom ($\kappa_i$), the two transverse degrees of freedom ($m_i,n_i$) and the spin degree of freedom ($\lambda_i$). $\Psi\left( \alpha_e,\alpha_{\bar{e}}\right) = \langle \alpha_e,\alpha_{\bar{e}} \mid P,\Lambda \rangle$ are the components of the eigenvectors generated from the diagonalization of the Hamiltonian. These single-particle momentum dependent wavefunctions, Eq.~(\ref{single}), contain CM excitations, which need to be factored out. After factorizing out the CM excitations we end up with corresponding components of the LFWFs in the relative coordinates $\Psi_{\lambda_e,\lambda_{\bar{e}}}^{\Lambda}\left(x_e,k_e^{\perp},x_{\bar{e}},k_{\bar{e}}^{\perp}\right) $ relevant for Eq.~(\ref{multi}). The unpolarized and the polarized TMDs can  be expressed as overlaps of the relative momentum dependent wavefunctions as,
	\be
	f^1_{\gamma}(x,(k^{\perp})^2) &=& \frac{1}{2}\int [de\bar{e}]\sum_{\Lambda,\lambda_e,\lambda_{\bar{e}}}
	\Psi_{\lambda_e,\lambda_{\bar{e}}}^{*\Lambda}\left(x_e,k_e^{\perp},x_{\bar{e}},k_{\bar{e}}^{\perp}\right)
	\Psi_{\lambda_e,\lambda_{\bar{e}}}^{\Lambda}\left(x_e,k_e^{\perp},x_{\bar{e}},k_{\bar{e}}^{\perp}\right), \\ \nn
	g^{1L}_{\gamma}(x,(k^{\perp})^2) &=& \frac{1}{2}\int [de\bar{e}]\sum_{\Lambda,\lambda_e,\lambda_{\bar{e}}} \lambda_e
	\Psi_{\lambda_e,\lambda_{\bar{e}}}^{*\Lambda}\left(x_e,k_e^{\perp},x_{\bar{e}},k_{\bar{e}}^{\perp}\right)
	\Psi_{\lambda_e,\lambda_{\bar{e}}}^{\Lambda}\left(x_e,k_e^{\perp},x_{\bar{e}},k_{\bar{e}}^{\perp}\right),
	\label{tmdeq}
	\ee
	where
	\be
	[de\bar{e}] = \frac{dx_{e}dx_{\bar{e}} d^2k_{e}^{\perp}  d^2k_{\bar{e}}^{\perp}}{2(2\pi)^3} \delta\left(x_{e} + x_{\bar{e}}-1\right) 
	\delta^2\left(k_{e}^{\perp} + k_{\bar{e}}^{\perp}\right) \delta \left(x - x_{e}\right) \delta^2\left( k^{\perp} - k^{\perp}_{e}\right).
	\ee
	
	On the other hand, the unpolarized and the polarized GPDs for the dressed photon can be defined through the following nonforward matrix  elements~\cite{Friot:2006mm}:
	\be
	F_{\gamma}(x,t) &=& \int \frac{dy^-}{8\pi}e^{\frac{-iP^+ y^-}{2}} \Big{\langle} \gamma(P',\Lambda) \Big{|} \bar{\psi}(0)\gamma^+ \psi(y^-) \Big{|} \gamma(P,\Lambda) \Big{\rangle}\Big|_{y^+=0,y^{\perp} =0} ,\\ \nn 
	\tilde{F}_{\gamma}(x,t) &=& \int \frac{dy^-}{8\pi}e^{\frac{-iP^+ y^-}{2}} \Big{\langle} \gamma(P',\Lambda)  \Big{|} \bar{\psi}(0)\gamma^+ \gamma_5 \psi(y^-)  \Big{|} \gamma(P,\Lambda) \Big{\rangle}\Big|_{y^+=0,y^{\perp} =0}.
	\label{gpdeqmain}\ee
	We choose a frame, where the initial and final four momenta of the photon are given by
	\be
	P &=& \left( P^+ , 0^{\perp} , 0 \right) ,\nn \\ 
	P' &=& \left( P^+ , -\Delta^{\perp} , \frac{\left(\Delta^{\perp}\right)^2}{P^+} \right).
	\ee
	Therefore, the momentum transferred to the photon is
	\be
	\Delta = P - P' = \left( 0, \Delta^{\perp} , \frac{t}{P^+} \right),
	\ee
	with  $t = -\left(\Delta^{\perp}\right)^2$ being the square of the momentum transfer in the transverse direction. 
	The overlap representation for the photon GPDs reads,
	\be
	F_{\gamma}(x,t) &=& \frac{1}{2}\int \{de\bar{e}\} \sum_{\lambda_e,\lambda_{\bar{e}}}
	\Psi_{\lambda_e,\lambda_{\bar{e}}}^{*\uparrow}\left(x_e,k_e^{\perp}-(1-x)\Delta^{\perp},x_{\bar{e}},k_{\bar{e}}^{\perp}+x\Delta^{\perp}\right)
	\Psi_{\lambda_e,\lambda_{\bar{e}}}^{\uparrow}\left(x_e,k_e^{\perp},x_{\bar{e}},k_{\bar{e}}^{\perp}\right), \nn \\
	\tilde{F}_{\gamma}(x,t) &=& \frac{1}{2}\int \{de\bar{e}\} \sum_{\lambda_e,\lambda_{\bar{e}}}
	\lambda_e\Psi_{\lambda_e,\lambda_{\bar{e}}}^{*\uparrow}\left(x_e,k_e^{\perp}-(1-x)\Delta^{\perp},x_{\bar{e}},k_{\bar{e}}^{\perp}+x\Delta^{\perp}\right)
	\Psi_{\lambda_e,\lambda_{\bar{e}}}^{\uparrow}\left(x_e,k_e^{\perp},x_{\bar{e}},k_{\bar{e}}^{\perp}\right) 
	\label{gpdeq},
	\ee
	where
	\be
	\{de\bar{e}\} &=& \frac{dx_{e}dx_{\bar{e}} d^2k_{e}^{\perp}  d^2k_{\bar{e}}^{\perp}}{2(2\pi)^3} \delta(x_{e} + x_{\bar{e}}-1) \delta^2\left(k_{e}^{\perp} + k_{\bar{e}}^{\perp}\right) \delta (x - x_{e}) 
	.\ee
	
	We now briefly describe how the perturbative results are obtained.  The results in perturbation theory are calculated using the two-component form of light-front field theory~\cite{Zhang:1993dd}, wherein the component $A^-$  
	of the photon field is eliminated by choosing $A^+ =0$ gauge. Thus, only the transverse component of the photon field $A^{\perp}$ survives. The $\psi^-$ fermionic field is constrained and written in terms of $\psi^+$~\cite{Zhang:1993dd}. The photon state can be written in terms of its Fock components~\cite{Friot:2006mm}. We truncate the Fock space expansion at the two particle sector :
	\be
	\label{fockpert}
	\Big{|} \gamma (P,\Lambda)  \Big{\rangle}&=& a^{\dagger}\left(P,\Lambda \right) \Big{|}  0 \Big{\rangle} \\ \nn &+& 
	\sum_{\lambda_e,\lambda_{\bar{e}}} \int {dp_{e}} \int {dp_{\bar{e}}} 
	\sqrt{2(2\pi)^3 P^+} \delta^3\left(P-p_{e}-p_{\bar{e}}\right) \Phi^{\Lambda}_{\lambda_e,\lambda_{\bar{e}}}\left(p_{e},p_{\bar{e}}\right) b^{\dagger}\left(p_{e},\lambda_e\right)d^{\dagger}\left(p_{\bar{e}},\lambda_{\bar{e}}\right) \Big{|}  0\Big{\rangle} ,
	\ee
	where we use the abbreviation ${dp_j} = \frac{dp_j^+ d^2p_j^{\perp}}{\sqrt{2(2\pi)^3p_j^+}}$ such that $j \in \{ e, \bar{e} \}$.  The two particle LFWF written in terms of the physical momenta, $\Phi^{\Lambda}_{\lambda_e,\lambda_{\bar{e}}}\left(p_{e},p_{\bar{e}}\right)$ represents the Fock component containing one electron and a positron;  the two-particle LFWF can be expressed in terms of  the Jacobi variables $x_j = \frac{k_j^+}{P^+}$ and $k_j^{\perp} = p_j^{\perp} - x_j P^{\perp}$ as follows \cite{kundu1}:
	\be
	\Psi_{\lambda_e,\lambda_{\bar{e}}}^{\Lambda}\left(
	x_e,k^{\perp}_e
	\right) &=& \frac{1}{m_{\gamma}^2+ \frac{m_e^2+\left(k^{\perp}_e\right)^2}{x_e(1-x_e)}}\frac{e}{\sqrt{2(2\pi)^3}}\chi_{\lambda_e}^{\dagger}\Bigg[
	\frac{\left(\sigma^{\perp}.k^{\perp}_e\right)}{x_e}\sigma^{\perp} -
	\sigma^{\perp} \frac{\left(\sigma^{\perp}.k^{\perp}_e\right)}{1-x_e} -i \frac{m_e}{x_e(1-x_e)}\sigma^{\perp} \Bigg]\chi_{-\lambda_{\bar{e}}}\epsilon_{\Lambda}^{\perp*},
	\label{tpwf}\ee
	
	such that $ x_e + x_{\bar{e}} =1 $ , $k_e^{\perp} + k_{\bar{e}}^{\perp}  = 0$ and the two particle LFWF with physical momenta shown in Eq.~\ref{fockpert} is related to the LFWF with relative Jacobi momenta in Eq.~\ref{tpwf} by the Jacobian, $\sqrt{P^+}$, of the transformation such that $\sqrt{P^+}\Phi^{\Lambda}_{\lambda_e,\lambda_{\bar{e}}}\left(p_{e},p_{\bar{e}}\right) =\Psi_{\lambda_e,\lambda_{\bar{e}}}^{\Lambda}\left(
	x_e,k^{\perp}_e \right)$.  The wavefunction in Eq.~\ref{tpwf} is written in the two component formalism~\cite{Zhang:1993dd,kundu1}. The two component fermionic spinor is denoted by $\chi_{\lambda_j}$ such that $\chi_{+1} =\frac{1}{\sqrt{2}} \begin{pmatrix} 1 \\ 0 \end{pmatrix}$ and $\chi_{-1} =\frac{1}{\sqrt{2}} \begin{pmatrix} 0 \\ 1 \end{pmatrix}$. $\epsilon_{\pm1}^{\perp} =\frac{1}{\sqrt{2}}\left( \mp 1 ,- i \right)$ is the polarization vector of the photon and $\sigma^{\perp}$ are the usual $2 \times 2$ complex Pauli matrices.

	The explicit analytical expressions for the photon observables using the perturbative wavefunction are given by
	\be
	\label{gpd_pert}
	F_{\gamma}(x,t) &=& \frac{e^2}{8\pi^3} \Big[
	\left(
	(1-x)^2+x^2
	\right)(I_1+I_2+LI_3) +2 m_e^2 I_3
	\Big],  \nn\\
	\tilde{F}_{\gamma}(x,t) &=& \frac{e^2}{8\pi^3} \Big[
	\left(
	x^2-(1-x)^2
	\right)(I_1+I_2+LI_3) +2 m_e^2 I_3
	\Big], 
	\ee
	and
	\be
	\label{perttmd}
	f^1_{\gamma}\left(x,\left(k^{\perp}\right)^2\right) &=& \frac{e^2}{8\pi^3} \frac{\left(m_e^2 + (k^{\perp})^2 (2x^2-2x+1)\right)}{\left( (k^{\perp})^2 +m_{\gamma}^2 x(1-x) +m_e^2\right)^2}, \nn \\
	g^{1L}_{\gamma}\left(x,\left(k^{\perp}\right)^2\right) &=& \frac{e^2}{8\pi^3} \frac{\left(m_e^2 + (k^{\perp})^2 (2x-1)\right)}{\left( (k^{\perp})^2 +m_{\gamma}^2 x(1-x) +m_e^2\right)^2},
	\ee 
	with the integrals
	\be
	I_1 &=& I_2 = \int \frac{d^2 k^{\perp}}{D} = \pi ~\mathrm{Log} \Bigg[
	\frac{\Lambda_{\mathrm{pert}}^2 +m_e^2 - m_{\gamma}^2x(1-x)}{\mu^2 +m_e^2- m_{\gamma}^2x(1-x)}\Bigg] ,\nn \\
	I_3 &=& \int \frac{d^2 k^{\perp}}{DD'} = \int_0^1 d\beta \frac{\pi}{P(x,\beta,t) },
	\label{int123}\ee
	where $L = -2m_e^2 + 2m_{\gamma}^2x(1-x) + t(1-x)^2$; $D = (k^{\perp})^2 - m_{\gamma}^2x(1-x) +m_e^2$; $D'=(k^{\perp})^2 -t (1-x)^2 -2k^{\perp}.\Delta^{\perp}(1-x) -m_{\gamma}^2x(1-x)+m_e^2$; 
	and $P(x,\beta,t) = -m_{\gamma}^2x(1-x)+m_e^2 -t\beta(1-\beta)(1-x)^2$. $\Lambda_{\mathrm{pert}}$ and $\mu$ are the ultraviolet ($\mathrm{UV}$) and infrared ($\mathrm{IR}$) cut-offs of the transverse momentum integral respectively. Here $x$ and $k^{\perp}$ denote the electron momenta such that $x = x_e$ and $k^{\perp} = k^{\perp}_e$.
 	
	The QED component of the photon structure function \cite{Berger:2014rva}, which is related to the GPD at the forward limit, is defined as
	\be
	F_{2,\mathrm{QED}}^{\gamma}(x) = 2 x F_{\gamma}(x,0).
	\label{psfeq}
	\ee

	\section{Numerical Results}
	\label{secnum}
	\begin{figure}[htp!]
		\centering
		\includegraphics[width=9.5cm,height=7.5cm,clip]{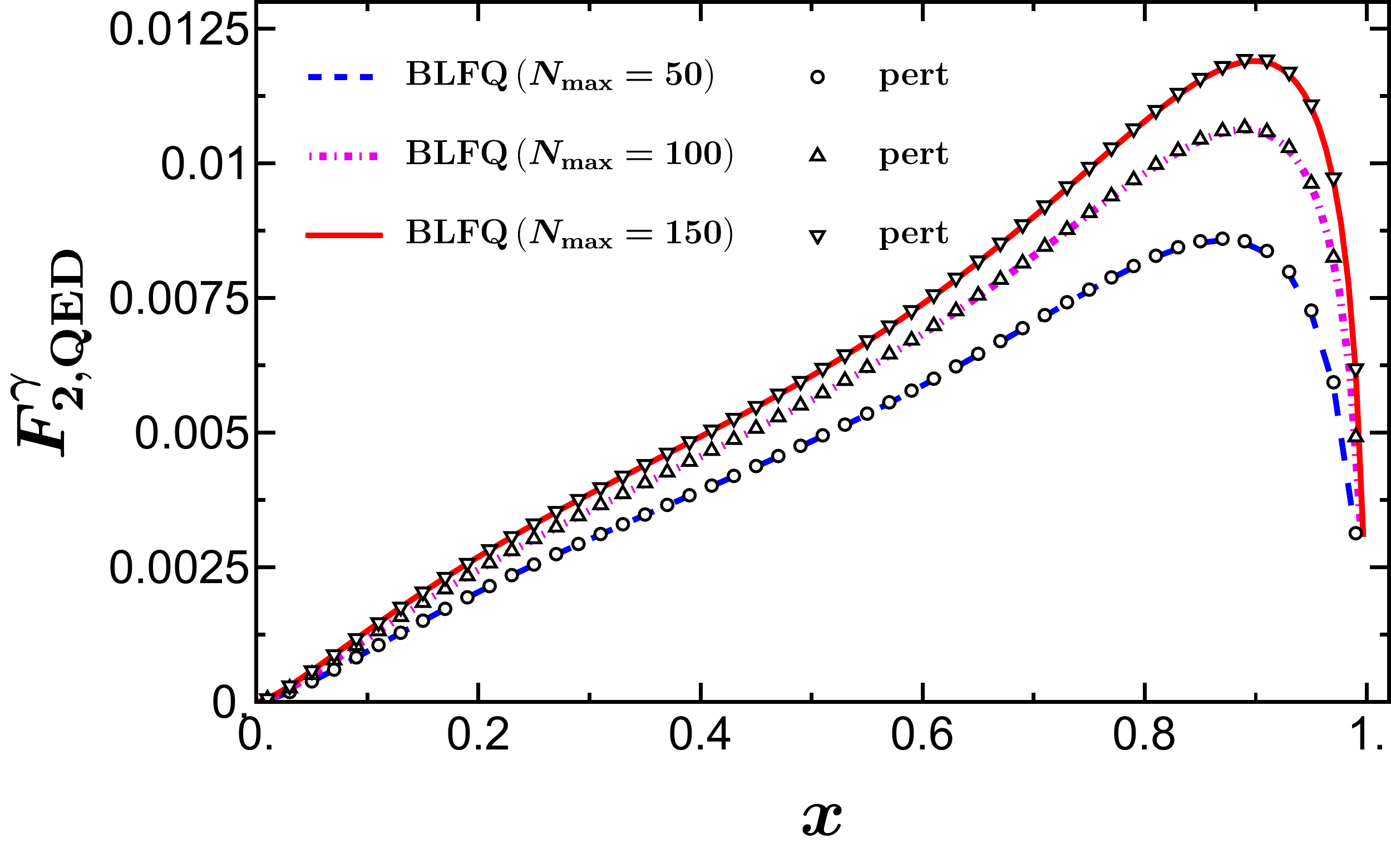}
		\caption{\label{fig2} The photon structure function $F_{2,\mathrm{QED}}^{\gamma}(x)$ for three different values of the basis truncation parameter $K=N_{\mathrm{max}} = (50,\,100,\,150)$. 
			The BLFQ results are shown with lines, while the different symbols represent the corresponding perturbative results.}
	\end{figure}
	For the numerical calculations, the values of parameters common to both the perturbative and the BLFQ computations are kept the same. These common parameters are the fermion mass in the two particle Fock sector $m_e = 0.511 ~\mathrm{MeV}$, vertex coupling constant $e = 0.3$, and the photon mass $m_{\mathrm{\gamma}} =0$. The energy scale parameter of the BLFQ basis is set to $b_0 = m_e = 0.511 ~\mathrm{MeV}$. The numerical codes for BLFQ were written in Fortran. 
	Single precision (32 bits) is inadequate to calculate the GPDs for large values of $N_{\mathrm{max}}$ and $K$ and hence, we  perform the numerical integration required to calculate the GPDs using the FM package, which is a FORTRAN package for Floating-point Multiple-precision arithmetic~\cite{fm}.
	
	Figure~\ref{fig2} shows the photon structure function as defined in Eq.~(\ref{psfeq}). $F_{2,\mathrm{QED}}^{\gamma}(x)$ is plotted as a function of $x$ and we compare our BLFQ computations with the perturbative results for three different values of the basis truncation parameter $N_{\mathrm{max}}$. Unless specified otherwise we show all results with the longitudinal truncation parameter 
	$K = N_{\mathrm{max}}$. 
	The perturbative UV cutoff $\Lambda_{\mathrm{pert}}$ in Eq.~(\ref{int123}) is related to the transverse basis truncation parameter $N_{\mathrm{max}}$ in BLFQ. Our choice of the $x$-dependent scale parameter $b$ translates to an $x$-dependent UV cutoff:  $\mathrm{UV_{co}} = b_0 \sqrt{x(1-x)2N_{\mathrm{max}}}$~\cite{Chakrabarti:2014cwa} , which is adopted as the UV cutoff in the perturbative calculations. This $x$ dependent UV cutoff convention has been used throughout in our calculation. Note that the perturbative results for the photon GPDs calculated previously in Ref.~\cite{Mukherjee:2011bn} and Ref.~\cite{Friot:2006mm} are for $x$ independent UV cutoff. We observe that the peak  of the structure function increases with increasing the UV cutoff. We find excellent agreement between our BLFQ computations  and the perturbative results.
	\begin{figure}[htp!]
		\centering
		\includegraphics[width=8.4cm,height=6.5cm,clip]{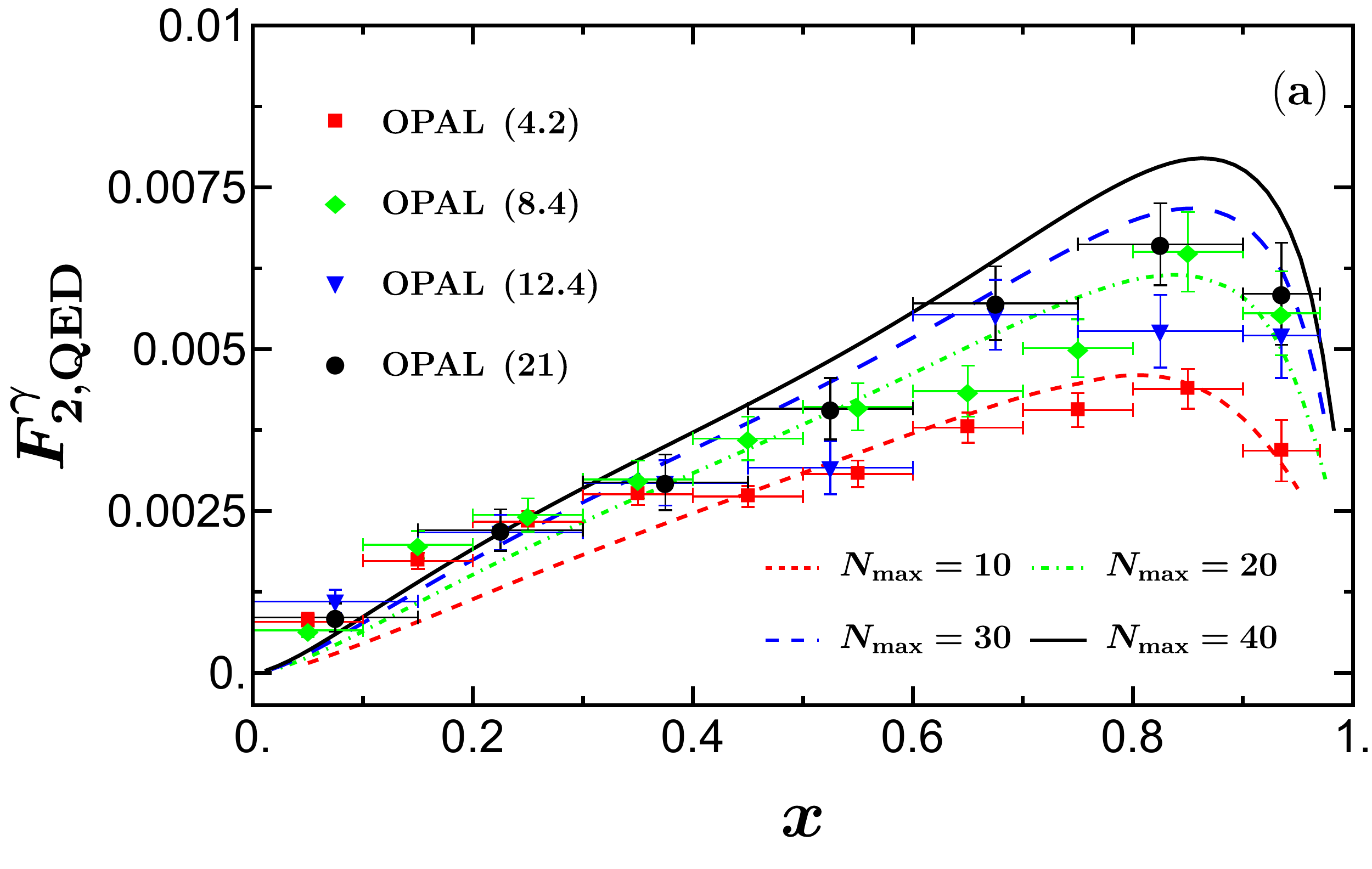}
		\includegraphics[width=8.4cm,height=6.5cm,clip]{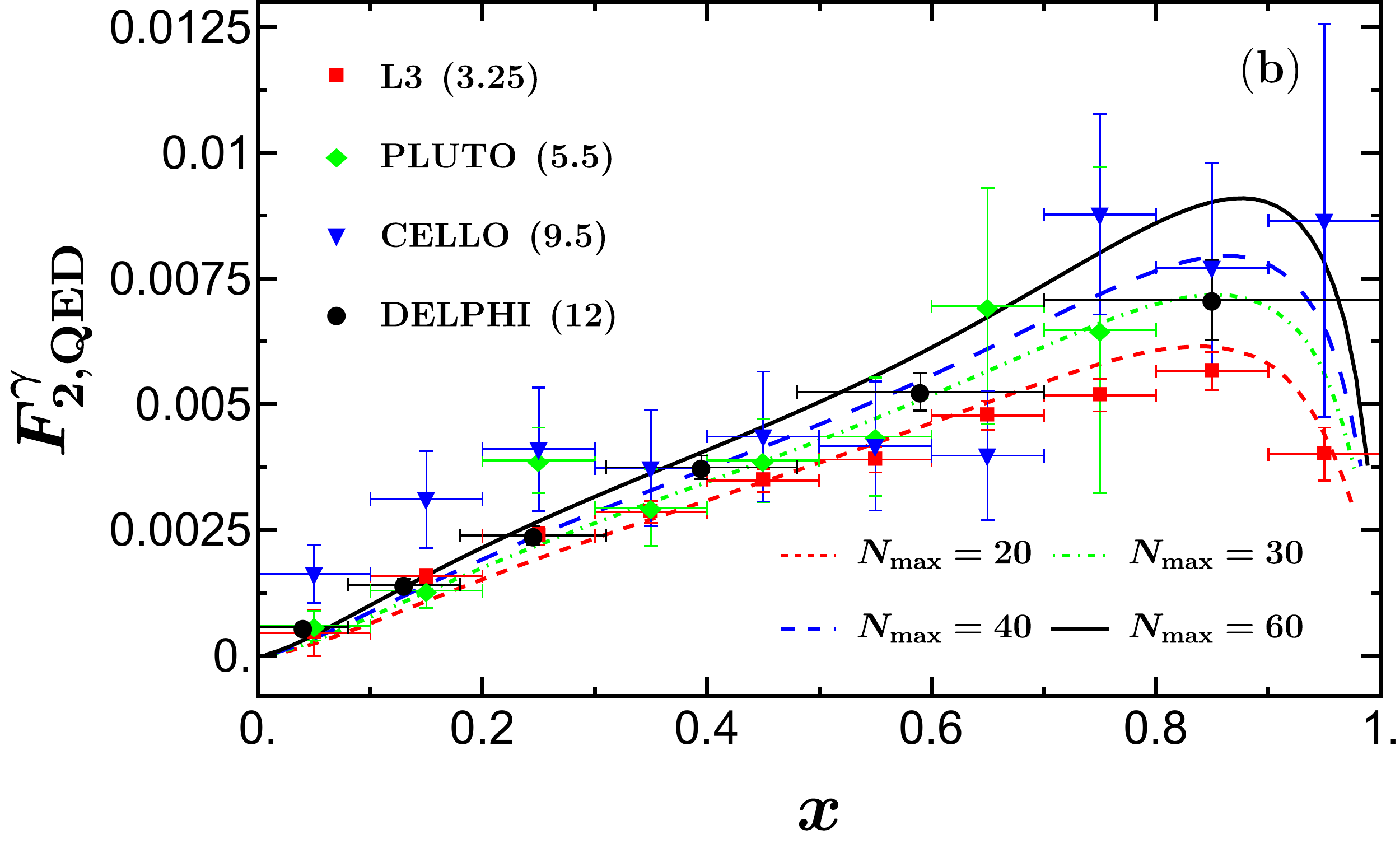}
		\caption{\label{fig3} We compare our BLFQ results for the photon structure function $F_{2,\mathrm{QED}}^{\gamma}(x)$ with the experimental data for QED photon structure function reported in Ref.~\cite{Nisius:1999cv}. Plot (a) shows the comparison with the results from the OPAL Collaboration~\cite{OPAL:1999rcd} and plot (b)   
		shows the comparison with the results from the	L3 Collaboration~\cite{L3:1998ijt}, PLUTO Collaboration~\cite{PLUTO:1984gmq}, CELLO Collaboration~\cite{CELLO:1983crq} and DELPHI Collaboration~\cite{DELPHI:1995fid}. The experimental data are shown with symbols and the numbers inside the parentheses indicate the experimental scale in $\mathrm{GeV}^2$. The experimental data are scaled by a factor of $\alpha =1/137$. The lines show BLFQ results for different values of $N_{\mathrm{max}}$ which are related to the experimental scales as discussed in the text. The horizontal error bars indicate the $x$ bin boundaries.}
	\end{figure}
	\begin{figure}[htp!]
		\centering
		\includegraphics[width=8cm,height=7cm,clip]{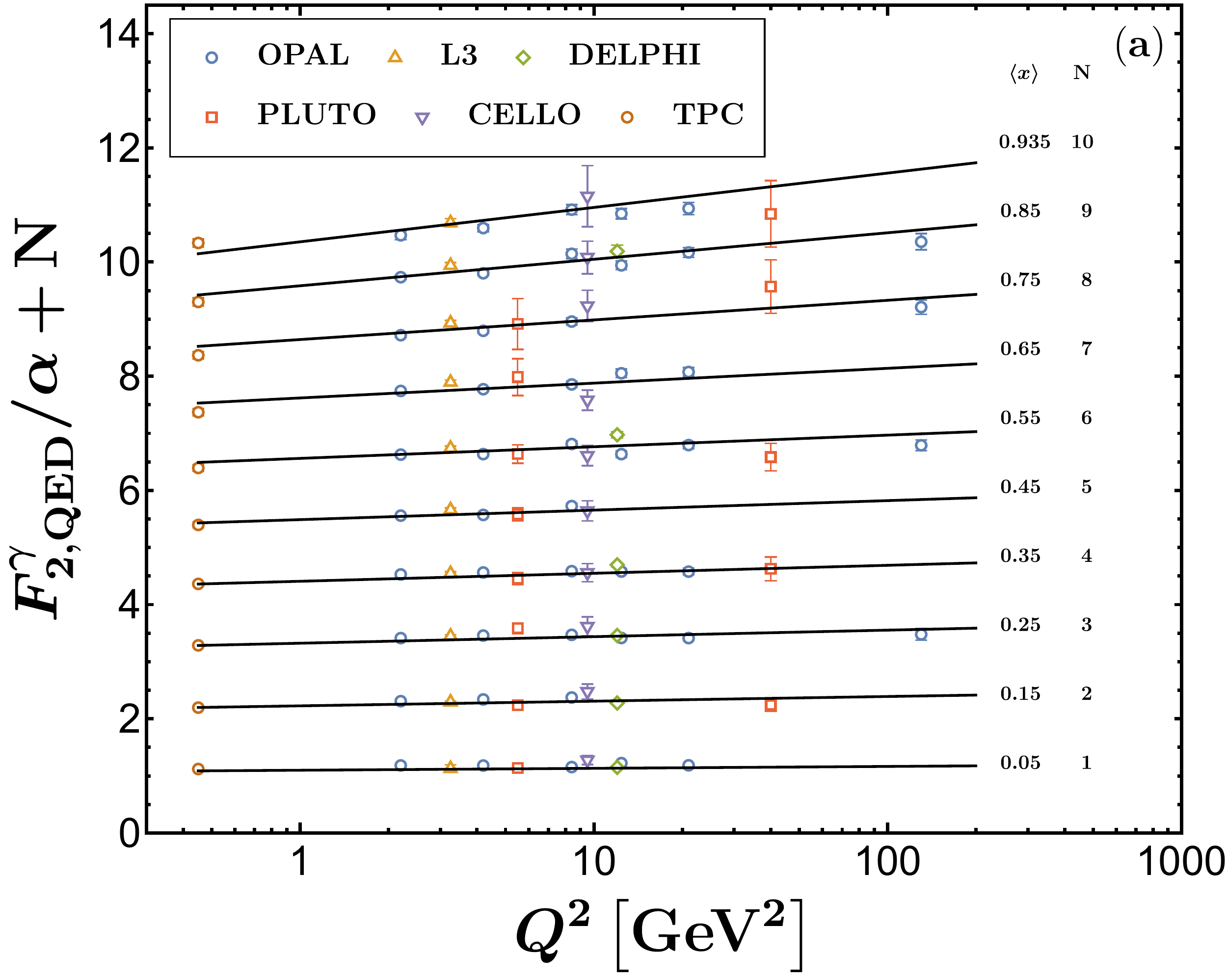}
		\includegraphics[width=8cm,height=7cm,clip]{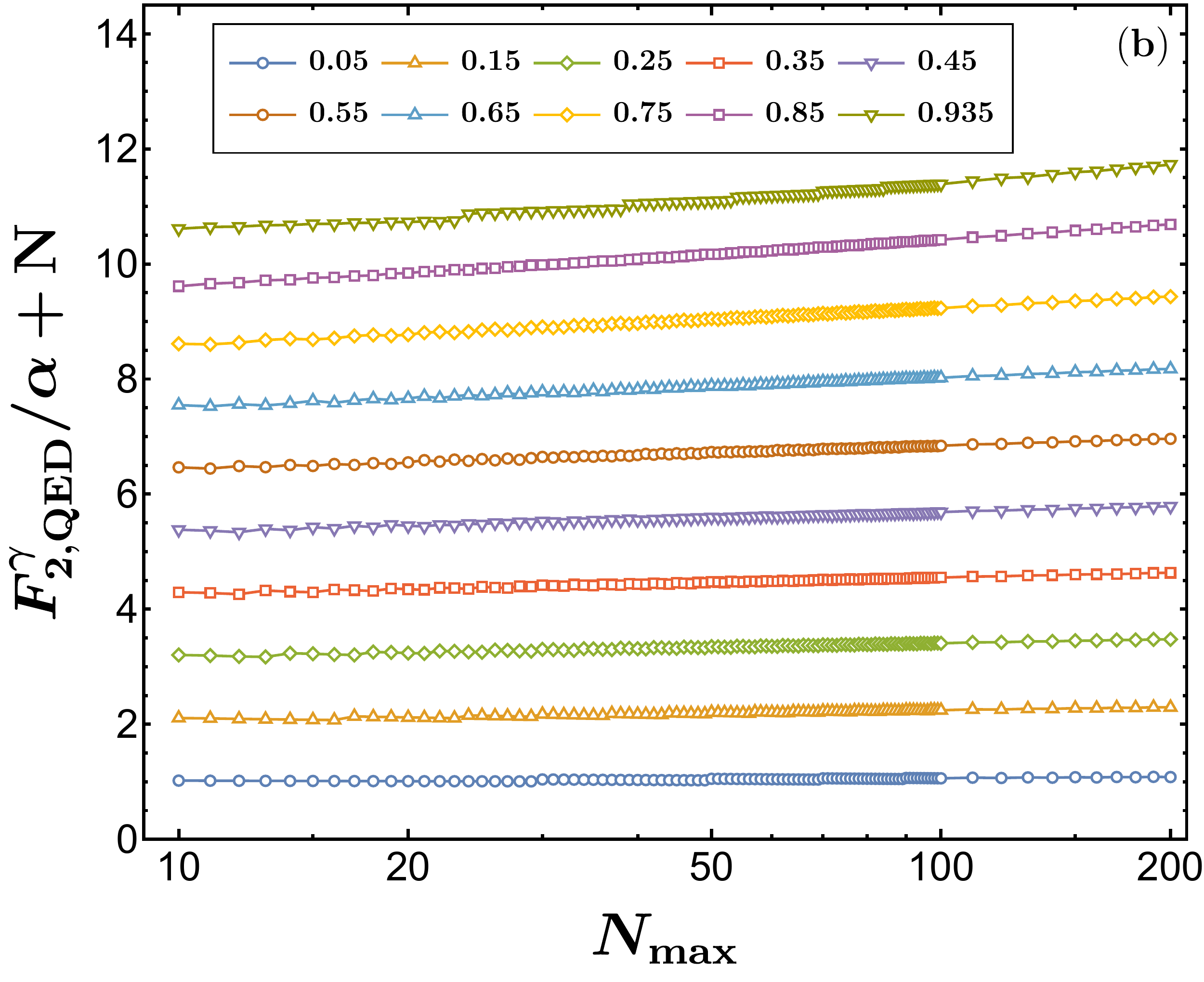}
		\caption{\label{fig4} Plot (a) shows the $F_{2,\mathrm{QED}}^{\gamma}(x)$ data from different experiments as a function of $Q^2$ for different values of $x$ as shown in Ref.~\cite{Nisius:1999cv}. The black solid lines in plot (a) correspond to the structure function in the leading logarithmic approximation as shown in Ref.~\cite{Nisius:1999cv}. An increasing integer value of $N$ is added to each fixed $x$ result of the structure function in order to separate them from each other. The top two data points for PLUTO at $Q^2= 40 ~\mathrm{GeV}^2$ belong to $N =8$ and $N =10$. Plot (b) shows our BLFQ results for $F_{2,\mathrm{QED}}^{\gamma}(x)$ as a function of $N_{\mathrm{max}}$ for similar values of $x$ where the numbers in the legend denote $x$ values.}
	\end{figure}
	
	In Fig.~\ref{fig3}, we compare the BLFQ results with the experimental data for the QED photon structure function reported in Ref.~\cite{Nisius:1999cv}. 
	The experimental data are from the OPAL Collaboration~\cite{OPAL:1999rcd}, 
	 L3 Collaboration~\cite{L3:1998ijt}, PLUTO Collaboration~\cite{PLUTO:1984gmq}, CELLO Collaboration~\cite{CELLO:1983crq} and DELPHI Collaboration~\cite{DELPHI:1995fid}. The error in the data points shows the total error, which includes the statistical and systematic errors combined in quadrature. The 
	horizontal error bar indicates the bin boundaries for $x$. The experimental data reported in Ref.~\cite{Nisius:1999cv} are for $F_{2,\mathrm{QED}}^{\gamma}(x)/\alpha$. So we rescale the experimental data by a factor of $\alpha = 1/137$ to compare them with our BLFQ results for $F_{2,\mathrm{QED}}^{\gamma}(x)$. We show our results with increasing values of $N_{\mathrm{max}}$ to study the cutoff dependence of our results. The $N_{\mathrm{max}}$ values are chosen in order to provide a reasonable comparison over the wide range of the experimental scale. Our intention here is to show the qualitative behavior of the UV cutoff determined by $N_{\mathrm{max}}$. We further compare our $N_{\mathrm{max}}$ dependence with the data by analyzing them in a different way as shown in Fig.~\ref{fig4}.
	In Fig.~\ref{fig4}(a), we show the experimental data on the QED photon structure function from Ref.~\cite{Nisius:1999cv}, where the data are shown as function of the experimental scale $Q^2$ for different central $x$ values. An integer value $N$ corresponding to the $x$ bin number is added to each measurement to have sufficient vertical separation. If the experimental data do not have the exact $x$ bin central value as indicated then the nearest $x$ bin central value is used. In Fig.~\ref{fig4}(b), we show our BLFQ result for $F_{2,\mathrm{QED}}^{\gamma}(x)$ as function of $N_{\mathrm{max}}$ for different $x$ values similar to  Fig.~\ref{fig4}(a). We observe from this comparison that as $x$ increases the slope also increases for both the data and our BLFQ results. The approximate relation between the experimental scale $Q^2$ and the transverse cut-of in BLFQ $N_{\rm max}$ is identified as $Q^2 \approx b_0^2\, x(1-x)\,2N_{\mathrm{max}}$.

	\begin{figure}[htp!]
		\centering
		\includegraphics[width=8.4cm,height=6.5cm,clip]{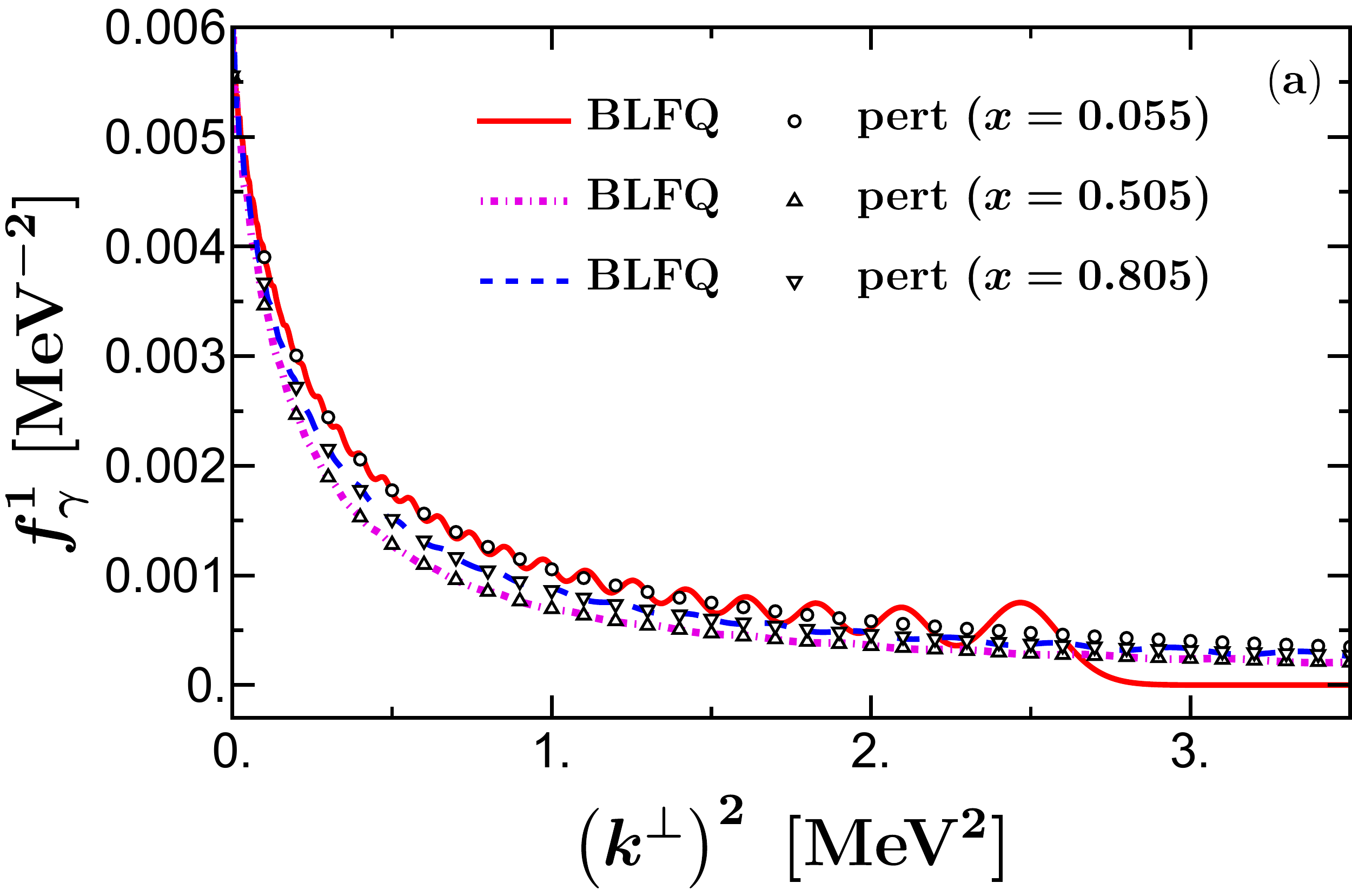}
		\includegraphics[width=8.4cm,height=6.4cm,clip]{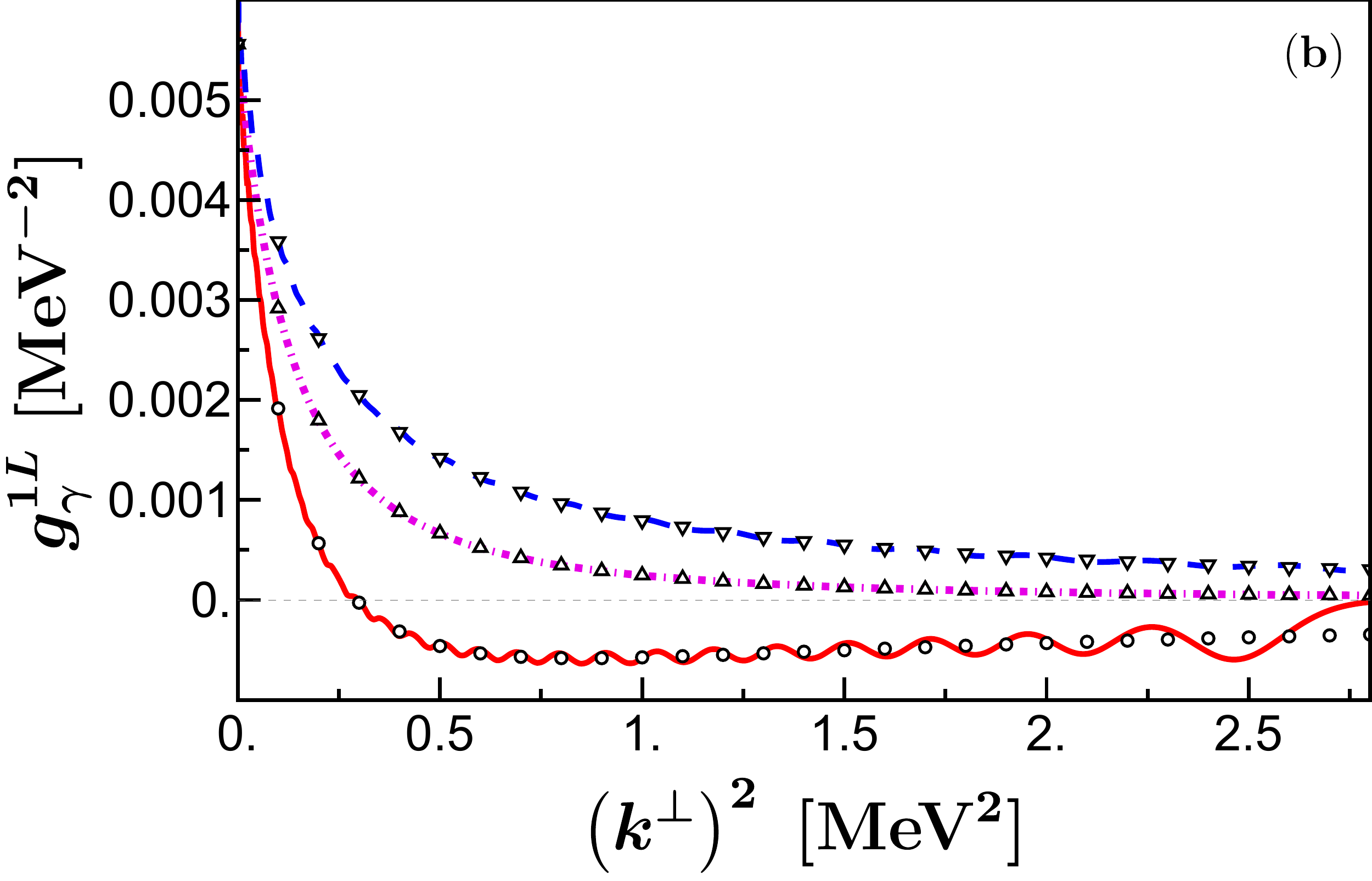}\\
		\includegraphics[width=8.4cm,height=6.5cm,clip]{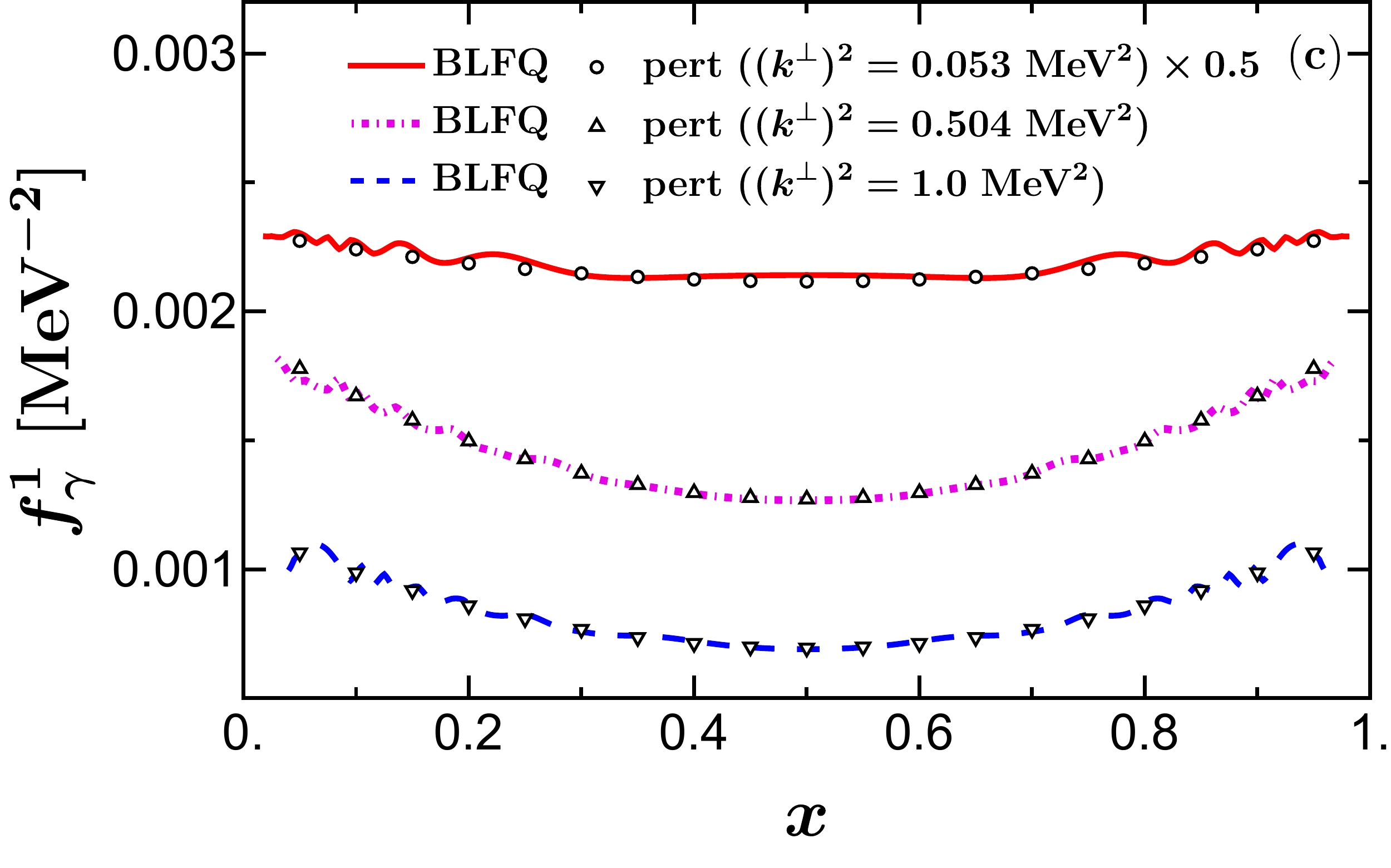}
		\includegraphics[width=8.4cm,height=6.5cm,clip]{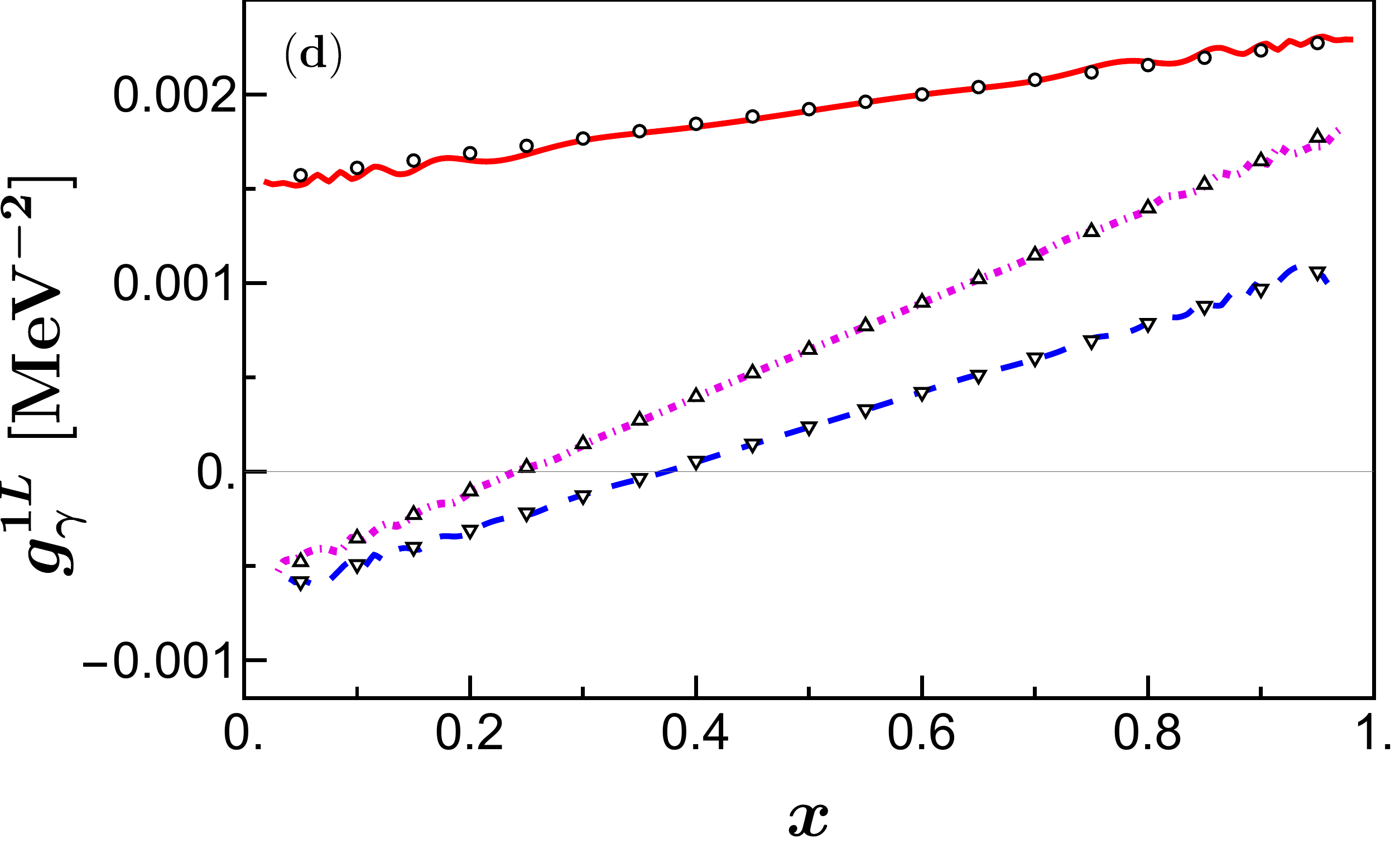}
		\caption{\label{fig5} Plots (a) and (b) show the photon unpolarized TMD  $f^1_{\gamma}(x,(k^{\perp})^2)$ and the photon helicity TMD $g^{1L}_{\gamma}(x,(k^{\perp})^2)$, respectively as functions of $(k^{\perp})^2$ for fixed $x$, whereas plots (c) and (d) present them as functions of $x$ for fixed $(k^{\perp})^2$. We compare our BLFQ results (lines) with the perturbative results (symbols). The BLFQ results are obtained by averaging over the BLFQ computations at $N_{\mathrm{max}} = \{100,102,104\}$ for $K = 100$. Both the BLFQ and the perturbative results for $(k^{\perp})^2 = 0.053 ~\mathrm{MeV}^2$ in plots (c) and (d) are scaled by a factor of $0.5$ to enhance visualization. }
	\end{figure}

	In Fig.~\ref{fig5}  we show our two-dimensional structure of the real photon's TMDs. The TMDs calculated in BLFQ have oscillations in the transverse direction resulting from the oscillatory nature of the HO basis functions employed in the transverse plane. In order to reduce these oscillations, we adopt an averaging scheme~\cite{Hu:2020arv}, wherein we average three results at different $N_{\mathrm{max}}$ with fixed $K$. The averaging scheme is achieved by taking an average of averages.  We first take the average of the results obtained at $N_{\mathrm{max}} = n$ and $N_{\mathrm{max}} = n + 2$. Furthermore, we take another average of the results obtained at $N_{\mathrm{max}} = n+2$ and $N_{\mathrm{max}} = n + 4$. The final result is obtained by taking average of the previous two averages. We denote this averaging method of three values as $N_{\mathrm{max}} = \{n, n + 2 ,n+4\}$. We set $K = n+4$, while calculating all the averages~\cite{Hu:2020arv}. 
	
	Figures~\ref{fig5} (a) and (b) show our results for the unpolarized TMD $f^1_{\gamma}$ and the photon helicity TMD $g^{1L}_{\gamma}$ as functions of $(k^{\perp})^2$ for three fixed values of $x = (0.055,\, 0.505,\, 0.805)$. We observe that the TMDs reach their maximum value at zero transverse momentum of the electron (positron) and at exactly $(k^{\perp})^2 = 0$, where the TMDs become independent of $x$. The unpolarized TMD $f^1_{\gamma}$ remains positive over the chosen range of $(k^{\perp})^2$, whereas the helicity TMD $g^{1L}_{\gamma}$ becomes negative over certain regions for particular values of $x$. 
	
	Figures~\ref{fig5} (c) and (d) show our results for the unpolarized TMD $f^1_{\gamma}$ and the photon helicity TMD $g^{1L}_{\gamma}$ as functions of $x$ for three fixed values of $(k^{\perp})^2 = (0.053,\,0.504,\,1.0)~ \mathrm{MeV^2}$. For a fixed value of $(k^{\perp})^2$, the unpolarized TMD $f^1_{\gamma}$ approaches its maximum value when either the electron or the positron carries most of the longitudinal momentum fraction, which happens at the end points of $x$. Meanwhile, $f^1_{\gamma}$ approaches its minimum value at $x =0.5$, i.e., when the electron-positron pair share exactly equal momenta. The polarized TMD $g^{1L}_{\gamma}$ breaks this symmetry over $x$ as can be seen in Fig.~\ref{fig5}(c) and  becomes negative for certain values of $x$ and $(k^{\perp})^2$. Overall, we observe that both the TMDs calculated in our BLFQ approach are in excellent agreement with the corresponding perturbative results. The averaging strategy used to curb the oscillatory behavior in the BLFQ computations has its limitations due to finite basis artifacts which become prominent at the endpoints of the $x$ range.
	\begin{figure}[htp!]
		\centering
		\includegraphics[width=8.4cm,height=7.5cm,clip]{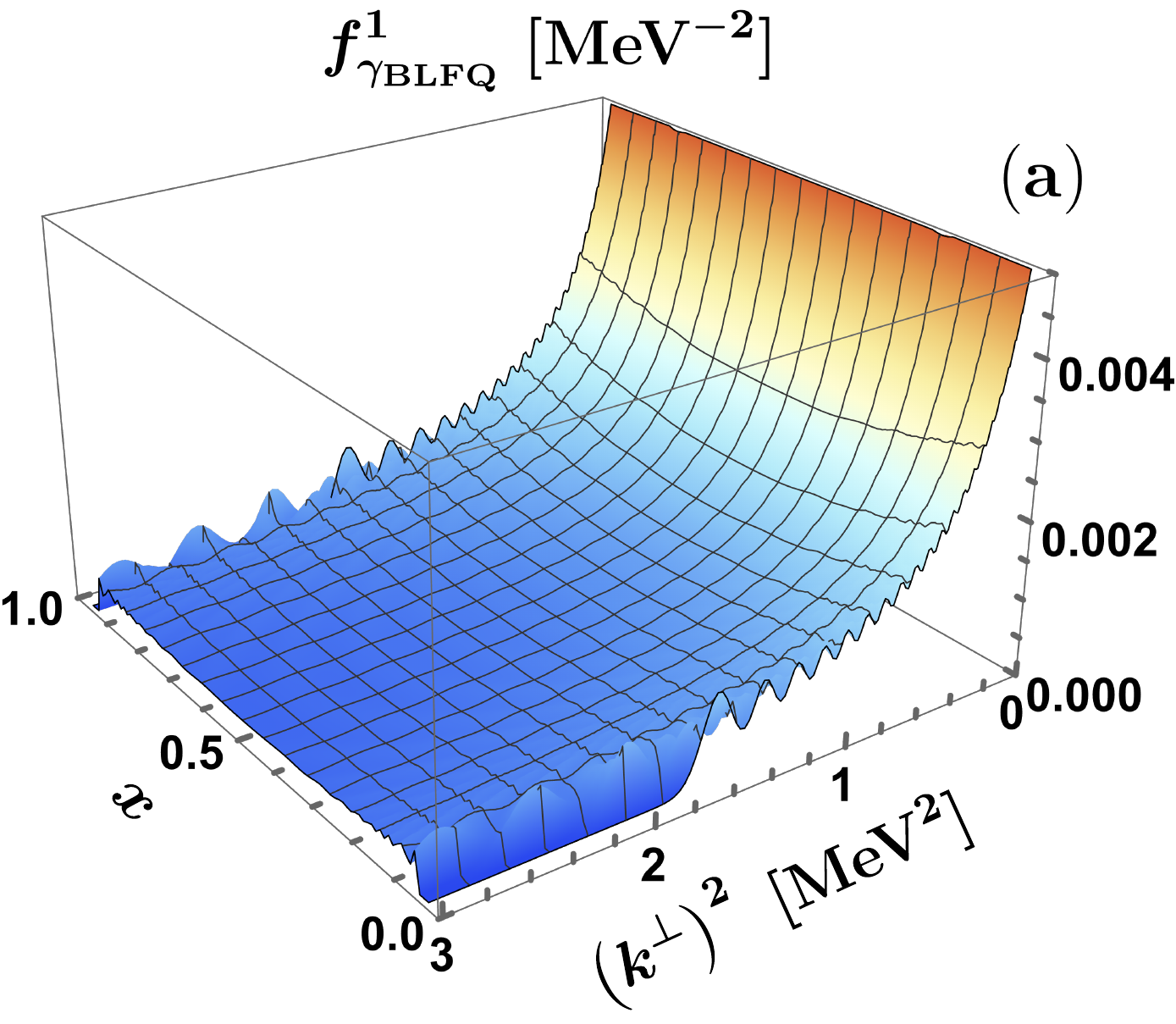}
		\includegraphics[width=8.4cm,height=7.5cm,clip]{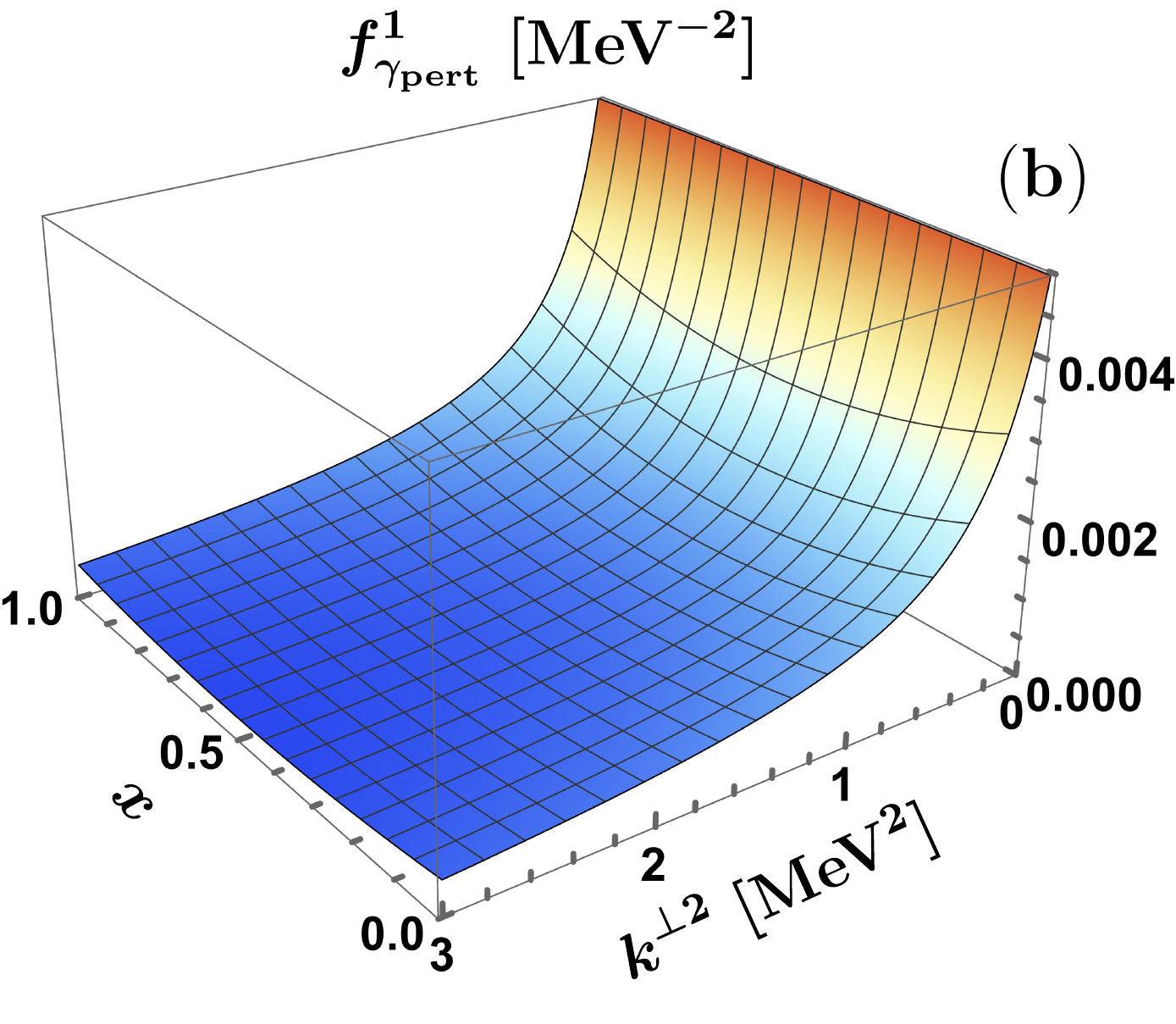}
		\includegraphics[width=8.4cm,height=7.5cm,clip]{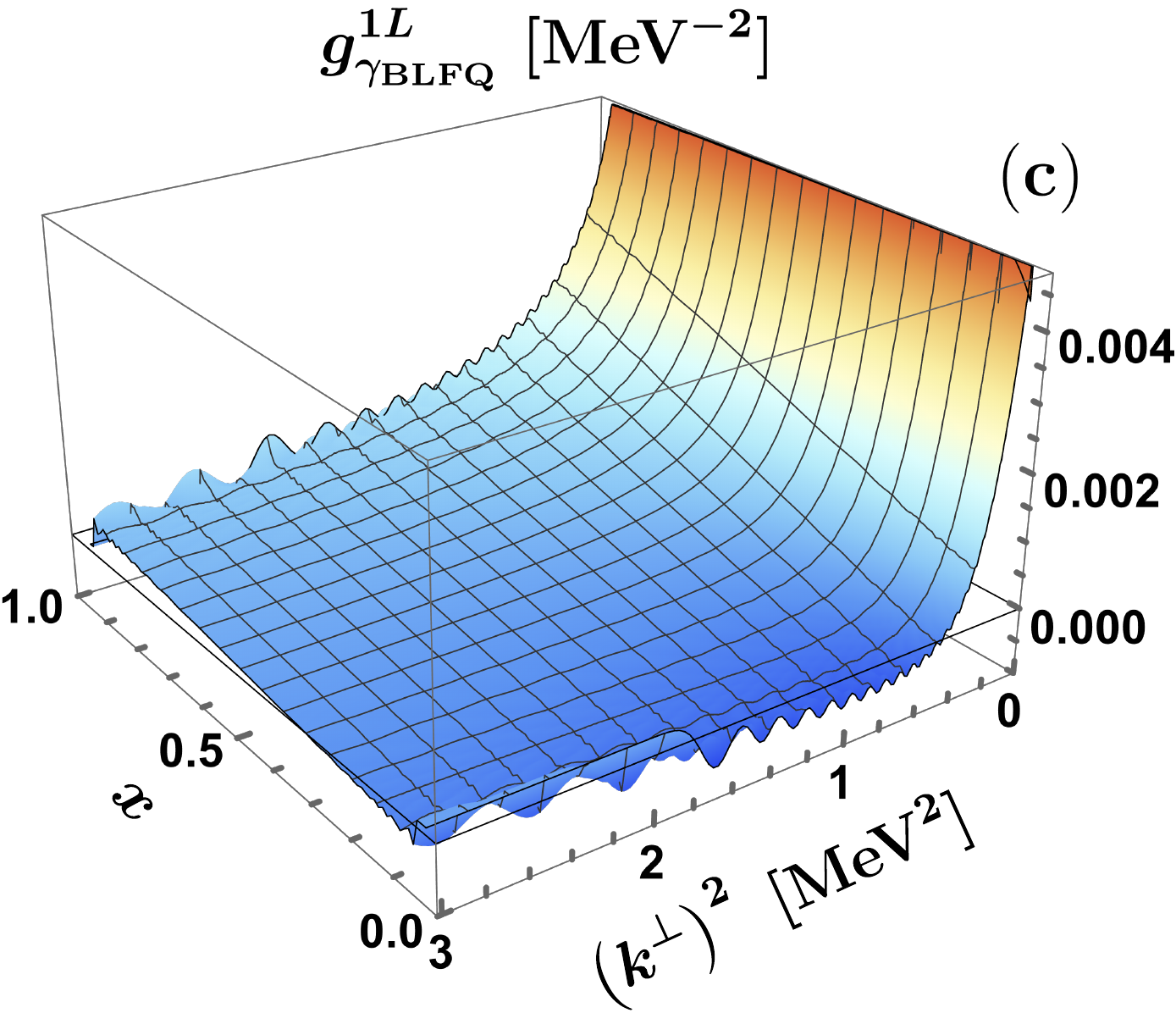}
		\includegraphics[width=8.4cm,height=7.5cm,clip]{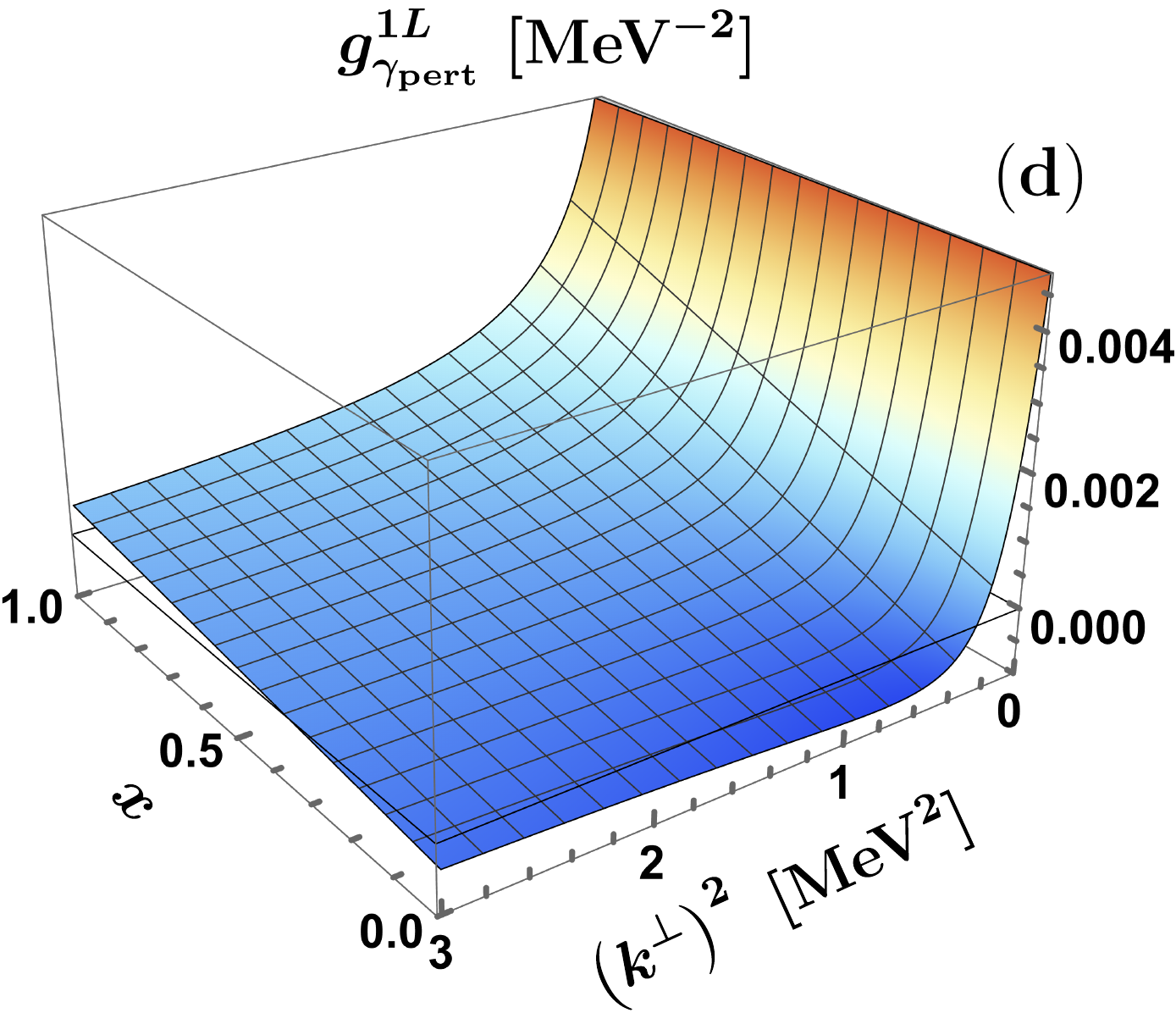}
		\caption{\label{fig6}
			3D plots for photon unpolarized TMD $f^1_{\gamma}(x,(k^{\perp})^2)$ and the  photon helicity TMD $g^{1L}_{\gamma}(x,(k^{\perp})^2)$. Plots (a) and (c) correspond to the BLFQ results and plots (b) and (d) represent the perturbative results.  The BLFQ results are obtained by averaging over the BLFQ computations at $N_{\mathrm{max}} = \{100,102,104\}$ for $K = 100$.}
	\end{figure}
	
	Figure~\ref{fig6}  compares the 3D plots for the unpolarized TMD $f^1_{\gamma}$ and the polarized TMD $g^{1L}_{\gamma}$, respectively with the corresponding plots from perturbation theory. The $f^1_{\gamma}$ is symmetric over $x$, while this symmetry is broken in the $g^{1L}_{\gamma}$. The peaks observed in the transverse $(k^{\perp})^2$ direction fall off as $(k^{\perp})^2$ increases and it falls off to negative values at certain $x$ range for the $g^{1L}_{\gamma}$. We observe good consistency between the BLFQ and the perturbative results. The oscillations seen around the endpoint of the $x$ range are the reflections of the finite basis artifacts.

	\begin{figure}[htp!]
		\centering
		\includegraphics[width=8.4cm,height=6.5cm,clip]{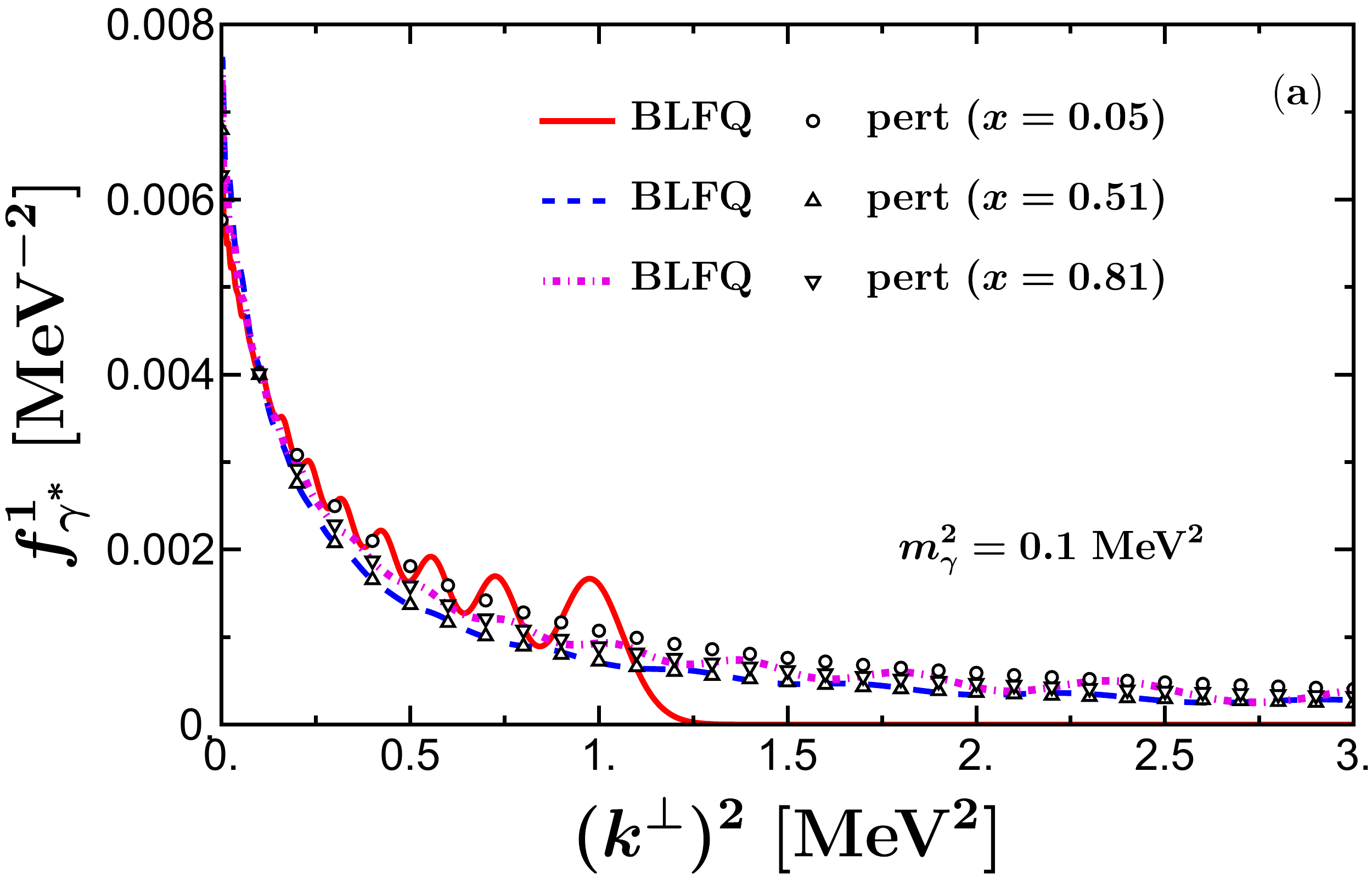}	
		\includegraphics[width=8.4cm,height=6.5cm,clip]{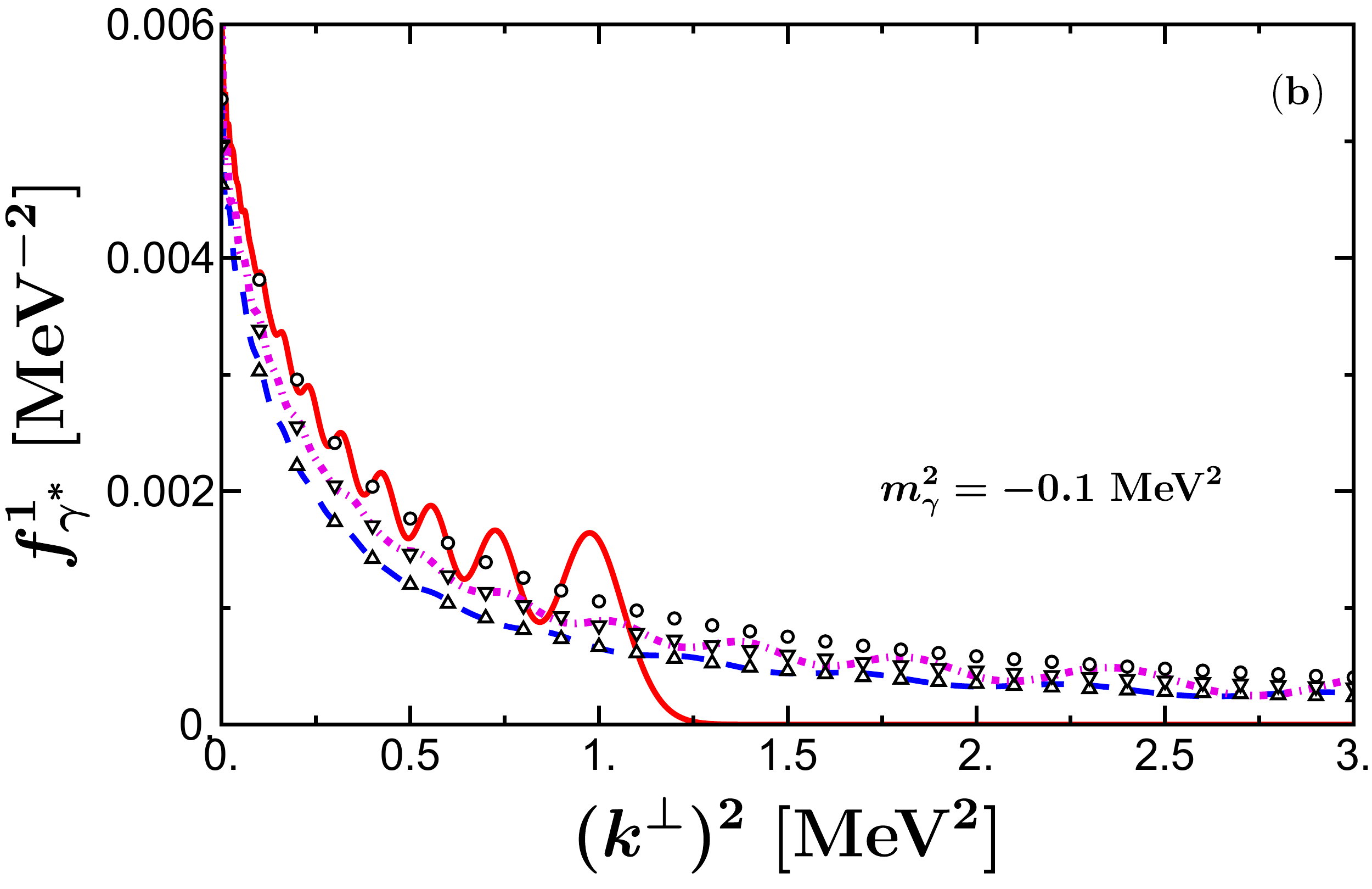}
		\includegraphics[width=8.4cm,height=6.5cm,clip]{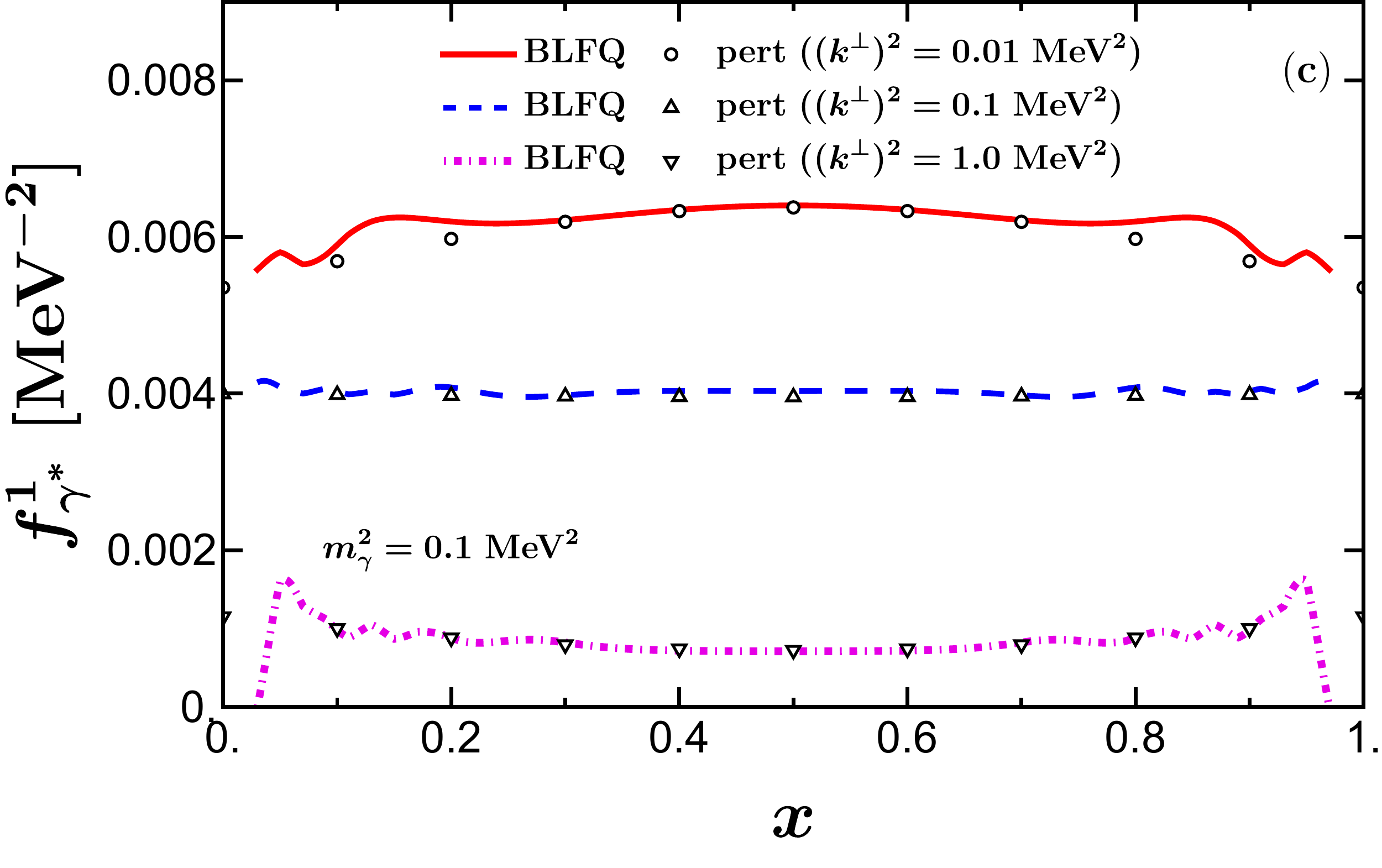}
		\includegraphics[width=8.4cm,height=6.5cm,clip]{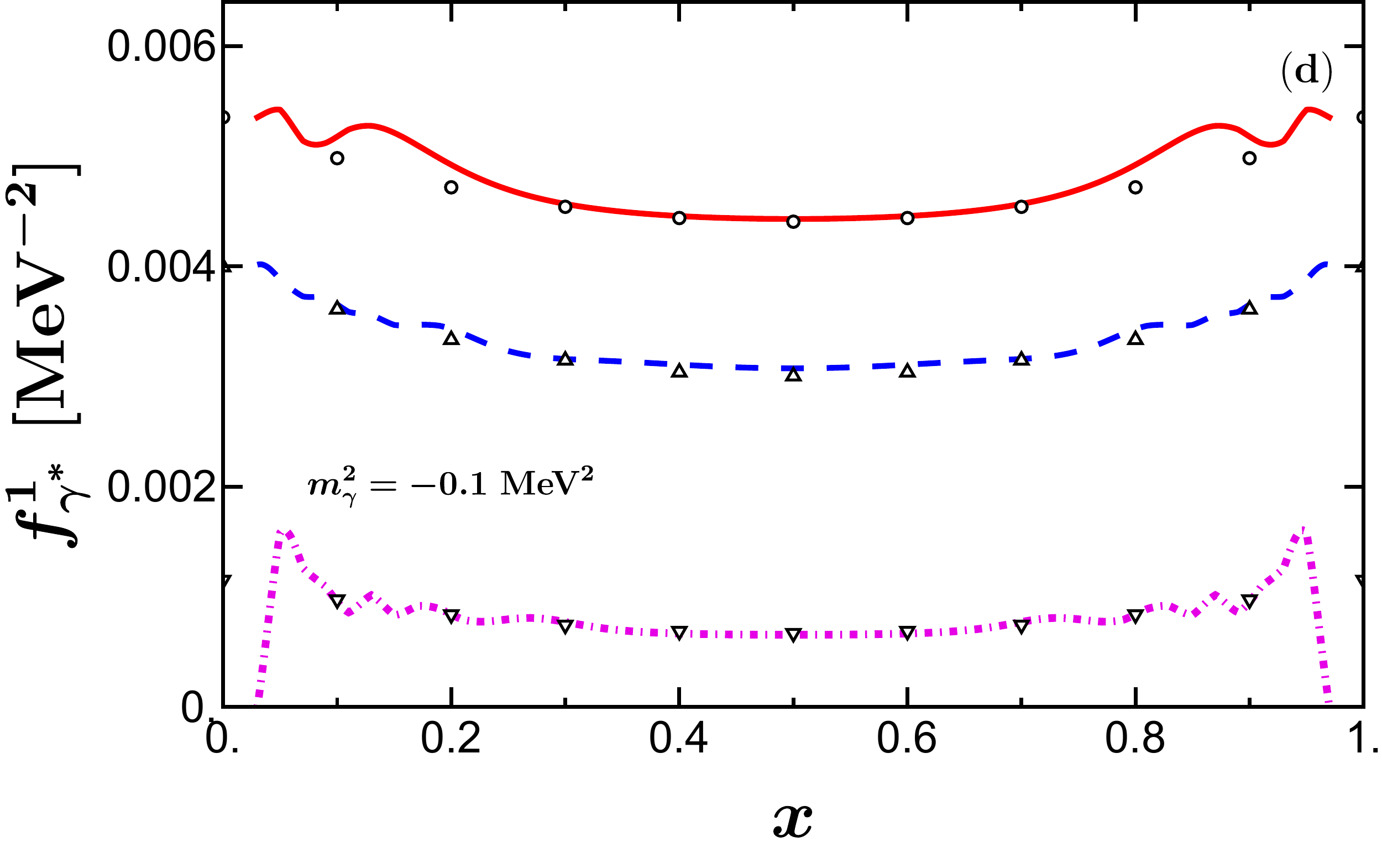}
		\caption{\label{fig7}  Plots for virtual photon unpolarized TMD $f^1_{\gamma^*}(x,(k^{\perp})^2)$, where (a) and (b) are for $f^1_{\gamma^*}$ vs. $(k^{\perp})^2$ at fixed $x$ . Plots (c) and (d) are for $f^1_{\gamma^*}$ vs. $x$ at fixed $(k^{\perp})^2$. 
			Plots (a) and (c) are for the time-like virtual photon with $m_{\gamma}^2 = 0.1 ~\mathrm{MeV}^2$, whereas
			plots (b) and (d) are for the space-like virtual photon with $m_{\gamma}^2 = - 0.1 ~\mathrm{MeV}^2$.
			We compare our results (lines) with the perturbative results (symbols).  The BLFQ results are obtained by averaging over the BLFQ computations at $N_{\mathrm{max}} = \{46,48,50\}$ for $K = 50$. }
	\end{figure}
	\begin{figure}[htp!]
		\centering
		\includegraphics[width=8.4cm,height=7.5cm,clip]{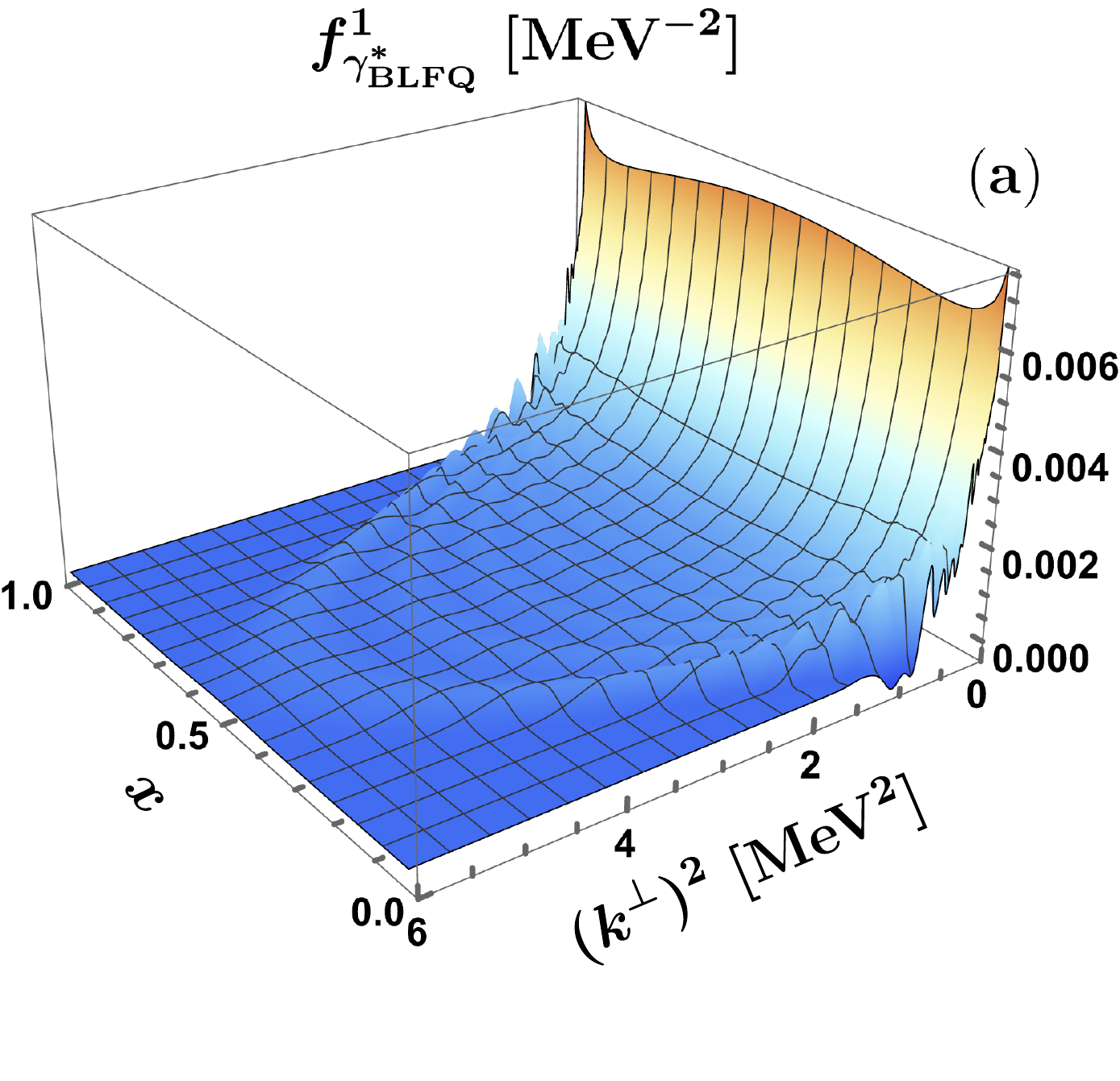}
		\includegraphics[width=8.4cm,height=7.5cm,clip]{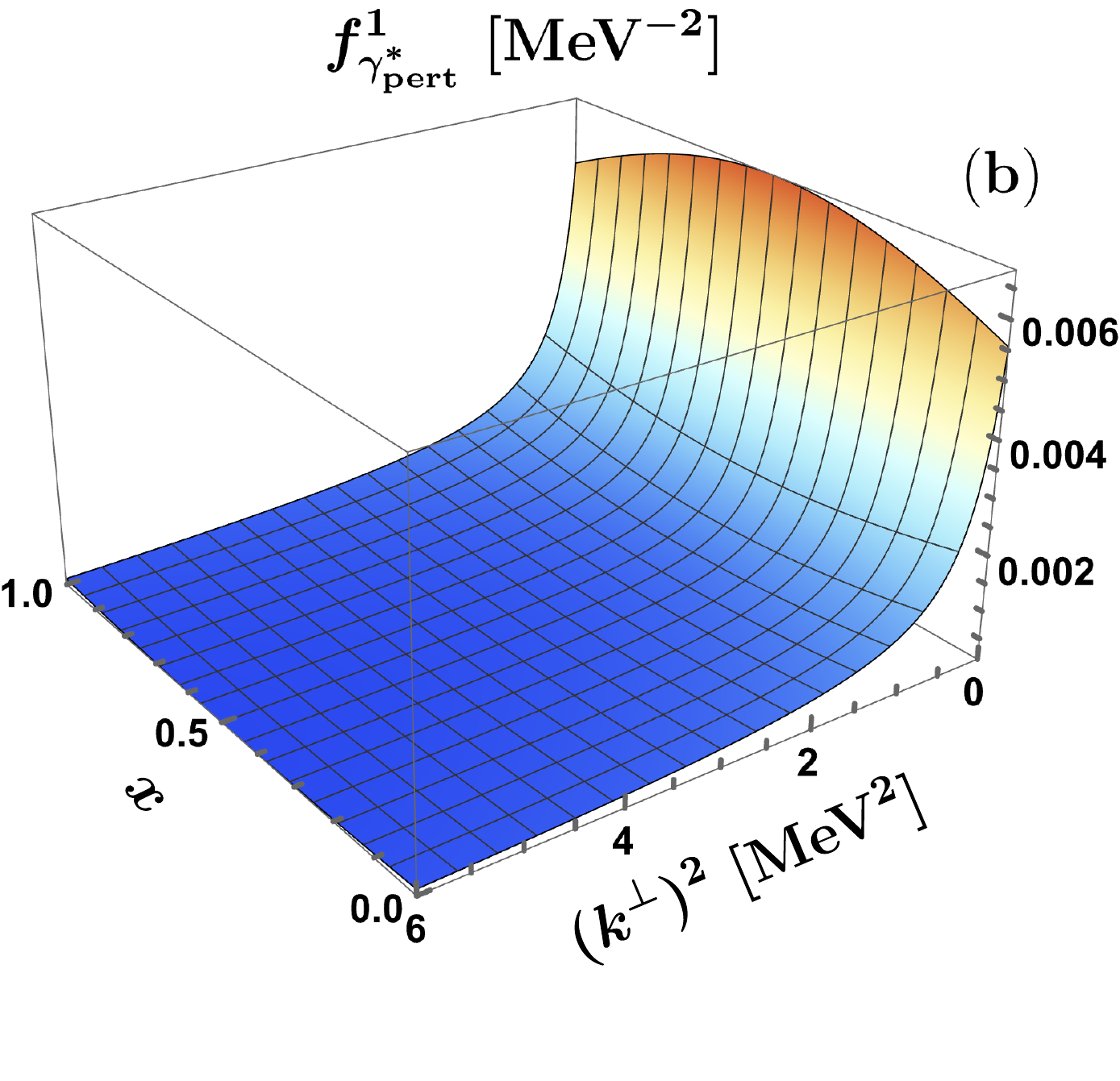}
		\includegraphics[width=8.4cm,height=7.5cm,clip]{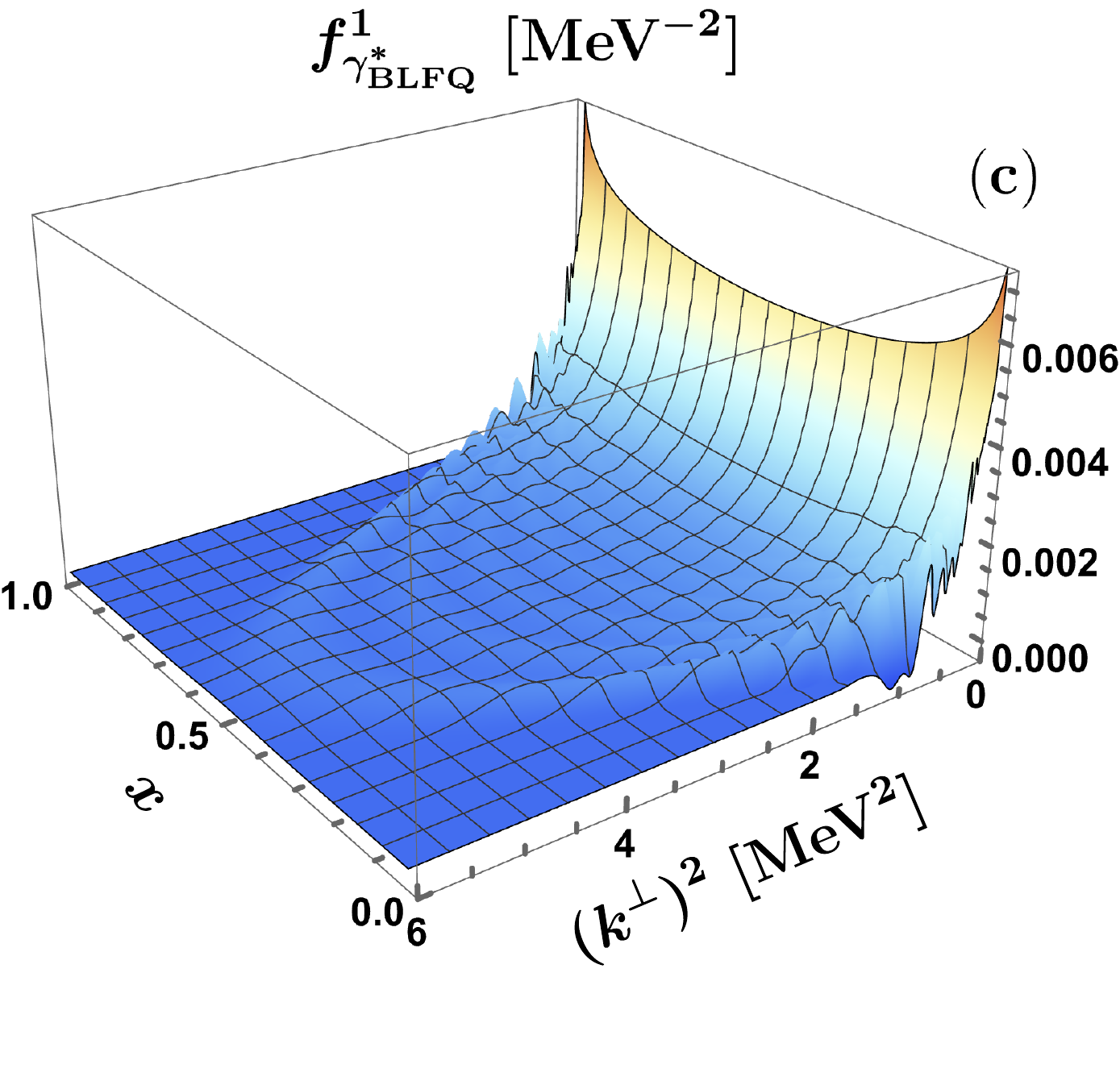}
		\includegraphics[width=8.4cm,height=7.5cm,clip]{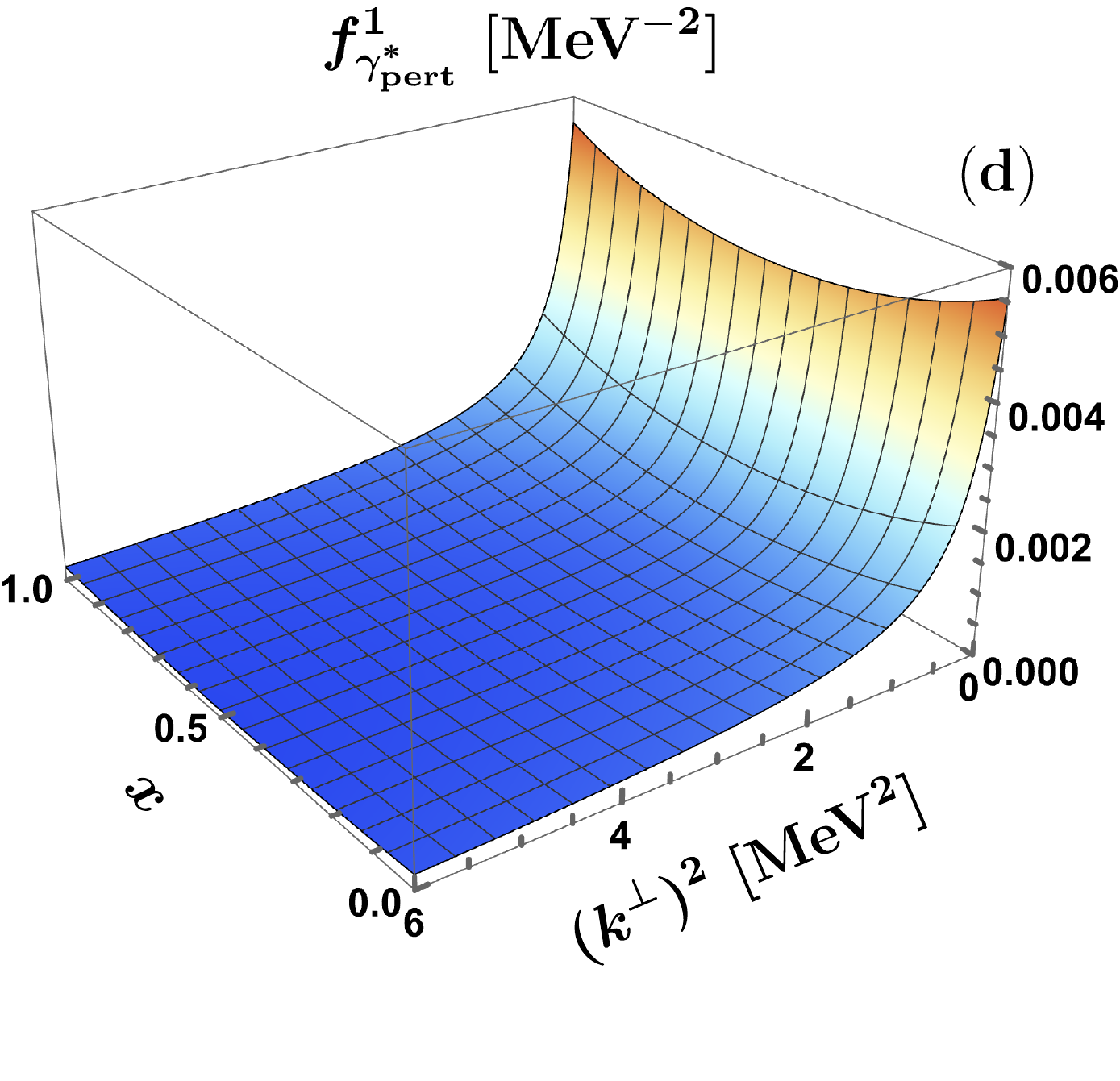}
		\caption{\label{fig8} 3D plots for virtual photon unpolarized TMD $f^1_{\gamma}(x,(k^{\perp})^2)$. Plots (a) and (b) are for $m_{\gamma}^2 = 0.1 ~\mathrm{MeV}^2$. Plots (c) and (d) are for $m_{\gamma}^2 = - 0.1 ~\mathrm{MeV}^2$. We compare our BLFQ computations with the perturbative results. The BLFQ results are obtained by averaging over the BLFQ computations at $N_{\mathrm{max}} = \{46,48,50\}$ for $K = 50$. }
	\end{figure}
	\begin{figure}[htp!]
		\centering
		\includegraphics[width=8.4cm,height=7.5cm,clip]{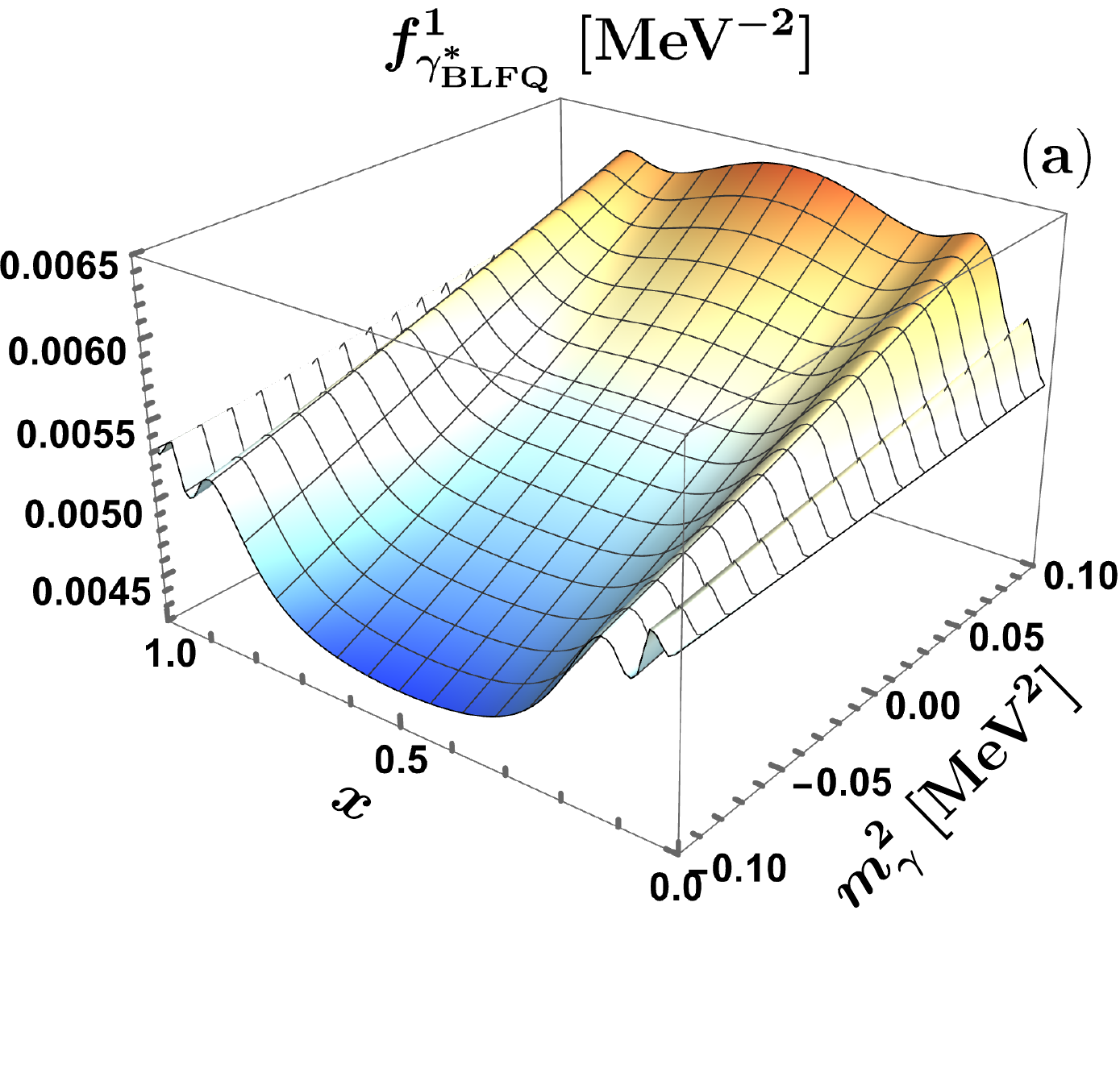}
		\includegraphics[width=8.4cm,height=7.5cm,clip]{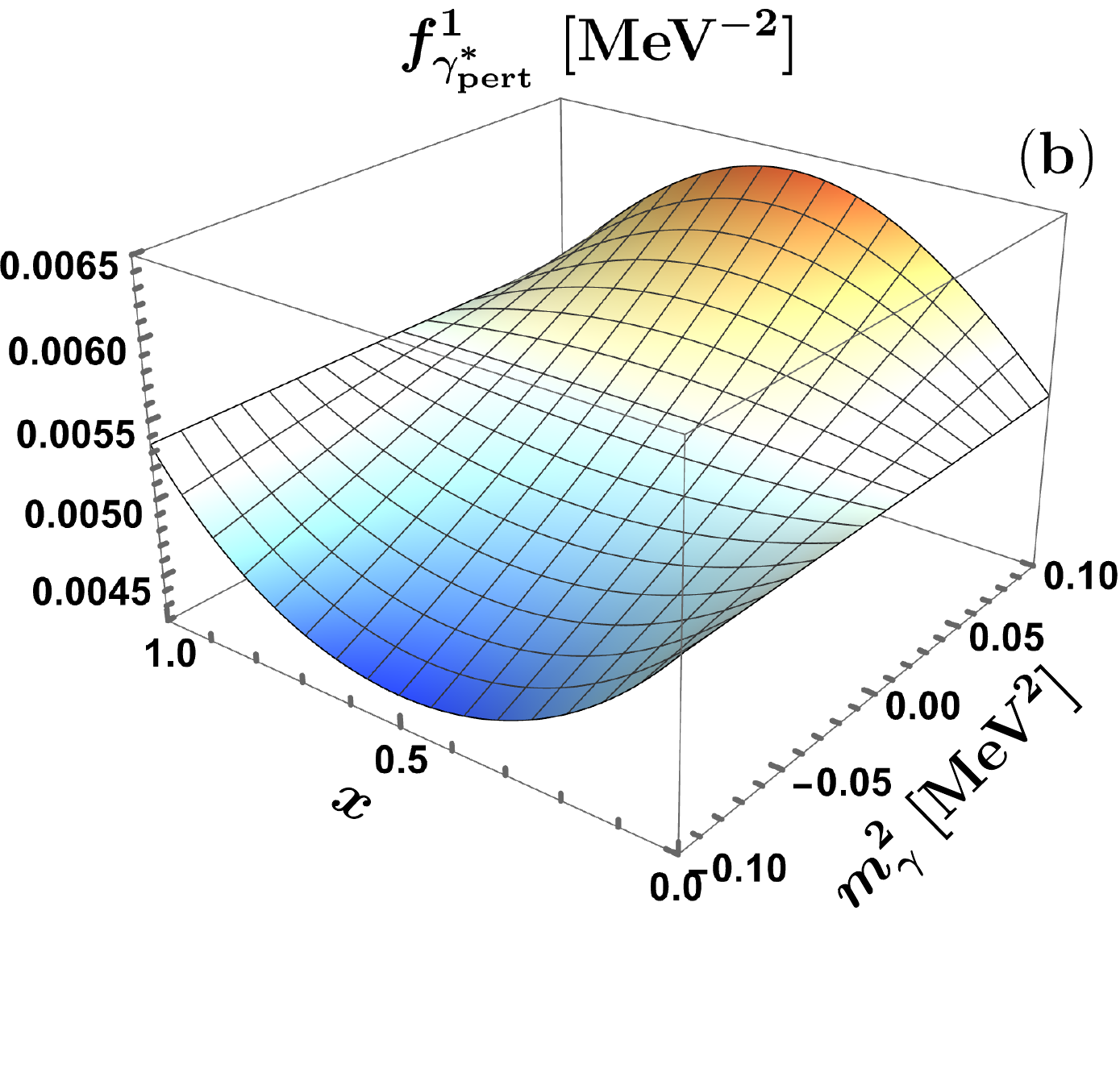}
		\includegraphics[width=8.4cm,height=7.5cm,clip]{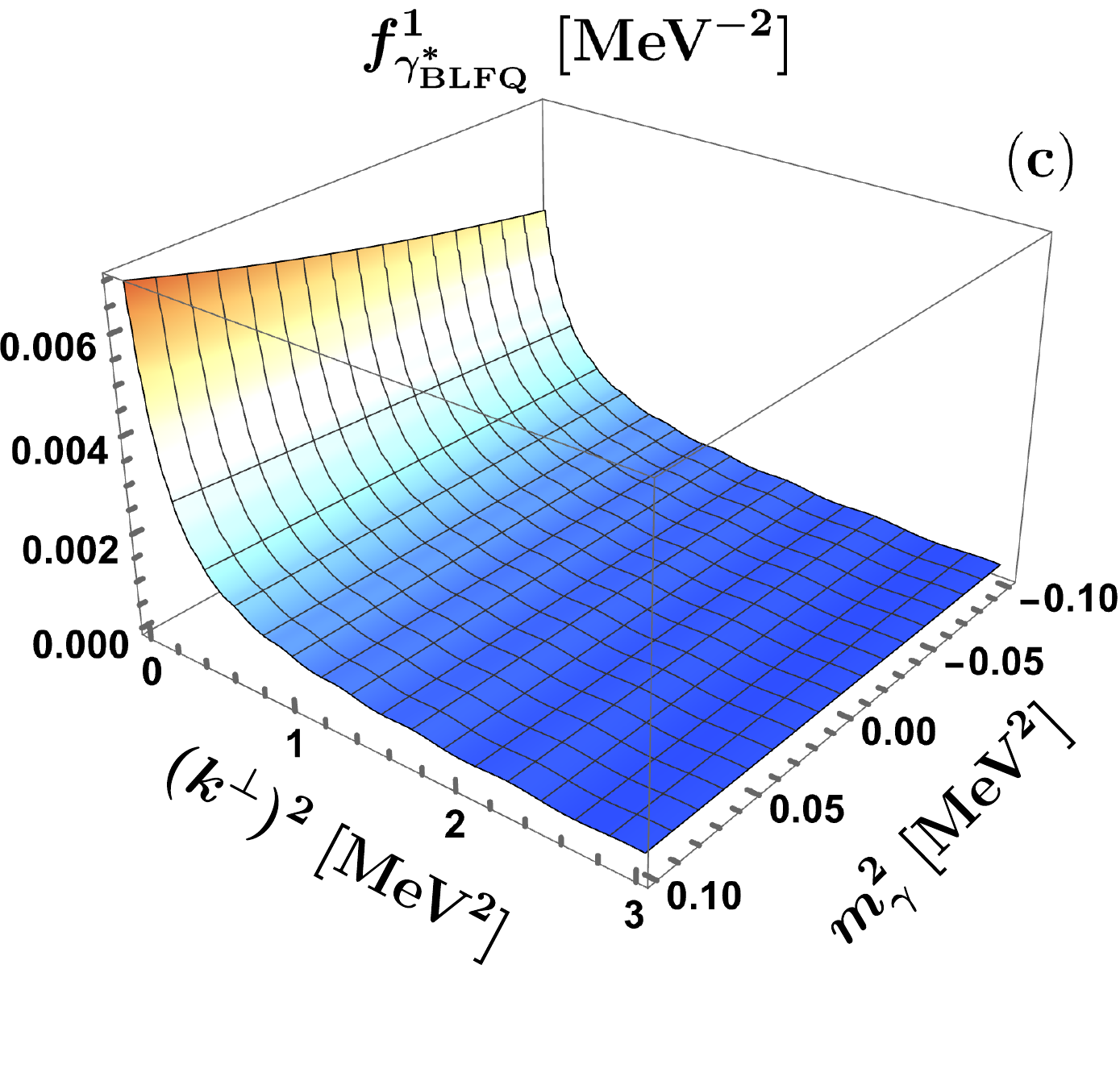}
		\includegraphics[width=8.4cm,height=7.5cm,clip]{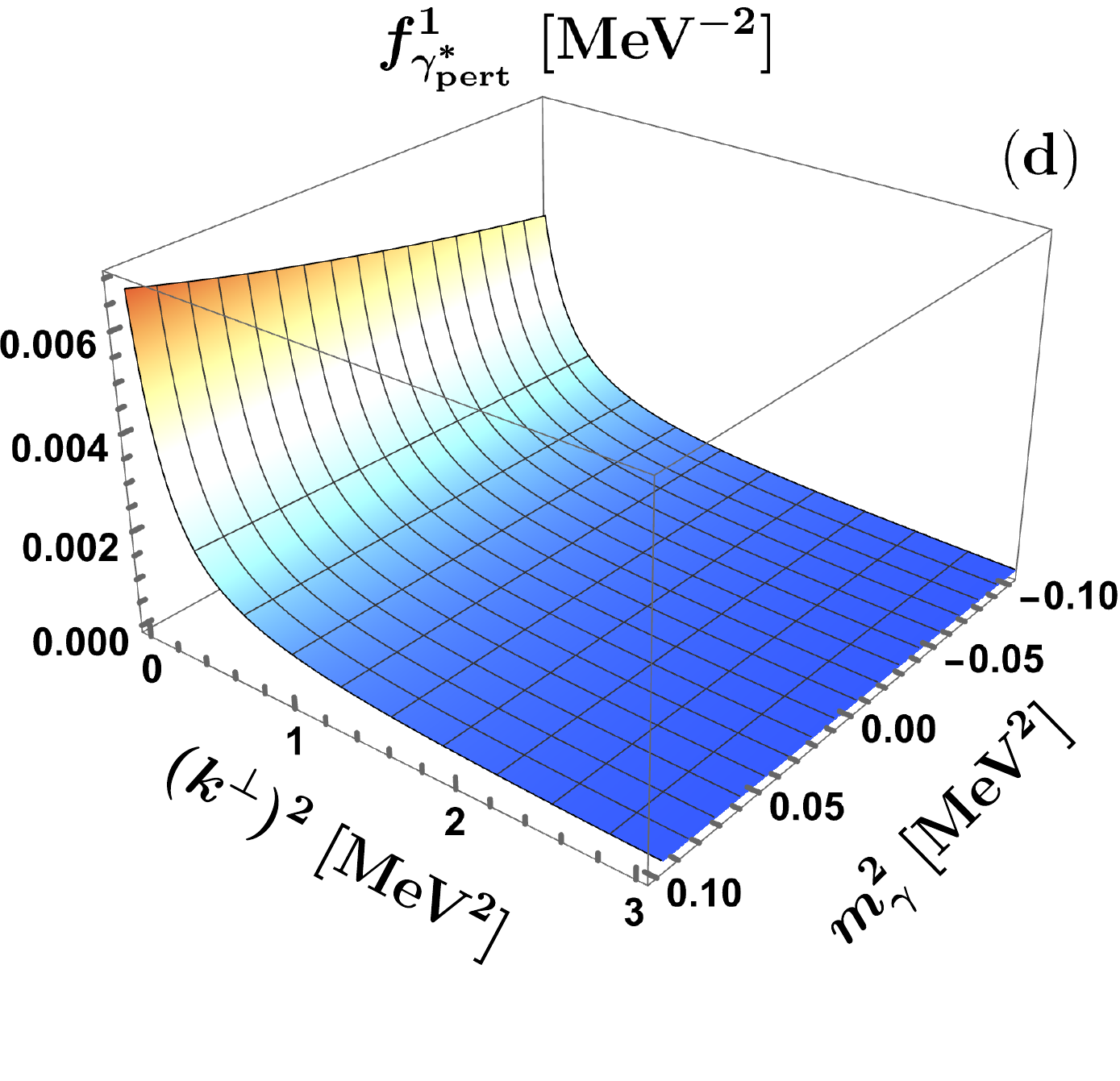}
		\caption{\label{fig9}  3D plots for virtual photon unpolarized TMD $f^1_{\gamma^{*}}(x,(k^{\perp})^2)$, where plots (a) and (b) are performed as functions of photon mass squared $m_{\gamma}^2$ and $x$ at $(k^{\perp})^2 = 0.01 ~\mathrm{MeV}^2$. Plots (c) and (d) are performed as functions of $m_{\gamma}^2$ and $(k^{\perp})^2$ at $x = 0.49$. We compare our BLFQ computations with the perturbative results. The BLFQ results are obtained by averaging over the BLFQ computations at $N_{\mathrm{max}} = \{46,48,50\}$ for $K = 50$. }
	\end{figure}
	
	It is interesting to employ our approach in order to obtain TMDs for both space-like and time-like virtual photons. Following the discussion in Sec.~\ref{secrenorm}, we therefore obtain BLFQ results for the virtual photon TMDs by renormalizing the bare photon mass squared to have a nonzero value. 
	
	Figures~\ref{fig7}(a) and \ref{fig7}(b) present the results for virtual photon unpolarized TMD $f^1_{\gamma}$ versus $(k^{\perp})^2$ for three fixed values of $x = (0.05,\, 0.51,\, 0.81)$. Figures~\ref{fig7}(c) and \ref{fig7}(d) are for $f^1_{\gamma}$ versus $x$ for three fixed values of $(k^{\perp})^2 = (0.01,\,0.1,\,1.0)~ \mathrm{MeV^2}$. Plots (a) and (c) are for $m_{\gamma}^2 = 0.1 ~\mathrm{MeV}^2$, whereas plots (b) and (d) are for $m_{\gamma}^2 = - 0.1 ~\mathrm{MeV}^2$.
	Unlike the real photon TMDs (Fig.~\ref{fig5}), the TMDs for the virtual photon at $(k^{\perp})^2 = 0$ are not independent of $x$. For fixed value of $(k^{\perp})^2$, we observe that the TMDs show a maximum (minimum) around $x=0.5$, when $m_{\gamma}^2 > 0$ ($m_{\gamma}^2 < 0$).

	We show the 3D plot for the virtual photon unpolarized TMD for $m_{\gamma}^2 = 0.1 ~\mathrm{MeV}^2$ in Figs.~\ref{fig8}(a) and \ref{fig8}(b), whereas Figs.~\ref{fig8}(c) and \ref{fig8}(d) are for $m_{\gamma}^2 = - 0.1 ~\mathrm{MeV}^2$. Our BLFQ results are compared with the corresponding perturbative results. The effect of nonzero mass squared is mostly concentrated around the  small $(k^{\perp})^2$ region.  When compared with the real photon unpolarized TMD (Fig.~\ref{fig6}), we observe that the $x$ dependence of the virtual photon TMDs behave differently near $(k^{\perp})^2 \sim 0$. This is due to the fact that the photon mass squared term is accompanied by an $x$-dependent factor as can be seen in the analytic expression of the perturbative results in Eq.~(\ref{perttmd}) and this mass squared term becomes dominant near $(k^{\perp})^2 \sim 0$. The oscillations observed in the BLFQ results are again indicators of the size of our finite basis artifacts.
	
	In Figs.~\ref{fig9}(a) and \ref{fig9}(b), we compare the BLFQ result with the perturbative result for the $f^1_{\gamma}$, where the TMDs are plotted as functions of $m_{\gamma}^2$ and $x$ at fixed value of $(k^{\perp})^2 = 0.01 ~\mathrm{MeV}^2$. We observe a minimum at $x=0.5$ for negative $m_{\gamma}^2$,  which transforms to a maximum for positive $m_{\gamma}^2$. The perturbative results exhibit similar behavior.  In Figs.~\ref{fig9}(c) and \ref{fig9}(d),  we compare the BLFQ result with the perturbative result, where the TMDs are plotted as functions of $m_{\gamma}^2$ and $(k^{\perp})^2$ at a fixed value of $x = 0.49$. Here, we observe that the effect of $m_{\gamma}^2$ is prominent around $(k^{\perp})^2 \approx 0 $. The maximum value of the TMDs decreases as the photon mass squared changes from positive to negative. Again, the perturbative results show excellent agreement with our BLFQ results.


	\begin{figure}[htp!]
		\centering
		\includegraphics[width=8.4cm,height=6.5cm,clip]{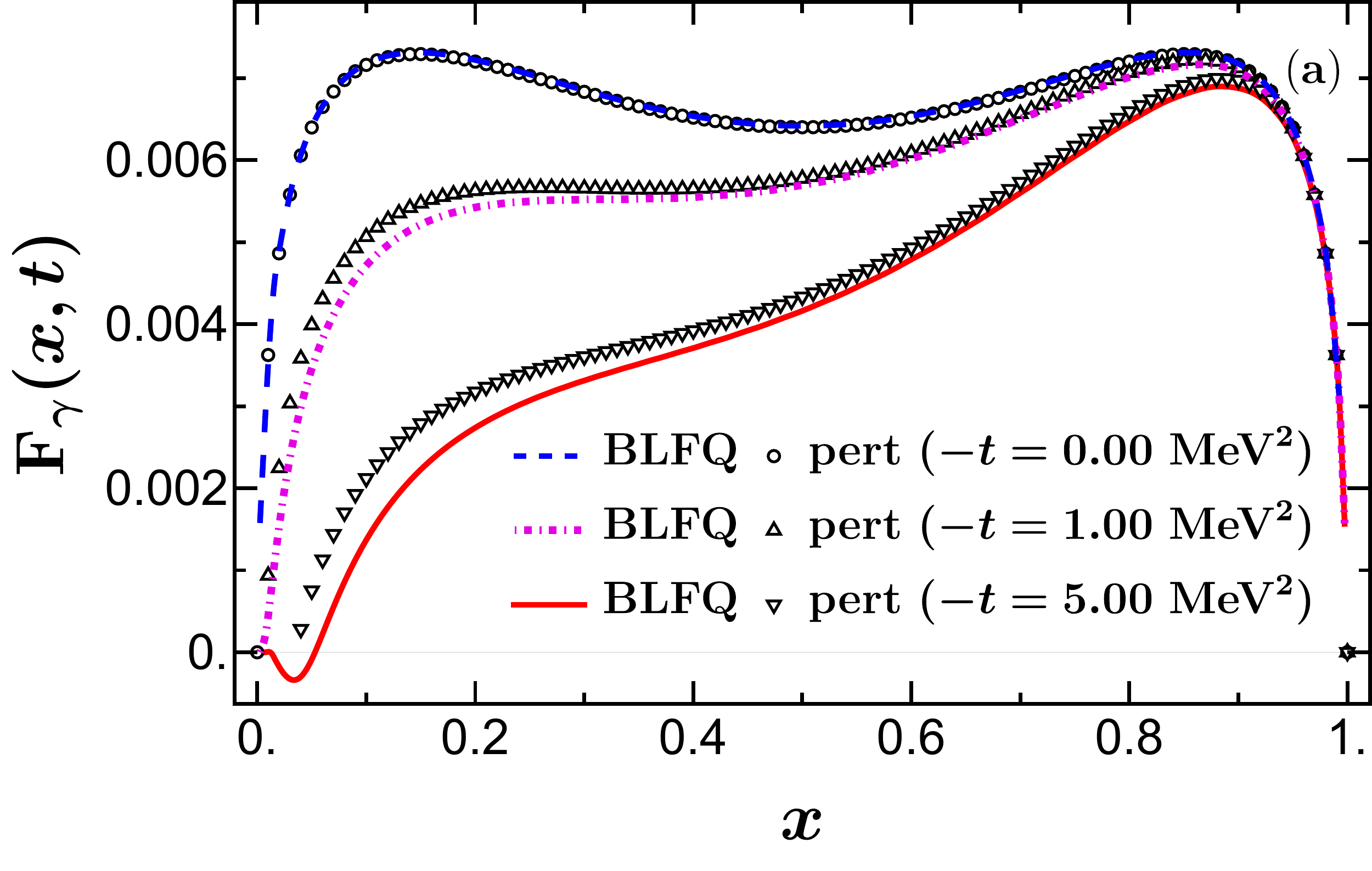}
		\includegraphics[width=8.4cm,height=6.5cm,clip]{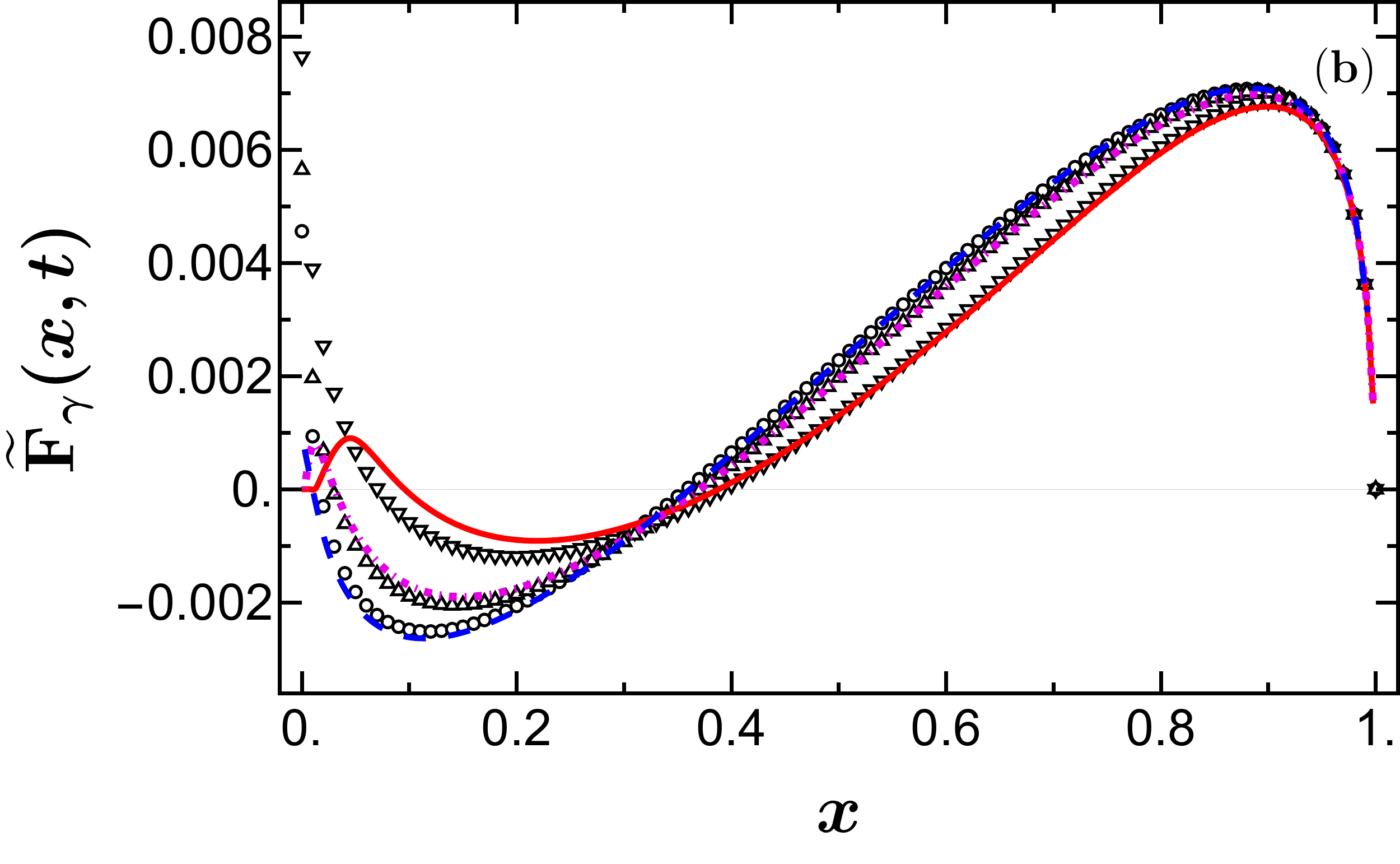}
		\includegraphics[width=8.4cm,height=6.5cm,clip]{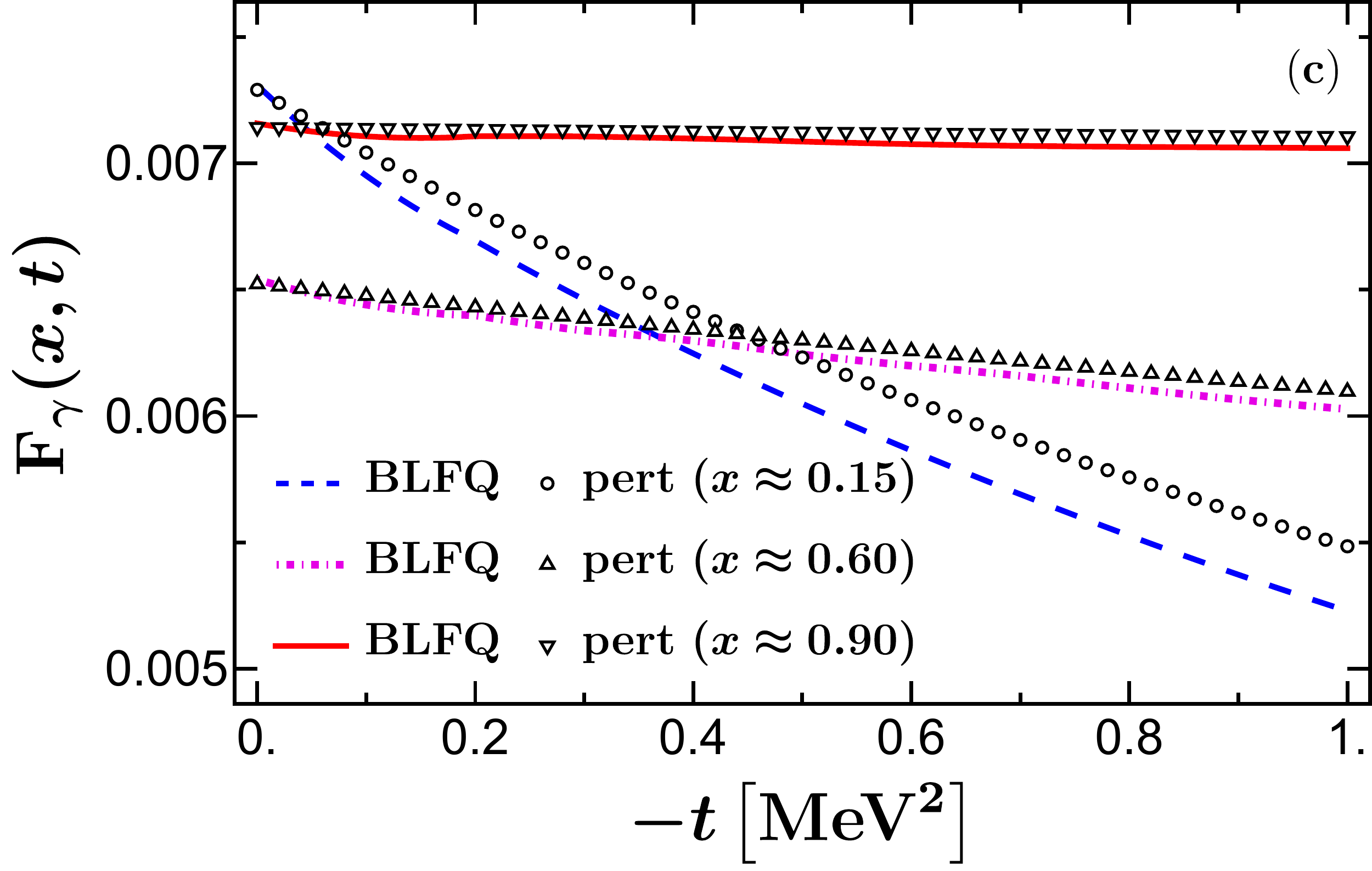}
		\includegraphics[width=8.4cm,height=6.5cm,clip]{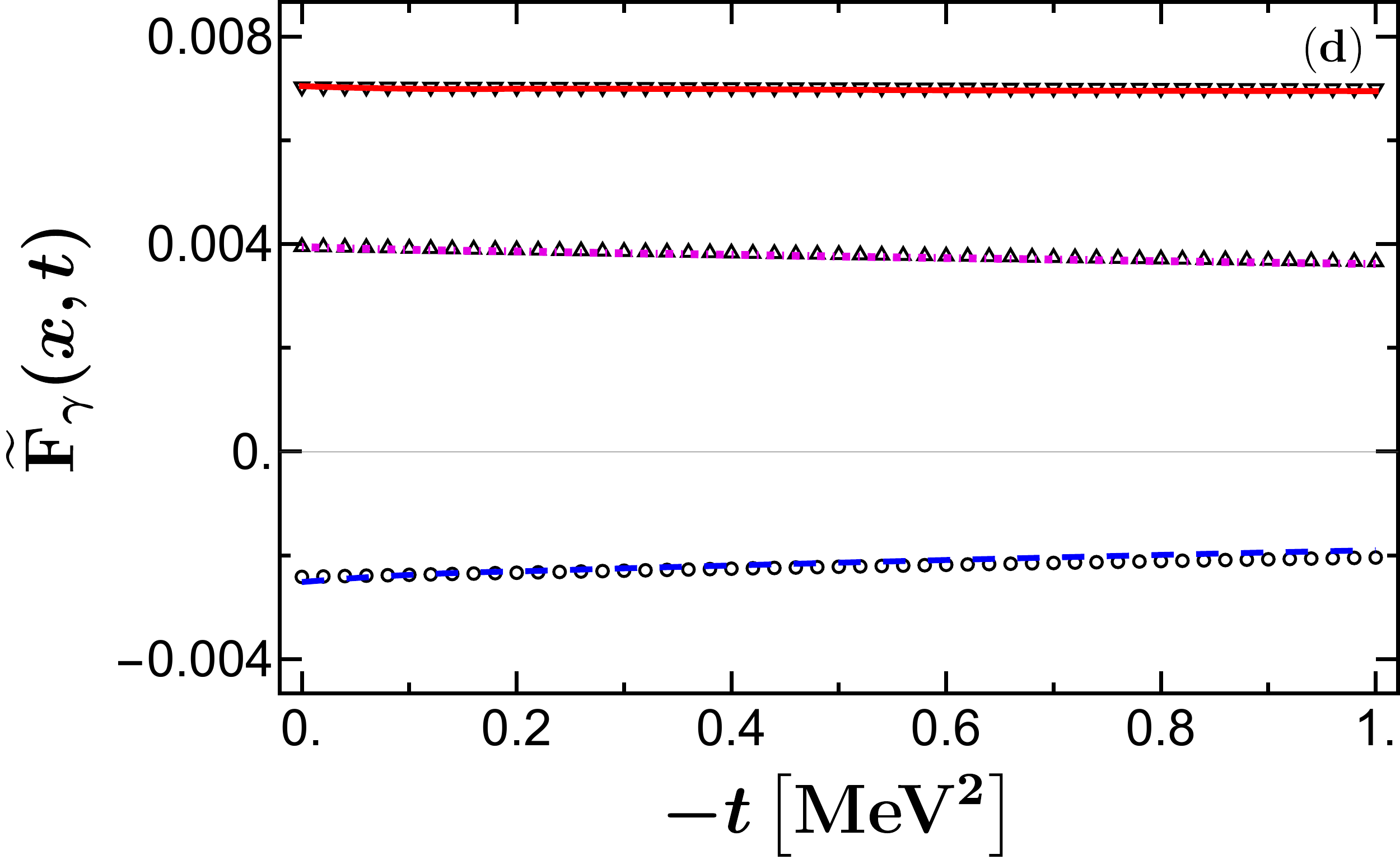}
		\caption{\label{fig10} 
			Plot (a) for the photon unpolarized GPD and plot (b) for the polarized  GPD vs $x$ for three fixed values of $-t = (0.00,1.00,5.00)~\mathrm{MeV}^2$. Plot (c) for the photon unpolarized GPD and plot (d) for the polarized GPD vs $-t ~(\mathrm{MeV}^2)$ for three fixed values of $x = (0.155,0.605,0.905)$. We compare our results (lines) with the perturbative results (symbols).  The BLFQ results are obtained at $N_{\mathrm{max}} = K = 200$.}
	\end{figure}

	In Figs.~\ref{fig10}(a) and \ref{fig10}(b), we show the results for the real photon unpolarized GPD $F(x,t)$ and polarized GPD $\tilde{F}(x,t)$ as functions of $x$ for three values of the momentum transfer $-t = (0.00,\, 1.00,\, 5.00)~\mathrm{MeV^2}$. The BLFQ computations are compared with the corresponding perturbative results. We observe that with increasing transverse momentum transfer $-t$, the BLFQ results in the low-$x$ region deviate from the perturbative results. For $-t = 0 ~\mathrm{MeV^2}$, we observe that the unpolarized photon GPD is symmetric over $x$, since both electron and positron are equally massive. For nonzero values of $-t$, this symmetry is broken and it becomes more asymmetric with increasing $-t$. The polarized GPD changes sign in the region $x < 0.4$. The change of sign observed in the polarized GPD depends on the transverse momentum transfer $-t$ and the electron mass $m_e$ as can be seen from the analytic expression of the perturbative result in Eq.~(\ref{gpd_pert}). Both the unpolarized and the polarized  GPDs become independent of  $-t$ when the momentum fraction carried by the electron approaches unity.
	
	In Figs.~\ref{fig10}(c) and \ref{fig10}(d), we present the results for the real photon unpolarized GPD $F_{\gamma}(x,t)$ and polarized GPD $\tilde{F}_{\gamma}(x,t)$ as functions of $-t~(\mathrm{MeV^2})$ for three values of $x = (0.155,\, 0.605,\ 0.905)$. The BLFQ computations for the unpolarized GPD $F_{\gamma}(x,t)$  deviate from the perturbative results as the value of $x$ decreases and $-t$ increases. This deviation is less prominent for the polarized GPD $\tilde{F}_{\gamma}(x,t)$. 
	
\begin{figure}[h]
		\centering
		\includegraphics[width=8.4cm,height=7.5cm,clip]{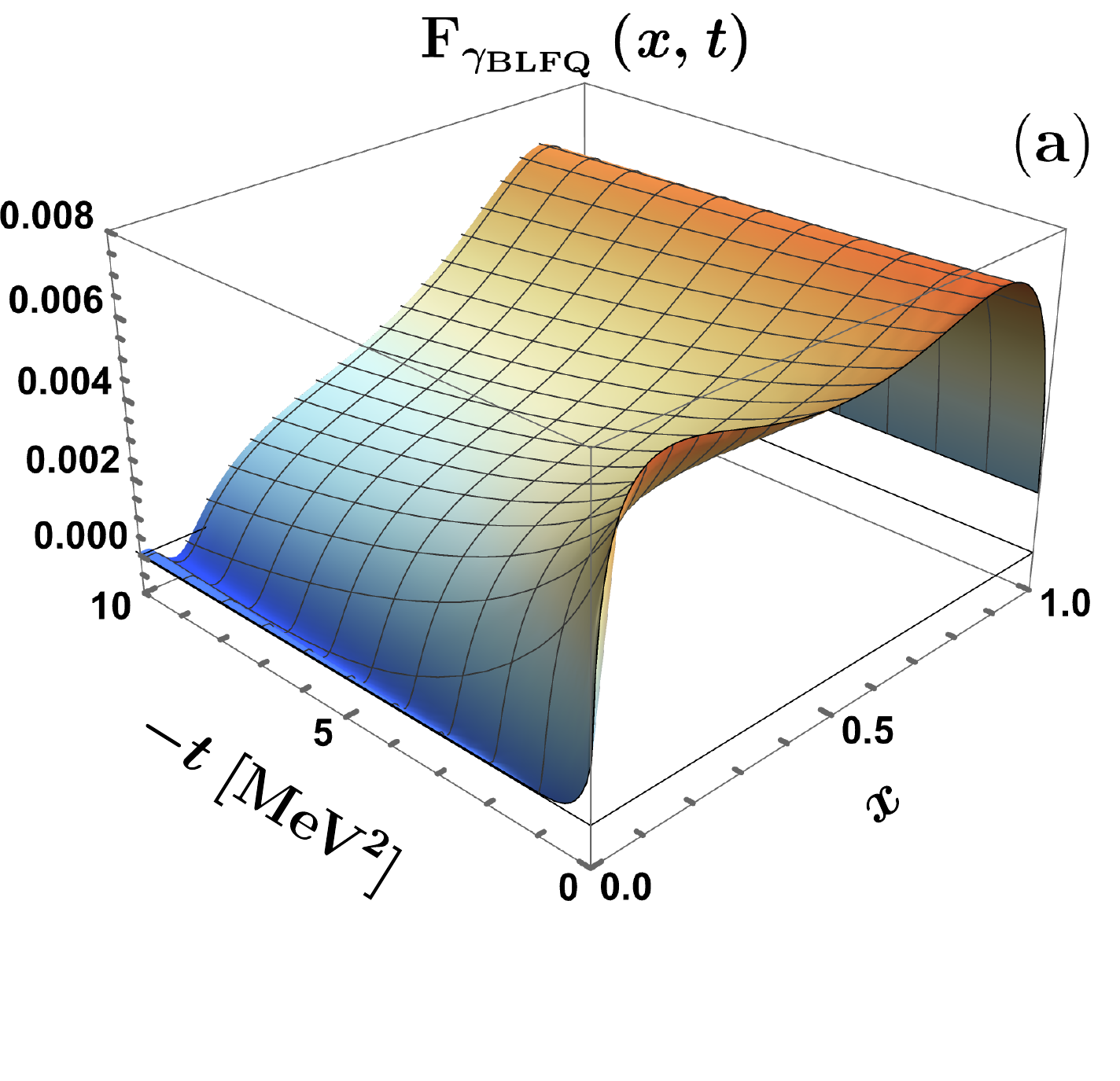}
		\includegraphics[width=8.4cm,height=7.5cm,clip]{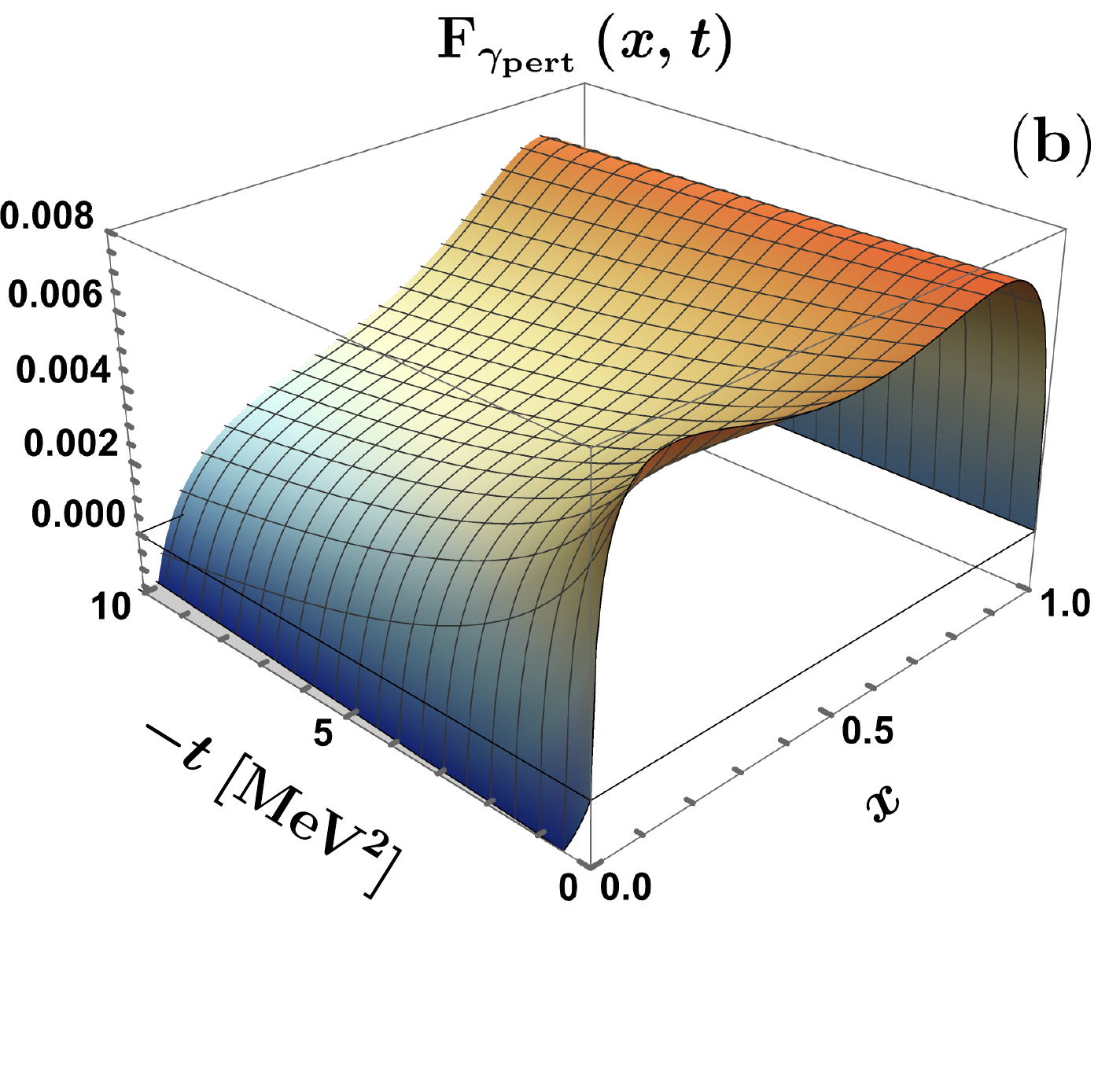}\\
		\includegraphics[width=8.4cm,height=7.5cm,clip]{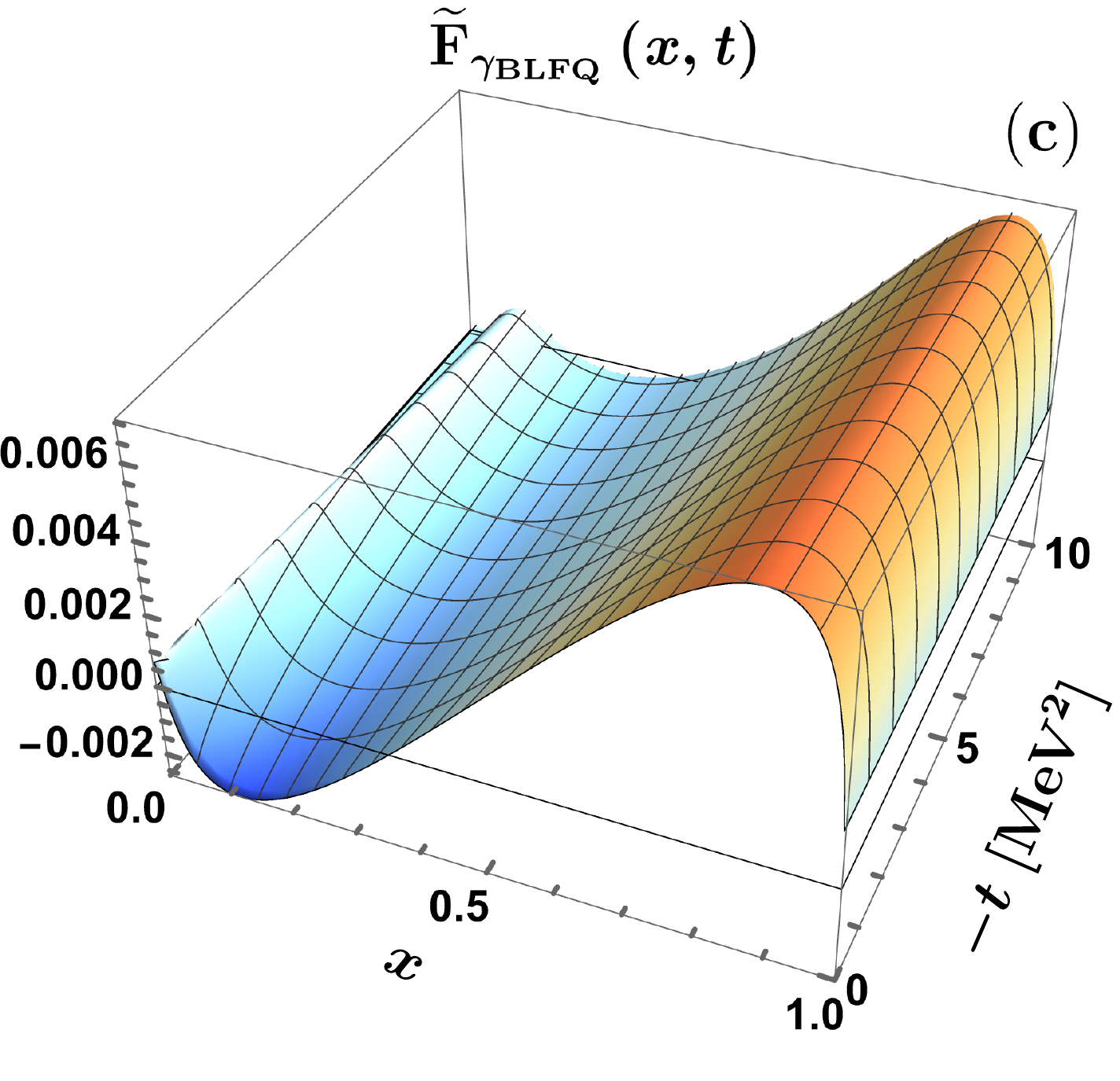}
		\includegraphics[width=8.4cm,height=7.5cm,clip]{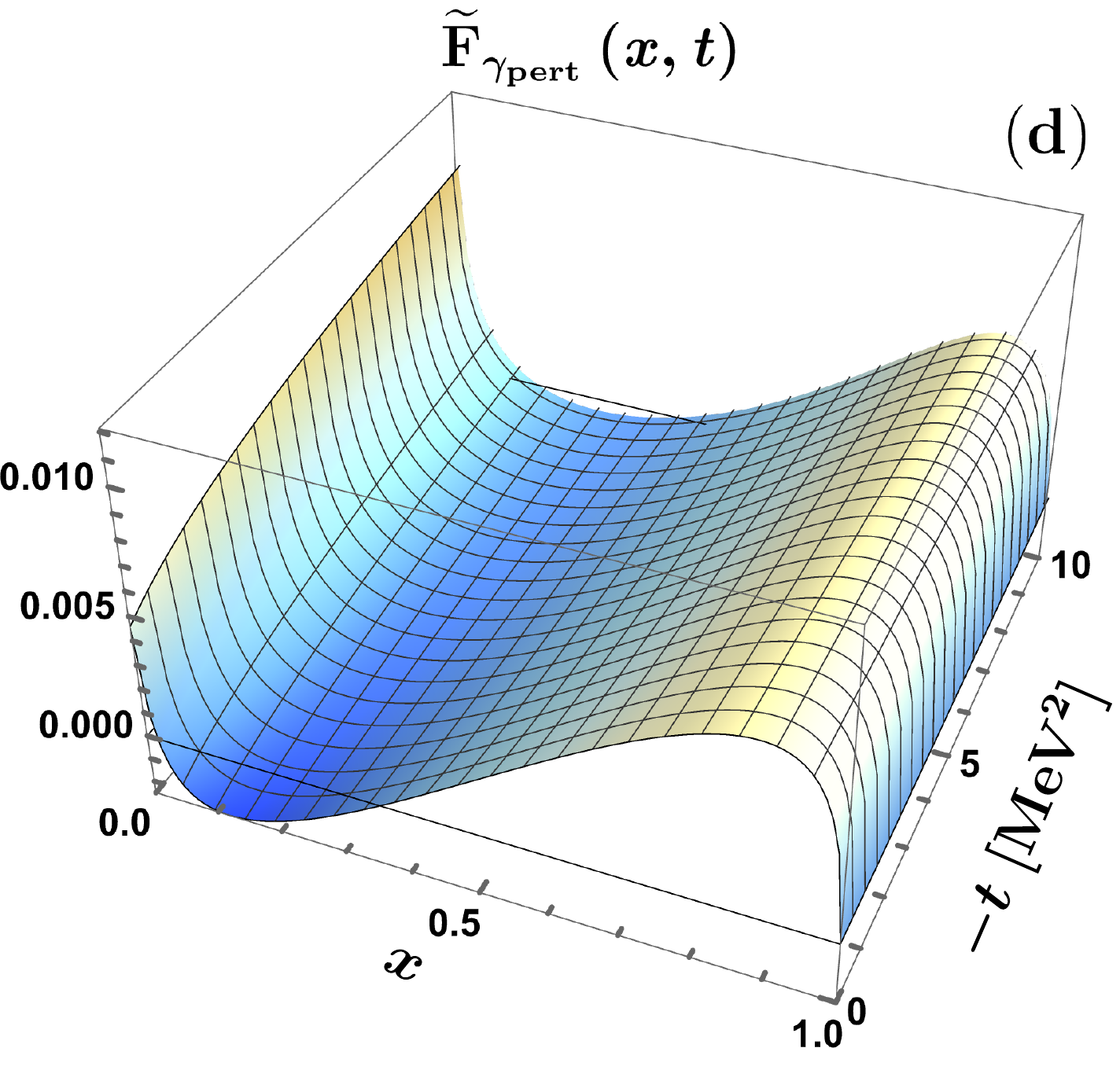}
		\caption{\label{fig11} 
			3D plots for the photon unpolarized and polarized GPDs.  Plots (a) and (b) are the unpolarized GPD from BLFQ and the perturbation theory, respectively. Plots (c) and (d) are the polarized GPD from BLFQ and the perturbation theory, respectively. Here $N_{\mathrm{max}} = K = 200$. }
	\end{figure}

	Figure~\ref{fig11} illustrates the 3D plots for the photon unpolarized and polarized GPDs calculated at $N_{\mathrm{max}} = K = 200$. The photon GPDs are nearly independent of $-t$  for large $x$ and the $-t$ dependence can be seen when $x$ decreases. This behavior can be understood from the analytic expression for the perturbative results shown in Eq.~(\ref{int123}), where the $-t$ dependence comes with the factor $(1-x)^2$. We observe that the qualitative behavior of the BLFQ computations for the photon GPDs is in agreement with the perturbative results.

\begin{figure}[h]
		\centering
		\includegraphics[width=7.5cm,height=6cm,clip]{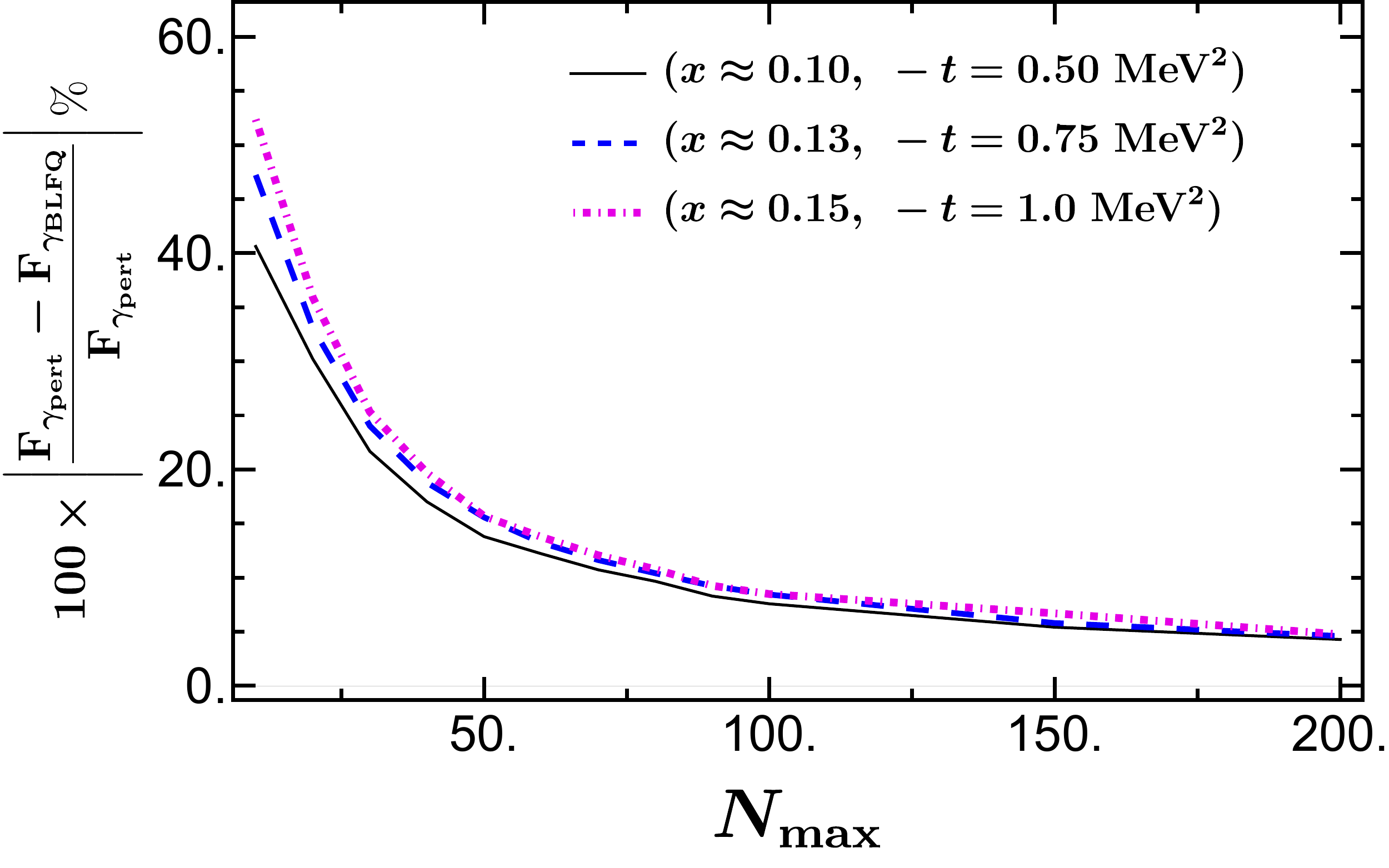}
		\caption{\label{fig12} 
			Plot for the relative percentage difference between the GPDs calculated in BLFQ ($F_{\gamma_{\mathrm{BLFQ}}}(x,t)$) and with perturbation theory ($F_{\gamma_{\mathrm{pert}}}(x,t)$) as a function of the basis truncation parameter $N_{\mathrm{max}} = K$.}
	\end{figure}

	The differences observed in Fig.~\ref{fig11} between the BLFQ and perturbative results for the GPDs are quantified in Fig.~\ref{fig12}, where we plot the relative percentage difference between $F_{\gamma_{\mathrm{BLFQ}}}(x,t)$ and $F_{\gamma_{\mathrm{pert}}}(x,t)$ as a function of $N_{\mathrm{max}}$. We choose three values for the pair $(x,t)$ such that the deviation is clearly visible at the scale of the plots shown in Fig.~\ref{fig11}.
	We observe that as $N_{\mathrm{max}}$ increases this deviation decreases and we expect that the two results will converge as $N_{\mathrm{max}} \rightarrow \infty$. At $N_{\mathrm{max}} = 200$, the deviation is less than $5 \%$ for the chosen values of the pair $(x,t)$.
	
	\begin{figure}[h]
		\centering
		\includegraphics[width=7.5cm,height=5.0cm,clip]{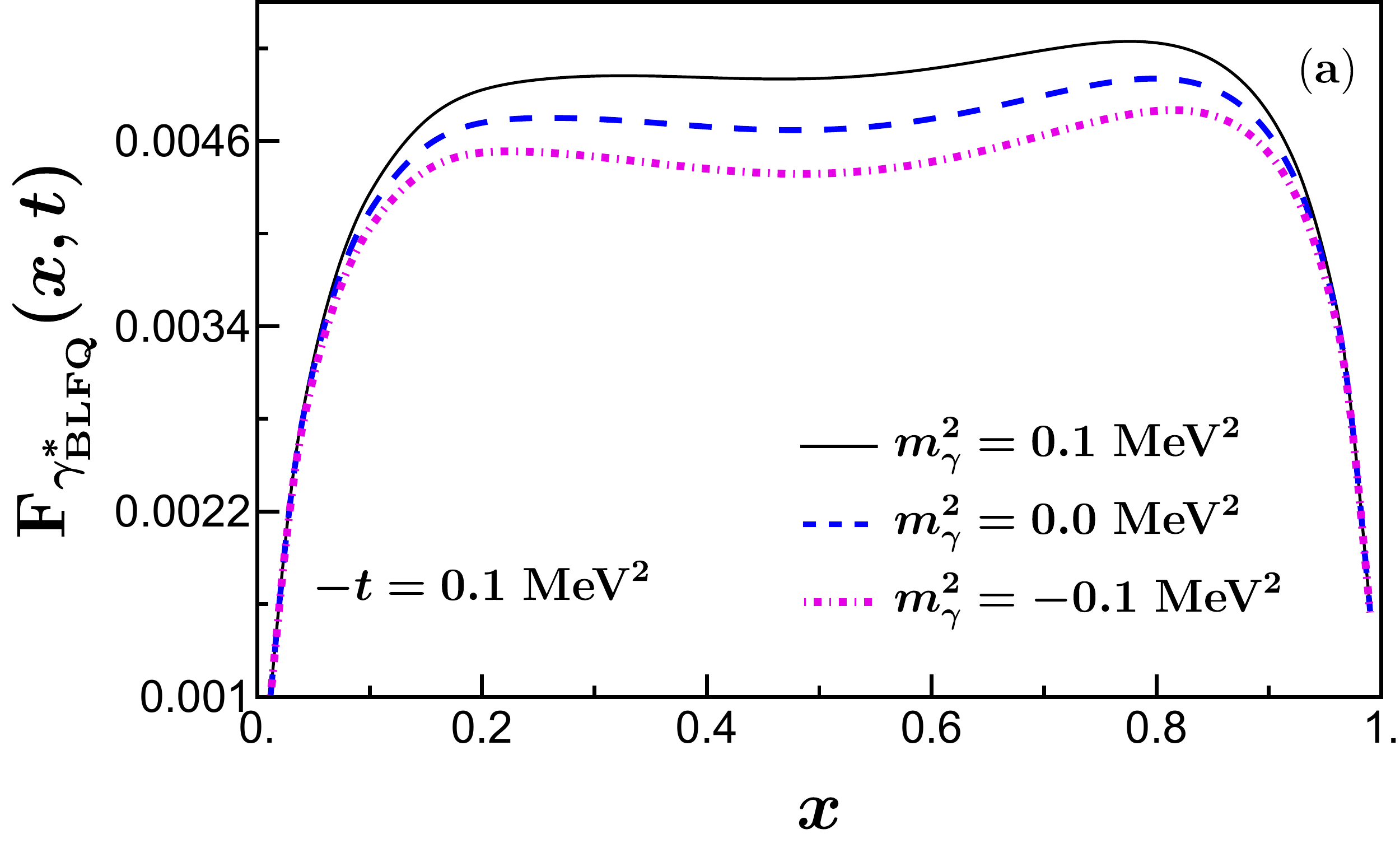}
		\includegraphics[width=7.5cm,height=5.0cm,clip]{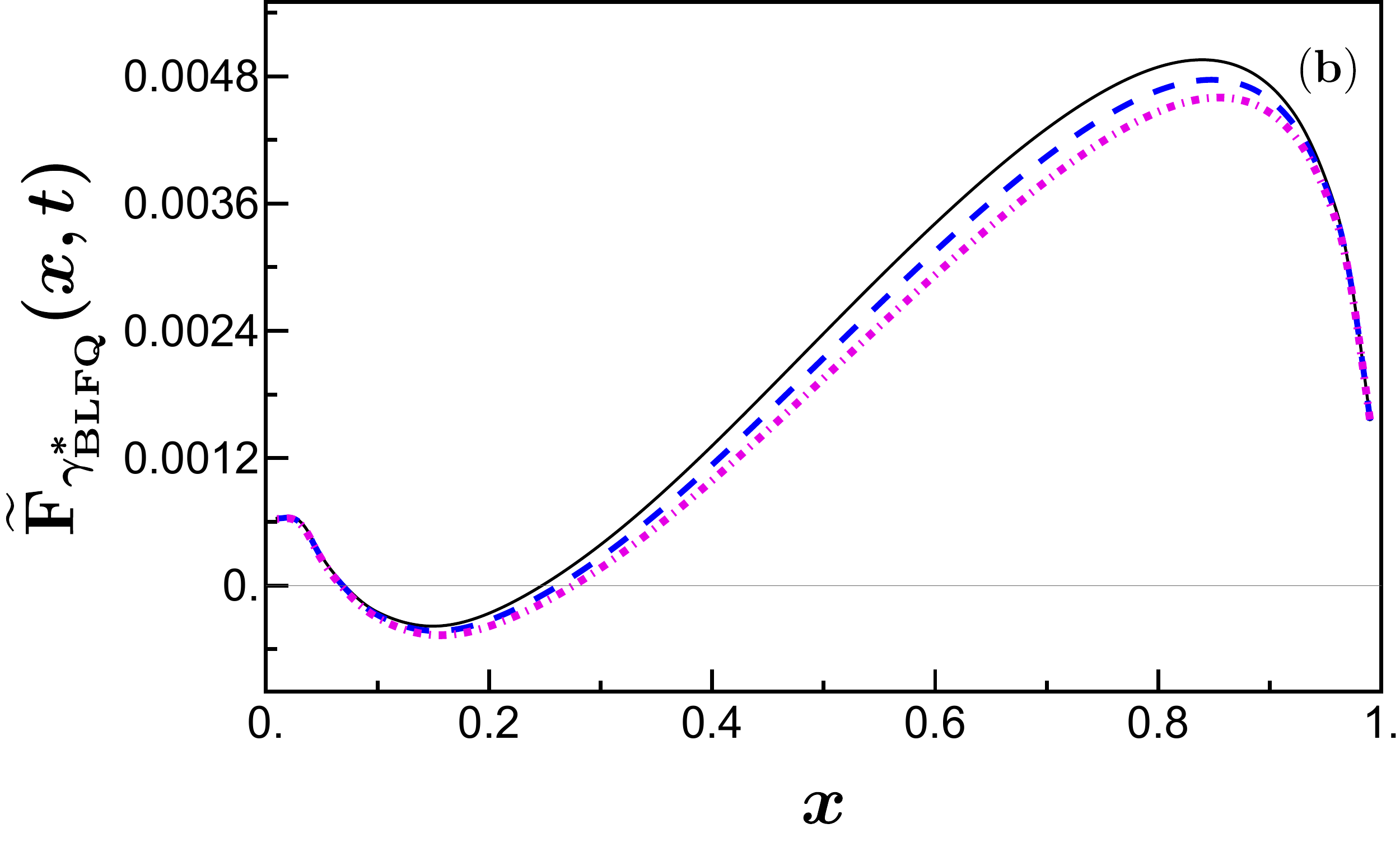}\\
		\includegraphics[width=7.5cm,height=5.0cm,clip]{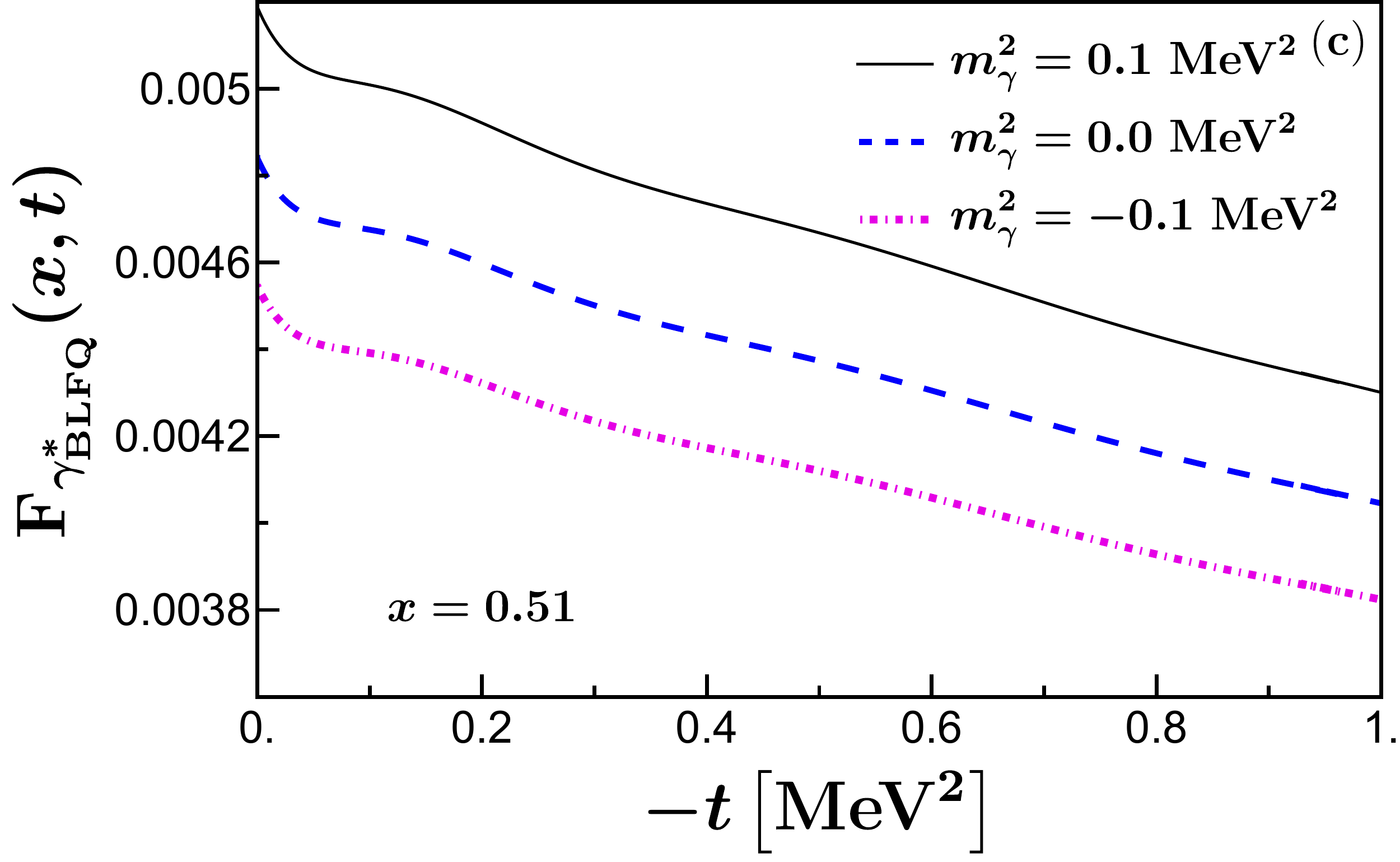}
		\includegraphics[width=7.5cm,height=5.0cm,clip]{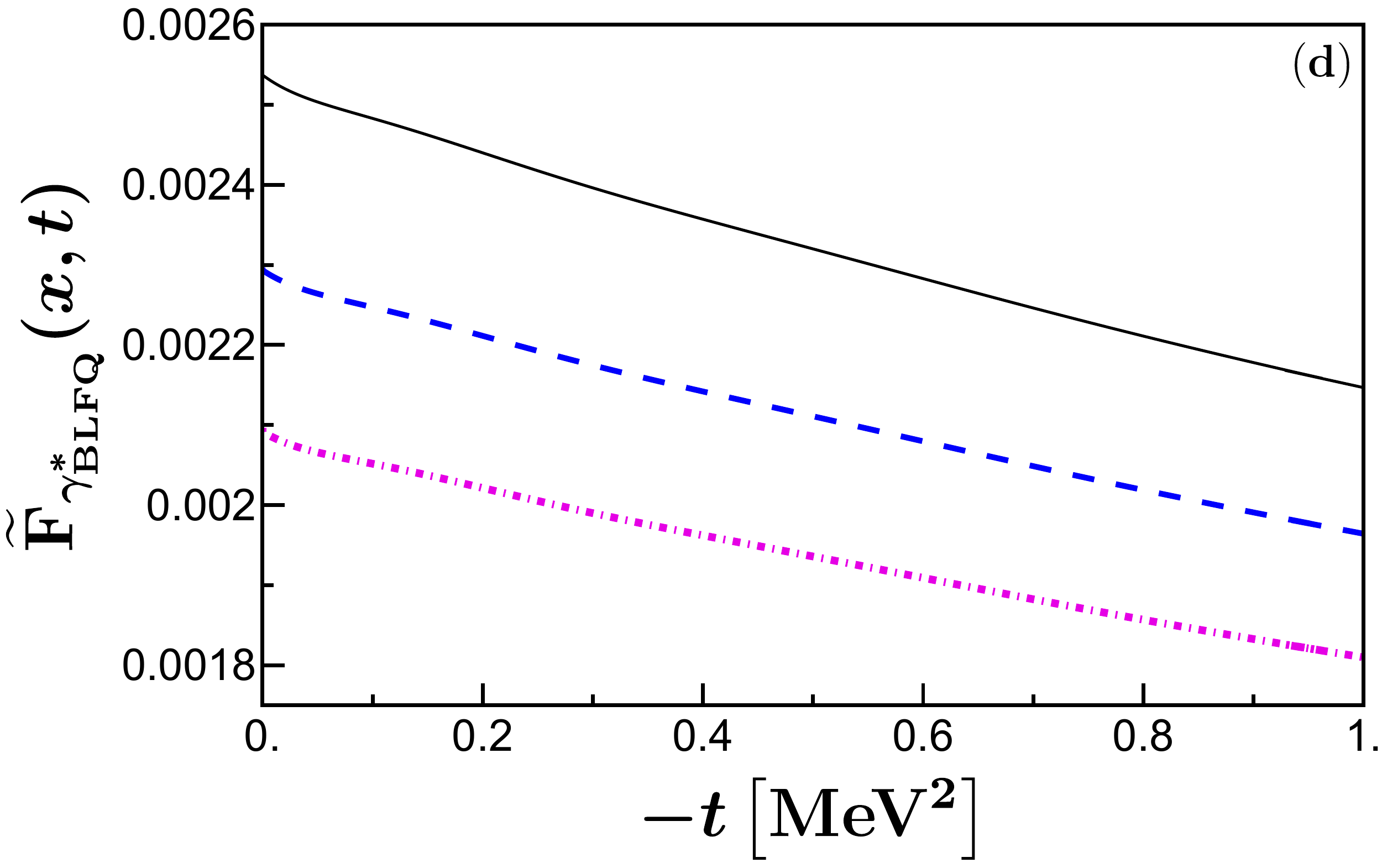}
		\caption{\label{fig13} 
			Comparison between the real and the virtual photon GPDs. Plots (a) and (b) are for the GPDs vs $x$ for a fixed value of $-t = 0.1~\mathrm{MeV}^2$ whereas plots (c) and (d) are vs $-t$ for a fixed value of $x = 0.51$.
			All plots are shown for three values of the photon mass $m_{\gamma}^2 = \left(-0.1,0.0,0.1\right) \mathrm{MeV}^2$. Here $N_{\mathrm{max}} = K = 50$.}
	\end{figure}

	\begin{figure}[h]
		\centering
		\includegraphics[width=8.4cm,height=7.5cm,clip]{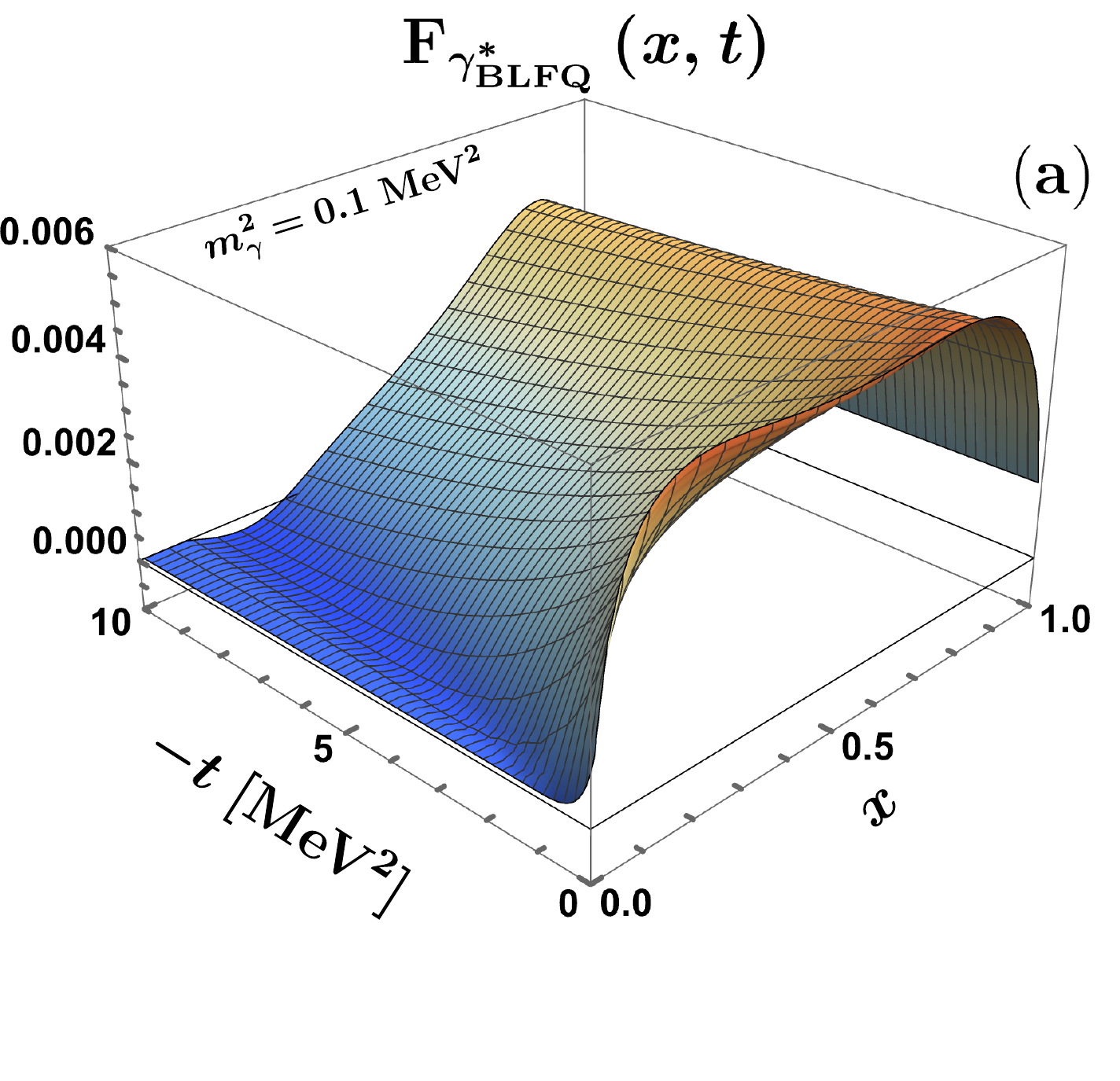}
		\includegraphics[width=8.4cm,height=7.5cm,clip]{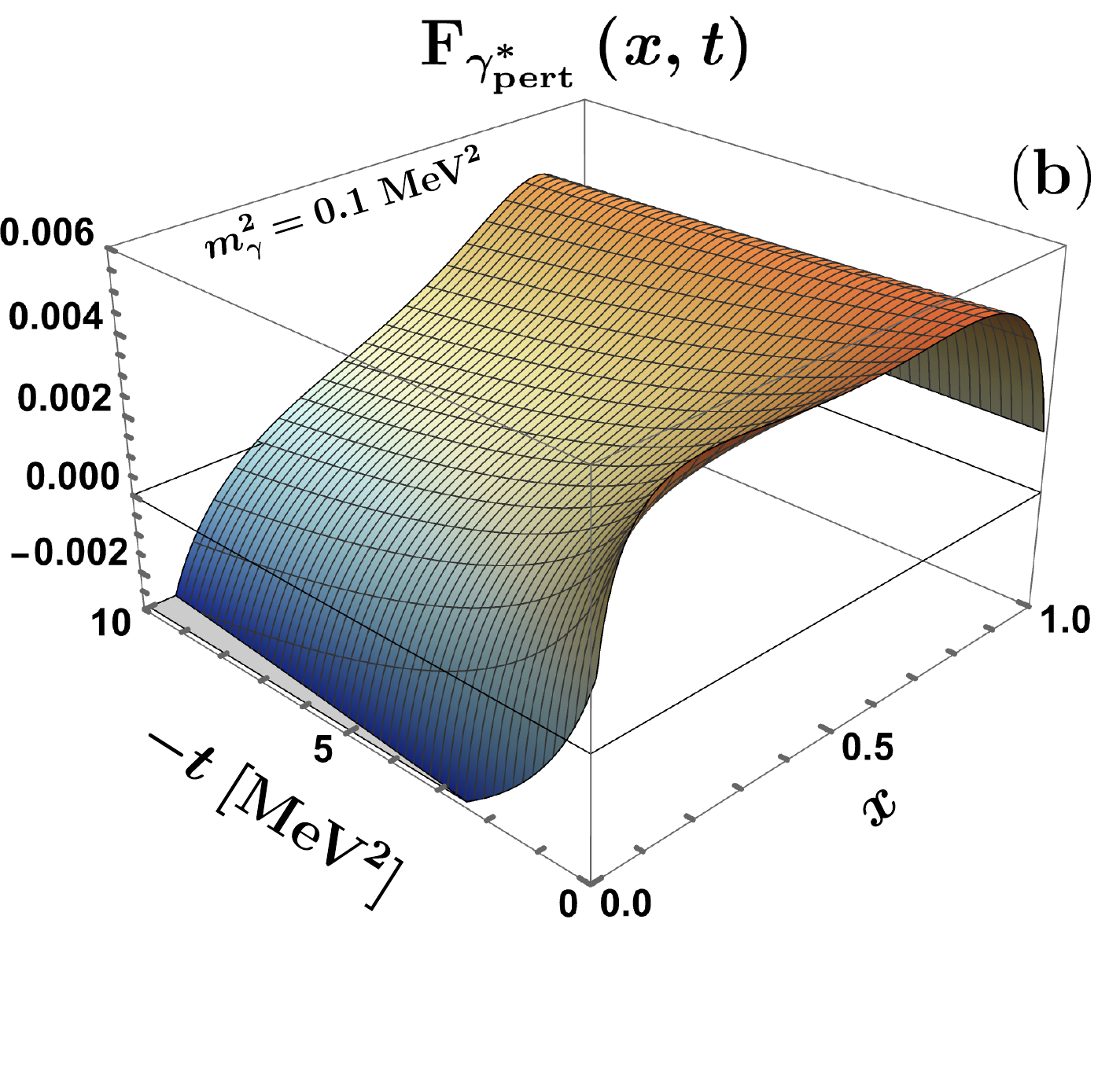}\\
		\includegraphics[width=8.4cm,height=7.5cm,clip]{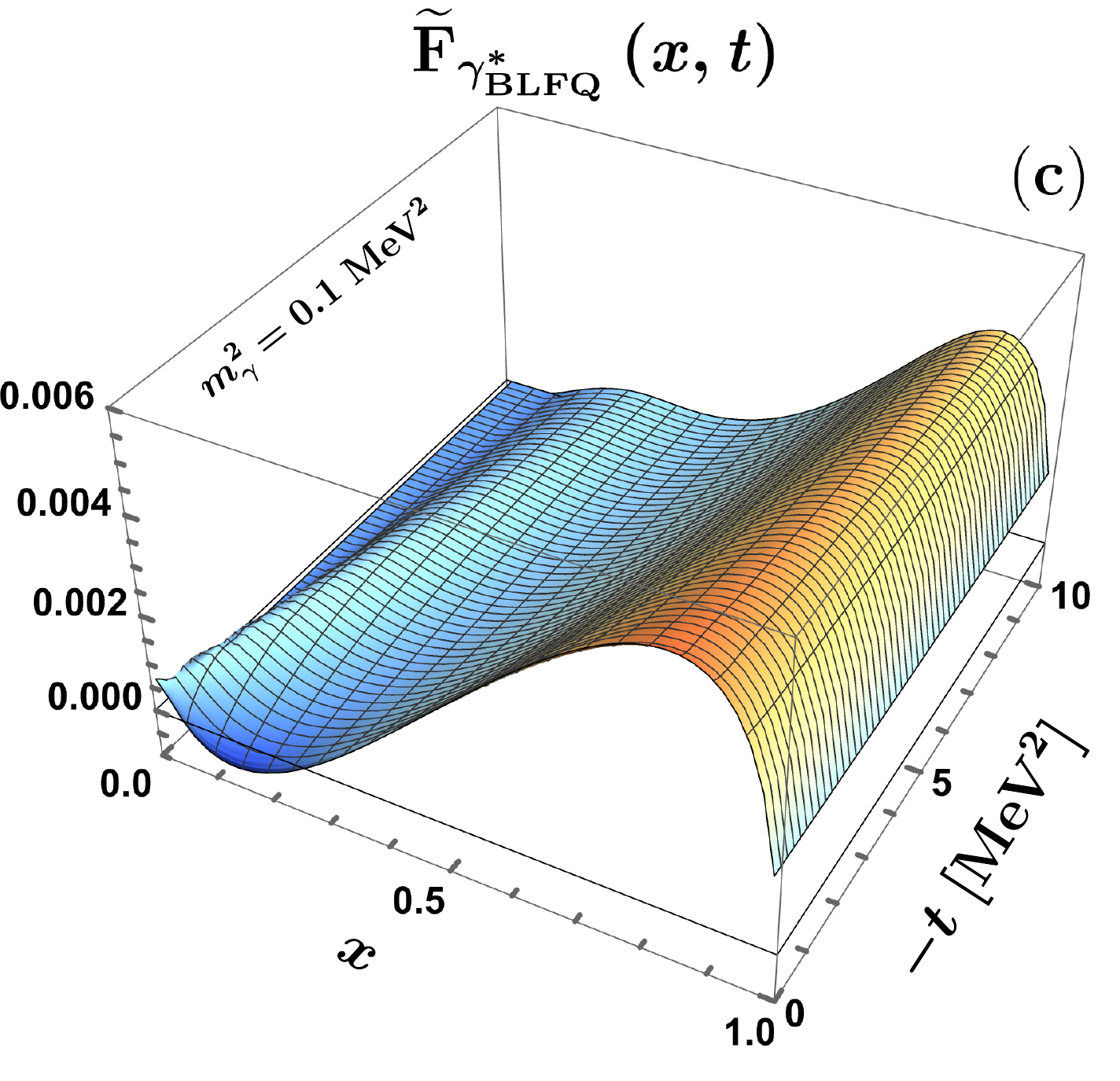}
		\includegraphics[width=8.4cm,height=7.5cm,clip]{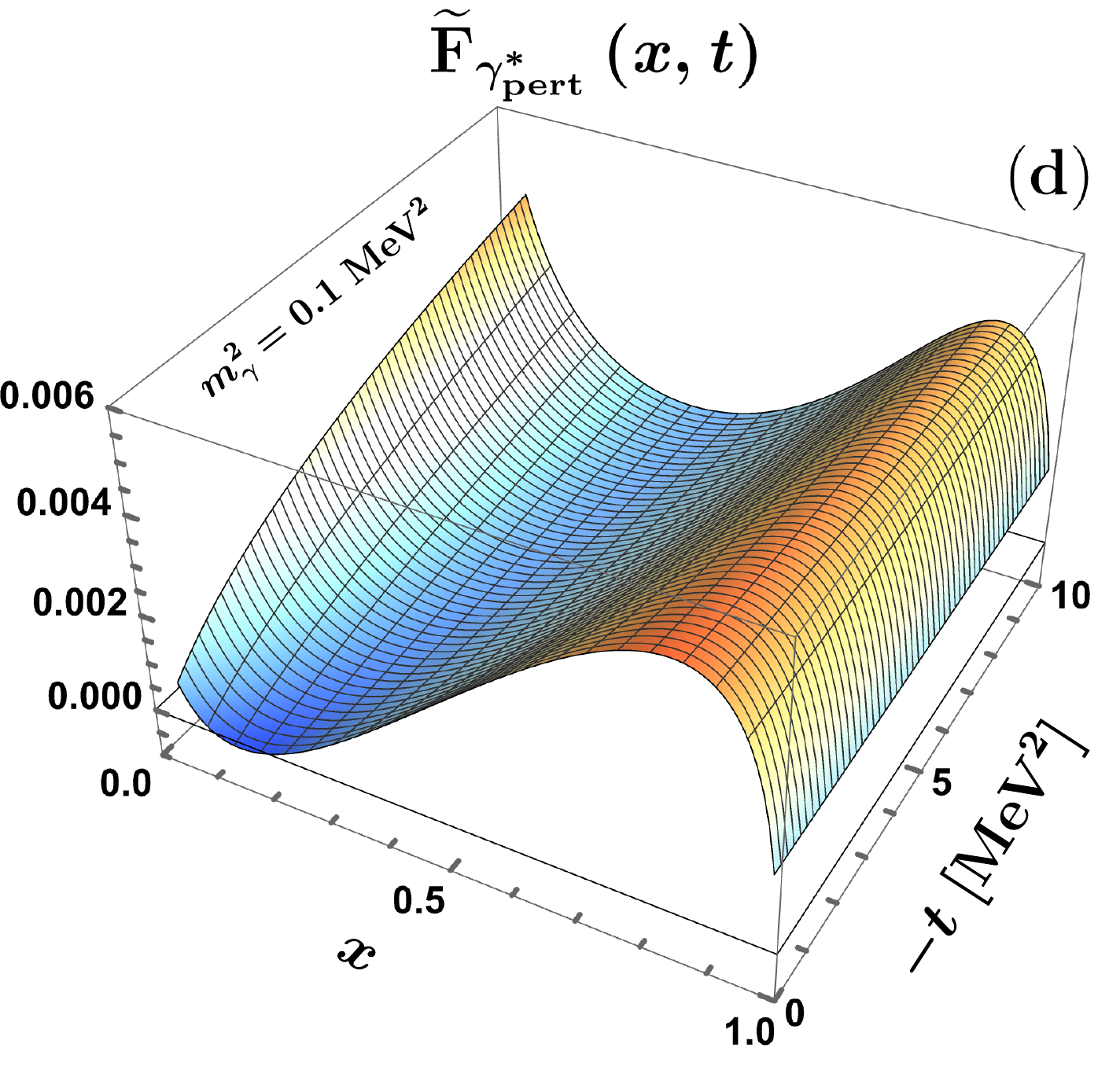}
		\caption{\label{fig14} 
			3D plots for the virtual photon GPDs. Plots (a) and (b) are the unpolarized GPD results from BLFQ and the
			perturbation theory, respectively. Plots (c) and (d) are the polarized GPD results from BLFQ and the
			perturbation theory, respectively. The photon mass is set to $m_{\gamma}^2 = 0.1 ~\mathrm{MeV}^2$ and
			$N_{\mathrm{max}} = K = 50$.}
	\end{figure}
	
	Figure~\ref{fig13} compares the BLFQ results for the virtual photon GPDs with those of the real photon. In plots \ref{fig13}(a) and \ref{fig13}(b), we present the photon GPDs as functions of $x$ for fixed value of $-t = 0.1~\mathrm{MeV}^2$ and in plots \ref{fig13}(c) and \ref{fig13}(d), we show them as function of $-t$ for fixed value of $x = 0.51$. Plots \ref{fig13}(a) and \ref{fig13}(c) are the results for the unpolarized GPDs, whereas plots \ref{fig13}(b) and \ref{fig13}(d) are the results for the polarized GPDs. We compute the virtual GPDs for two values of the photon mass $m_{\gamma}^2 = 0.1 ~\mathrm{MeV}^2$ and  $m_{\gamma}^2 = -0.1 ~\mathrm{MeV}^2$.  
	We observe that the GPDs of the time-like virtual photon, i.e., with positive $m_{\gamma}^2$ have a higher magnitude, whereas the one with negative $m_{\gamma}^2$, i.e., for the space-like virtuality, has a lower magnitude when compared to the real photon. As observed in the  TMDs (see Fig.~\ref{fig6} and Fig.~\ref{fig9}), the difference between the real and the virtual photon TMDs are mostly localized around the low $k^{\perp}$ region. Since the $k^{\perp}$ direction is integrated out in GPDs, this difference is translated in the form of a change in magnitude as observed in Fig.~\ref{fig13}.
	
	In Fig.~\ref{fig14}, we illustrate the 3D structure for the virtual photon  unpolarized and polarized GPDs. The BLFQ computations are compared with the corresponding results from perturbation theory. As observed for the real photon GPDs in Fig.~\ref{fig11}, the difference between BLFQ results and the perturbative results lies in regions close to $x=0$. The qualitative behavior of the virtual photon GPDs is similar to those of the real photon.
	
	\section{Conclusion}
	\label{con}
	In this work, we obtained the real and virtual photon LFWFs
	as the eigenvectors of the QED Hamiltonian in the light-cone gauge by solving its mass
	eigenstates using the theoretical framework of Basis Light-Front Quantization (BLFQ). We investigated the structure of the photons by considering them within the constituent bare photon and electron-positron Fock sectors. We employed the resulting LFWFs to compute various observables of the photon such as the structure functions, TMDs, and GPDs. The BLFQ results were compared with those from leading-order perturbation theory. Our calculations also employed sector-dependent renormalization and the rescaling of the wavefunction, which is required to compensate for the artifacts coming from the Fock sector truncation. We found good consistency between the BLFQ results and the perturbative calculations. Thus, this work supports the reliability of the BLFQ approach in solving such stationary state problems. The virtual photon LFWFs can be further utilized to study exclusive vector meson production in virtual photon-proton or photon-nucleus scattering.

	\section{ACKNOWLEDGMENTS}
	S. N and C. M. thank the Chinese Academy of Sciences Presidents International Fellowship Initiative for their support via Grants No. 2021PM0021 and 2021PM0023, respectively. C. M. is supported by new faculty start-up funding from the Institute of Modern Physics, Chinese Academy of Sciences, Grant No. E129952YR0.  X. Z. is supported by new faculty startup funding by the Institute of Modern Physics, Chinese Academy of Sciences, by Key Research Program of Frontier Sciences, Chinese Academy of Sciences, Grant No. ZDB-SLY-7020, by the Natural Science Foundation of Gansu Province, China, Grant No. 20JR10RA067, by the Central Funds Guiding the Local Science and Technology Development of Gansu Province, Grant No. 22ZY1QA006 and by the Strategic Priority Research Program of the Chinese Academy of Sciences, Grant No. XDB34000000. J. P. V. is supported in part by the Department of Energy under Grants No. DE-FG02-87ER40371, No. DE-SC0018223 (SciDAC4/NUCLEI), and DE-SC0023495 (SciDAC5/NUCLEI).


\begin{thebibliography}{80}
		
		\bibitem{walsh} T.F. Walsh, P.M. Zerwas, Phys. Lett. B {\bf 44} (1974) 95.
		
		
		\bibitem{Nisius:1999cv}
		R.~Nisius,
		Phys. Rept. \textbf{332}, 165-317 (2000).
		
		\bibitem{Peterson:1982tt}
		C.~Peterson, P.~M.~Zerwas and T.~F.~Walsh,
		Nucl. Phys. B \textbf{229}, 301-316 (1983).
		
		
		
		\bibitem{Witten:1977ju}
		E.~Witten,
		Nucl. Phys. B \textbf{120}, 189-202 (1977).
		
		\bibitem{Berger:1981bh}
		C.~Berger \textit{et al.} [PLUTO],
		Phys. Lett. B \textbf{107}, 168-172 (1981).
		
		
		
		\bibitem{Friot:2006mm}
		S.~Friot, B.~Pire and L.~Szymanowski,
		Phys. Lett. B \textbf{645}, 153-160 (2007).
		
		
		
		\bibitem{Ji:1996nm}
		X.~D.~Ji,
		Phys. Rev. D \textbf{55}, 7114-7125 (1997).
		
		
		
		\bibitem{Radyushkin:1996nd}
		A.~V.~Radyushkin,
		Phys. Lett. B \textbf{380}, 417-425 (1996).
		
		\bibitem{Mueller:1998fv}
		D.~M\"uller, D.~Robaschik, B.~Geyer, F.~M.~Dittes and J.~Ho\v{r}ej\v{s}i,
		Fortsch. Phys. \textbf{42}, 101-141 (1994).
		
		
		\bibitem{gpd_all}
		For reviews on generalized parton distributions, and DVCS, see 
		M.~Diehl, Phys. Rep. \textbf{388} (2003) 41;
		A.~V.~Belitsky, A.~V. ~Radyushkin, Phys. Rep. \textbf{418} (2005) 1;
		K. ~Goeke, M.~V. ~Polyakov, M. ~Vanderhaeghen, Prog. Part. Nucl. Phys. \textbf{47} (2001) 401.
		
		
		
		\bibitem{Mukherjee:2011bn}
		A.~Mukherjee and S.~Nair,
		Phys. Lett. B \textbf{706}, 77-81 (2011).
		
		
		\bibitem{Mukherjee:2011an}
		A.~Mukherjee and S.~Nair,
		Phys. Lett. B \textbf{707}, 99-106 (2012).
		
		
		\bibitem{Mukherjee:2013yf}
		A.~Mukherjee, S.~Nair and V.~Kumar Ojha,
		Phys. Lett. B \textbf{721}, 284-289 (2013).
		
		
		\bibitem{Angeles-Martinez:2015sea}
		R.~Angeles-Martinez \textit{et al.}
		Acta Phys. Polon. B \textbf{46}, no.12, 2501-2534 (2015).
		
		
		\bibitem{Barone:2001sp}
		V.~Barone, A.~Drago and P.~G.~Ratcliffe,
		Phys. Rept. \textbf{359}, 1-168 (2002).
		
		
		\bibitem{Accardi:2012qut}
		A.~Accardi \textit{et al.}
		Eur. Phys. J. A \textbf{52}, no.9, 268 (2016).
		
		
		
		\bibitem{Brodsky:2002cx}
		S.~J.~Brodsky, D.~S.~Hwang and I.~Schmidt,
		Phys. Lett. B \textbf{530}, 99-107 (2002).
		
		
		
		\bibitem{Bacchetta:2017gcc}
		A.~Bacchetta, F.~Delcarro, C.~Pisano, M.~Radici and A.~Signori,
		JHEP \textbf{06}, 081 (2017)
		[erratum: JHEP \textbf{06}, 051 (2019)].
		
		
		
		\bibitem{Ralston:1979ys}
		J.~P.~Ralston and D.~E.~Soper,
		Nucl. Phys. B \textbf{152}, 109 (1979).
		
		\bibitem{Donohue:1980tn}
		J.~T.~Donohue and S.~A.~Gottlieb,
		Phys. Rev. D \textbf{23}, 2577-2580 (1981).
		
		
		\bibitem{Tangerman:1994eh}
		R.~D.~Tangerman and P.~J.~Mulders,
		Phys. Rev. D \textbf{51}, 3357-3372 (1995).
		
		
		
		
		
		\bibitem{Hoodbhoy:1988am}
		P.~Hoodbhoy, R.~L.~Jaffe and A.~Manohar,
		Nucl. Phys. B \textbf{312}, 571-588 (1989).
		
		
		\bibitem{Hino:1999qi}
		S.~Hino and S.~Kumano,
		Phys. Rev. D \textbf{60}, 054018 (1999).
		
		\bibitem{Bacchetta:2000jk}
		A.~Bacchetta and P.~J.~Mulders,
		Phys. Rev. D \textbf{62}, 114004 (2000).
		
		
		
		\bibitem{Ninomiya:2017ggn}
		Y.~Ninomiya, W.~Bentz and I.~C.~Clo\"et,
		Phys. Rev. C \textbf{96}, no.4, 045206 (2017).
		
		
		
		\bibitem{blfq1}
		J.~P.~Vary, H.~Honkanen, J.~Li, P.~Maris, S.~J.~Brodsky, A.~Harindranath, G.~F.~de Teramond, P.~Sternberg, E.~G.~Ng and C.~Yang,
		Phys. Rev. C \textbf{81}, 035205 (2010).
		
		
		
		\bibitem{brodsky1}
		S.~J.~Brodsky, H.~C.~Pauli and S.~S.~Pinsky, Phys. Rept. {\bf 301} (1998) 299.
		
		
		
		\bibitem{maris}
		H. Honkanen, P. Maris, J. P. Vary and S. J. Brodsky, Phys. Rev. Lett. {\bf 106} (2011) 061603.
		
		\bibitem{zhao}
		X. Zhao, H. Honkanen, P. Maris, J. P. Vary and S. J. Brodsky, Phys. Lett. B {\bf 737} (2014) 65.
		
		\bibitem{li}
		P. Wiecki, Y. Li, X. Zhao, P. Maris and J. P. Vary, Phys. Rev. D {\bf 91} (2015) 105009.
		
		\bibitem{zhao2}
		D. Chakrabarti, X. Zhao, H. Honkanen, R. Manohar, P. Maris and J. P. Vary, Phys. Rev. D {\bf 89} (2014) 116004.
		
		
		
		
		\bibitem{Hu:2020arv}
		Z.~Hu, S.~Xu, C.~Mondal, X.~Zhao and J.~P.~Vary,
		Phys. Rev. D \textbf{103}, 036005 (2021).
		
		
		\bibitem{Nair:2022evk}
		S.~Nair , C.~Mondal, X.~Zhao, A.~Mukherjee and J.~P.~Vary,
		Phys. Lett. B \textbf{827}, 137005 (2022).
		
		
		
		
		\bibitem{Jia:2018ary}
		S.~Jia and J.~P.~Vary,
		Phys. Rev. C \textbf{99}, no.3, 035206 (2019).
		
		
		
		\bibitem{Lan:2019vui}
		J.~Lan, C.~Mondal, S.~Jia, X.~Zhao and J.~P.~Vary,
		Phys. Rev. Lett. \textbf{122}, no.17, 172001 (2019).
		
		
		\bibitem{Lan:2019rba}
		J.~Lan, C.~Mondal, S.~Jia, X.~Zhao and J.~P.~Vary,
		Phys. Rev. D \textbf{101}, no.3, 034024 (2020).
		
		
		\bibitem{Adhikari:2021jrh}
		L.~Adhikari, C.~Mondal, S.~Nair, S.~Xu, S.~Jia, X.~Zhao and J.~P.~Vary,
		Phys. Rev. D \textbf{104}, no.11, 114019 (2021).
		
		\bibitem{Lan:2021wok}
		J.~Lan, K.~Fu, C.~Mondal, X.~Zhao and J.~P.~Vary,
		Phys. Lett. B \textbf{825}, 136890 (2022).
		
		
		\bibitem{Mondal:2021czk}
		C.~Mondal, S.~Nair, S.~Jia, X.~Zhao and J.~P.~Vary,
		Phys. Rev. D \textbf{104}, no.9, 094034 (2021).
		
		
		\bibitem{Li:2015zda}
		Y.~Li, P.~Maris, X.~Zhao and J.~P.~Vary,
		Phys. Lett. B \textbf{758}, 118-124 (2016).
		
		
		\bibitem{Li:2017mlw}
		Y.~Li, P.~Maris and J.~P.~Vary,
		Phys. Rev. D \textbf{96}, 016022 (2017).
		
		
		\bibitem{Li:2018uif}
		M.~Li, Y.~Li, P.~Maris and J.~P.~Vary,
		Phys. Rev. D \textbf{98}, no.3, 034024 (2018).
		
		
		\bibitem{Lan:2019img}
		J.~Lan, C.~Mondal, M.~Li, Y.~Li, S.~Tang, X.~Zhao and J.~P.~Vary,
		Phys. Rev. D \textbf{102}, no.1, 014020 (2020).
		
		
		\bibitem{Tang:2018myz}
		S.~Tang, Y.~Li, P.~Maris and J.~P.~Vary,
		Phys. Rev. D \textbf{98}, no.11, 114038 (2018).
		
		
		\bibitem{Tang:2019gvn}
		S.~Tang, Y.~Li, P.~Maris and J.~P.~Vary,
		Eur. Phys. J. C \textbf{80}, no.6, 522 (2020).
		
		
		\bibitem{Xu:2019xhk}
		C.~Mondal, S.~Xu, J.~Lan, X.~Zhao, Y.~Li, D.~Chakrabarti and J.~P.~Vary,
		Phys. Rev. D \textbf{102}, no.1, 016008 (2020).
		
		\bibitem{Xu:2021wwj}
		S.~Xu, C.~Mondal, J.~Lan, X.~Zhao, Y.~Li and J.~P.~Vary,
		Phys. Rev. D \textbf{104}, no.9, 094036 (2021).
		
		\bibitem{Liu:2022fvl}
		Y.~Liu, S.~Xu, C.~Mondal, X.~Zhao and J.~P.~Vary,
		Phys. Rev. D \textbf{105}, no.9, 094018 (2022).
		
		\bibitem{Hu:2022ctr}
		Z.~Hu, S.~Xu, C.~Mondal, X.~Zhao and J.~P.~Vary,
		Phys. Lett. B \textbf{833}, 137360 (2022).
		
		
		\bibitem{Peng:2022lte}
		T.~Peng, Z.~Zhu, S.~Xu, X.~Liu, C.~Mondal, X.~Zhao and J.~P.~Vary,
		Phys. Rev. D \textbf{106} (2022) no.11, 114040.
		
		\bibitem{Kuang:2022vdy}
		Z.~Kuang, K.~Serafin, X.~Zhao and J.~P.~Vary,
		Phys. Rev. D \textbf{105}, no.9, 094028 (2022).
		
		
		
		
		\bibitem{Karmanov:2008br}
		V.~A.~Karmanov, J.~F.~Mathiot and A.~V.~Smirnov,
		Phys. Rev. D \textbf{77}, 085028 (2008).
		
		
		\bibitem{Karmanov:2012aj}
		V.~A.~Karmanov, J.~F.~Mathiot and A.~V.~Smirnov,
		Phys. Rev. D \textbf{86}, 085006 (2012).
		
		
		
		
		\bibitem{Wiecki:2014ola}
		P.~Wiecki, Y.~Li, X.~Zhao, P.~Maris and J.~P.~Vary,
		Phys. Rev. D \textbf{91}, 10, 105009 (2015).
		
		
		
		
		
		\bibitem{Zhao:2014hpa}
		X.~Zhao,
		Few Body Syst. \textbf{56}, no.6-9, 257-265 (2015).
		
		
		\bibitem{Zhao:2014xaa}
		X.~Zhao, H.~Honkanen, P.~Maris, J.~P.~Vary and S.~J.~Brodsky,
		Phys. Lett. B \textbf{737}, 65-69 (2014).
		
		
		\bibitem{Chakrabarti:2014cwa}
		D.~Chakrabarti, X.~Zhao, H.~Honkanen, R.~Manohar, P.~Maris and J.~P.~Vary,
		Phys. Rev. D \textbf{89}, no.11, 116004 (2014).
		
		
		
		\bibitem{Brodsky:2004cx}
		S.~J.~Brodsky, V.~A.~Franke, J.~R.~Hiller, G.~McCartor, S.~A.~Paston and E.~V.~Prokhvatilov,
		Nucl. Phys. B \textbf{703}, 333-362 (2004).
		
		
		
		\bibitem{Brodsky:2000xy}
		S.~J.~Brodsky, M.~Diehl and D.~S.~Hwang,
		Nucl. Phys. B \textbf{596}, 99-124 (2001).
		
		
		
		
		
		\bibitem{Zhang:1993dd}
		W.~M.~Zhang and A.~Harindranath,
		Phys. Rev. D \textbf{48}, 4881-4902 (1993).
		
		
		
		
		\bibitem{kundu1}
		A.~Harindranath, R.~Kundu, W.~M.~Zhang, Phys. Rev. D \textbf{59} (1999) 094013.
		
		
		
		
		
		\bibitem{Berger:2014rva}
		C.~Berger,
		J. Mod. Phys. \textbf{6}, 1023-1043.
		
		
		
		\bibitem{fm}
		D.~Smith,
		ACM Trans. Math. Softw., \textbf{17}, 273–283, (1991).
		
		
		\bibitem{OPAL:1999rcd}
		G.~Abbiendi \textit{et al.} [OPAL],
		Eur. Phys. J. C \textbf{11}, 409-425 (1999).
	
	
		
		
		\bibitem{L3:1998ijt}
		M.~Acciarri \textit{et al.} [L3],
		Phys. Lett. B \textbf{438}, 363-378 (1998).
		
	
		\bibitem{PLUTO:1984gmq}
		C.~Berger \textit{et al.} [PLUTO],
		Z. Phys. C \textbf{27}, 249 (1985).
		
		
	\bibitem{CELLO:1983crq}
	H.~J.~Behrend \textit{et al.} [CELLO],
	Phys. Lett. B \textbf{126}, 384-390 (1983).
	
	
	
		\bibitem{DELPHI:1995fid}
		P.~Abreu \textit{et al.} [DELPHI],
		Z. Phys. C \textbf{69}, 223-234 (1996).
		
		
		
	
		
		%
		%
		%
		%
		%
		%
		%
		%
		%
		%
		%
		%
		%
		
		
		
		
	\end{thebibliography}
\end{document}